\documentclass[nofootinbib,aps,twocolumn,epsf]{revtex4-1}
\usepackage{graphicx}
\usepackage{amsmath}
\usepackage{mathrsfs}
\usepackage{amsfonts}
\usepackage{amssymb}
\usepackage[percent]{overpic}

\def\y{\'{\i}}

\begin{document}

\title{Two-particle correlations at high-energy nuclear collisions, peripheral-tube model revisited}
\author
{Yogiro Hama$^1$, Takeshi Kodama$^{2,3}$ and  Wei-Liang Qian$^{4,5,6}$}
\affiliation{
$^1$ Instituto de F\y sica, Universidade de S\~ao Paulo, C.P. 66318, 05315-970, S\~ao Paulo-SP, Brazil \\
$^2$ Instituto de F\y sica, Universidade Federal do Rio de Janeiro,
C.P. 68528, 21945-970, Rio de Janeiro-RJ , Brazil \\
$^3$ Instituto de F\y sica, Universidade Federal Fluminense, 24210-346, Niter\'oi-RJ, Brazil \\
$^4$ Escola de Engenharia de Lorena, Universidade de S\~ao Paulo, 12602-810, Lorena-SP, Brazil \\
$^5$ Faculdade de Engenharia de Guaratinguet\'a, Universidade Estadual Paulista, 12516-410, Guaratinguet\'a-SP, Brazil \\
$^6$ Center for Gravitation and Cosmology, School of Physical Science and Technology, Yangzhou University, 225002, Yangzhou, Jiangsu, China
}

\begin{abstract}
In this paper, we give an account of the peripheral-tube model, which has been developed to give an intuitive and dynamical description of the so-called ridge effect in two-particle correlations in high-energy nuclear collisions. 
Starting from a realistic event-by-event fluctuating hydrodynamical model calculation, we first show the emergence of ridge + shoulders in the so-called two-particle long-range correlations, reproducing the data. 
In contrast to the commonly used geometric picture of the origin of the anisotropic flow, we can explain such a structure dynamically in terms of the presence of high energy-density peripheral tubes in the initial conditions.
These tubes violently explode and deflect the near radial flow coming from the interior of the hot matter, which in turn produces a two-ridge structure in single-particle distribution, with approximately two units opening in azimuth. 
When computing the two-particle correlation, this will result in characteristic three-ridge structure, with a high near-side ridge and two symmetric lower away-side ridges or shoulders. 
Several anisotropic flows, necessary to producing ridge + shoulder structure, appear naturally in this dynamical description. 
Using this simple idea, we can understand several related phenomena, such as centrality dependence and trigger-angle dependence.
\vspace*{1.5cm}
\end{abstract}
%}}
\maketitle
%\iffalse

\section{Introduction}

One of the main characteristics of the high-energy nuclear collisions is the existence of large event-by-event fluctuations: the fluctuation of impact parameter, of multiplicity, of particle species, of momentum distribution, among others~\cite{hydro-review-04, hydro-review-05, hydro-review-06, hydro-review-08, hydro-review-09, hydro-review-10}.
This is quite natural because of the lumpy structure of the systems in high-energy collisions. 
Sometimes, one tries to describe an average trend of such a phenomenon, by using, for example, hydrodynamic models, starting from just a single set of average initial conditions. 
However, the importance of event-by-event fluctuations can be demonstrated, for instance, by using a straightforward one-dimensional Landau model~\cite{Landau1, Landau2, Landau-collection} and considering the energy fluctuation through the Interacting Gluon Model (IGM)~\cite{igm1,igm2}.
In~\cite{Paiva1997, Hama1997}, it is observed that the multiplicity rapidity distribution produced from the average initial conditions is quite different from the average of the fluctuating distributions. 
It is noted that individual fluctuating events possess similar rapidity distributions, and the differences are related by the scaling in energy.

Stimulated by this result, efforts have been made to search for realistic initial conditions (IC) and to construct an improved version of hydrodynamic code that is capable of faithfully accounting for the complexity of heavy-ion collisions. 
As for the IC, the Regge theory based NEXUS generator~\cite{NEXUS}\footnote{
The NEXUS code was updated recently to EPOS~\cite{epos-1} to improve the particle production feature, but in the present discussion, the NEXUS version is sufficient.} is employed. 
Regarding the hydrodynamics code, we started from a variational formulation~\cite{Elze1,Elze2}, and used a method known as {\it Smoothed Particle Hydrodynamics} (SPH), first introduced for astrophysical applications~\cite{sph1,sph2}.
Moreover, this numerical method is extended to heavy-ion collisions~\cite{spherio1,spherio2}, and is denominated as (SPheRIO) {\bf S}{\it moothed} {\bf P}{\it article} {\bf h}{\it ydrodynamical} {\bf e}{\it volution of} {\bf R}{\it elativistic heavy} {\bf IO}{\it n collisions}.
As one combines the IC generated by NEXUS mentioned above with the hydrodynamical computation by SPheRIO code, it is referred to as NeXSPheRIO.

Throughout the present paper, we are going to use the by now well accepted event-by-event fluctuating hydrodynamics. 
Before proceeding further, we make a few comments. 
As one employs a hydrodynamical approach, the description of the system is realized in terms of thermodynamical quantities such as energy density, temperature, and entropy density.
Therefore, even when a set of smooth IC is adopted, one still encounters fluctuations originated from the statistical ensemble~\cite{Paiva1997, Hama1997}.
In the present paper, however, we do not explicitly consider this kind of fluctuations.
What will be focused on, as mentioned above, are those differences which appear ``macroscopically" in each collision of the incident objects, described by a set of different IC. 
In this regard, each fluctuating event evolves dynamically, in its own way, while obeying the same general and causal laws. 
Moreover, for each event, if specific singular but similar local objects appeared initially, they would persist during its independent evolution.
Subsequently, one might expect the relevant underlying physics to be captured through a careful analysis of the collectivity in final state observables of individual events.

Concerning NeXSPheRIO, as one may notice immediately (see FIGS.~\ref{fic} and \ref{fic2} and the corresponding text in the next section), that the matter distribution produced by NEXUS IC is usually lumpy, featured by many longitudinal high-energy tubes.
These high-energy regions are not color flux tubes, but fluctuations in the transverse plane {\it extruded out} in the longitudinal direction.
As one proceeds to investigate the physical quantities by using NeXSPheRIO on an event-by-event basis, we noticed the importance of the above mentioned lumpy structure in certain aspects of the data.
In particular, those appear in the two-particle interferometry~\cite{sph-hbt-01} and elliptical flows~\cite{granular}. 
For the latter, it is understood that the role of such tubes, or hot spots in the transverse plane, becomes essential if they are peripheral, or located close to the surface. 
Indeed, one could reproduce~\cite{Jun-Nu}, by using NeXSPheRIO, also the so-called {\it ridge effect} in the two-particle correlation data~\cite{Putschke, McCumber}. 
Subsequently, in an attempt to address how ridges are created in NeXSPheRIO, we recognize the dynamical significance of such peripheral tubes.

The primary purpose of this paper is to give an account of the {\it peripheral-tube model}.
Within the scope of the hydrodynamic picture, the manuscript is developed to give a dynamical and global vision of the so-called ridge phenomenon in the two-particle correlation at high-energy.
Our plan for the presentation is the following. 
In the next section, we start by giving a short account of NeXSPheRIO together with some relevant numerical results.
Then, in Section III, we further argue that hydrodynamical simulations' apparent success can be attributed to the importance of high energy tubes in the IC, mainly peripheral ones.
In Section IV, we first show that NeXSPheRIO can reproduce the so-called long-range two-particle correlation, which appears as ridges in $\eta$ direction and three characteristic peaks in $\phi$, when projected in the transverse plane. 
Then, in order to understand how such a structure appears dynamically in high energy collisions, we introduce~\cite{sph-corr-02} a simplified {\it single-peripheral-tube model}.
It is aiming to single out the main characteristic of the IC, minimally needed to producing the ridge+shoulders feature. 
The parameter dependence of the model is also discussed.
Subsequently, we extend it to {\it peripheral multitube model} by showing that the physical origin for producing the final three-ridge structure is each of the random tubes rather than the IC's global aspect.
Besides, some comparisons are given regarding the difference between the peripheral-tube model and the widely used eccentricities$\rightarrow$flow-harmonics interpretation.
Section V is devoted to the applications of the model, such as trigger-angle dependence and centrality dependence. 
Conclusions and outlooks are given in the last section. 
Throughout this paper, we primarily limited ourselves to the case of ideal hydrodynamics to avoid complexity.
Also, for the comprehensive reason, NEXUS and SPheRIO are employed for most of the present study.
Evidently, other hydrodynamical models based on different IC should produce similar results, so that the present findings hold in a more general context.
This aspect is partly explored in the Appendix.

\section{NeXSPheRIO as a prototype of realistic fluctuating hydrodynamic model}

In this section, we describe the basic tool, NeXSPheRIO code.
We then demonstrate that it is well suited for realistic high-energy nuclear collision simulations by discussing some preliminary results.
In a previous survey~\cite{topics}, we accounted for in detail how we obtain event-by-event fluctuating IC, starting from events generated by NEXUS~\cite{NEXUS}, so here we will just show how such IC look like.
In FIGS.~\ref{fic} and \ref{fic2} below, we take as an example, a central Au+Au collision at 130 A GeV and plot the energy-density distribution at mid-rapidity $\eta=0$ plane, at the initial time $\sqrt{t^2-z^2}\sim1$fm, in two different ways. 
As mentioned in the Introduction, and remarked by the authors of Ref.~\cite{NEXUS},
one sees many hot spots, typically of nucleon size.
\begin{figure}[!htb]
\begin{center}
\includegraphics*[angle=-90, width=8.5cm]{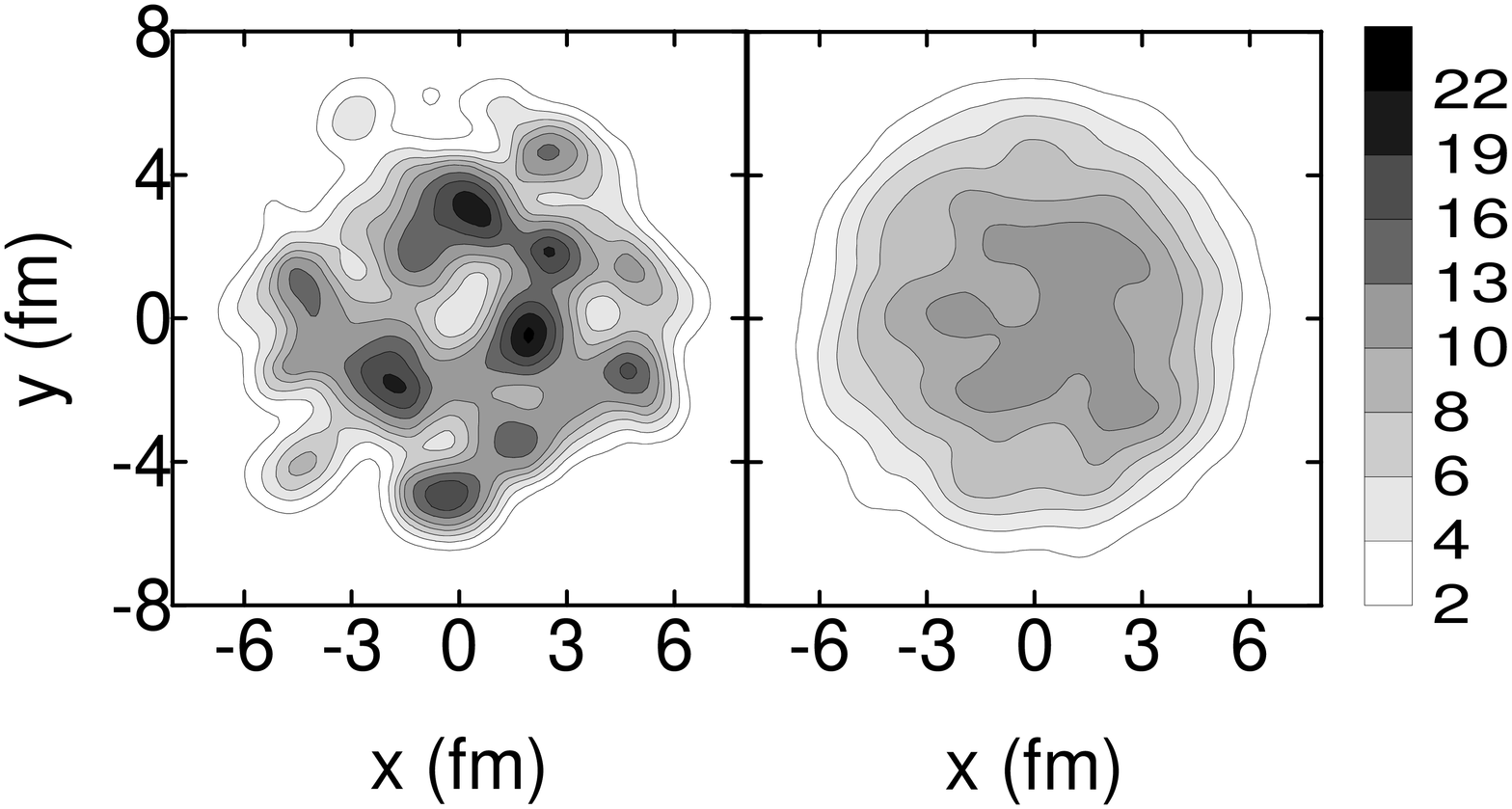}
\end{center}
\vspace*{-.5cm}
\caption{Examples of initial conditions for central Au+Au collisions at 130 A GeV given by NEXUS at mid-rapidity plane. 
The energy density is plotted in units of GeV/fm$^3$. 
Left: one random event. 
Right: average over 30 random events (corresponding to the smooth initial conditions as often done in hydrodynamical approach).}
\label{fic}
\end{figure}

%\begin{figure}[!htb]
%\vspace*{-2.cm}
%\begin{center}
%\includegraphics*[angle=270, hoffset=-10, hscale=33, vscale=33]{fig-sec2-condin3}
%\end{center}
%\vspace{5.cm}
%\caption{A different representation of the same IC shown in FIG.~\ref{fic}, at mid-rapidity plane. 
%The vertical axis represents the energy density in units of GeV/fm$^3$. }
%\label{fic2}
%\end{figure}
\begin{figure}[thb]
\vspace*{-2.cm}
\begin{center}
\includegraphics{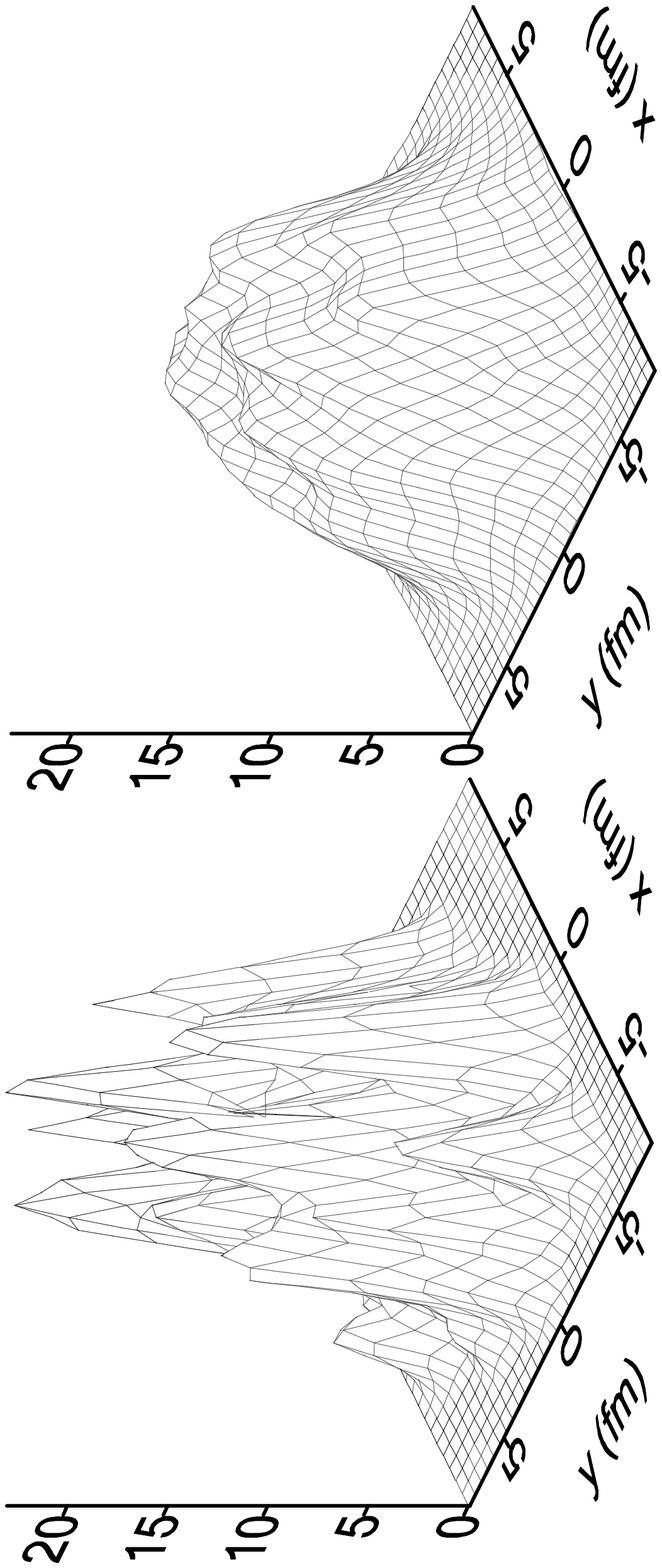}
\end{center}
\vspace{5.cm}
\caption{A different representation of the same IC shown in FIG.~\ref{fic}, at mid-rapidity plane. 
The vertical axis represents the energy density in units of GeV/fm$^3$. }
\label{fic2}
\end{figure}
We note that these hot spots are actually hot tubes, and fluctuations are also present in the initial velocity distribution.  
A more detailed discussion of these features of the initial state will be given in Section IV.
Moreover, if the thermalization is attained at a very early time, it is quite natural that the IC be very lumpy because the colliding objects (nuclei) are made of blobs (nucleons) separated from each other by transverse distances comparable or larger than their own sizes.

Once the IC are given, we solve the hydrodynamic equations by using the SPheRIO code, as mentioned in the Introduction. 
It is a code based on the so-called SPH method, which parametrizes the flow in terms of discrete  Lagrangian coordinates attached to small volumes (called ``particles'') with some conserved quantities. 
These quantities, in the case of ideal hydrodynamics, consists of entropy, strangeness, and baryon number. 
The algorithm's main advantage is that any geometry in the initial conditions can be incorporated while achieving the desired precision.
The details of this method, adapted for our purpose, have been described in Ref.~\cite{topics}.

Now, we have to specify some equation of state (EoS) describing the locally equilibrated matter. 
Here, in accordance with Ref.~\cite{qm05}, we will adopt a phenomenological implementation of EoS, giving a critical endpoint in the QGP-hadron gas transition line, as suggested by the lattice QCD~\cite{lattice-01,lattice-02,lattice-12}.
As mentioned in the Introduction, any dissipative effect is being neglected, and also usual sudden freeze-out at a constant temperature will be applied for simplicity.
Among various freeze-out criteria, such as ``continuous emission"~\cite{sph-ce-01} and chemical freeze-out~\cite{sph-cfo-01}, the Cooper-Frye formula is mostly adopted.
For both the EoS and hadronization, one considers resonance with a mass up to 2.459 GeV  (P wave D meson).
In the study of particle correlation, the hadrons are sampled in a Monte Carlo fashion.
Subsequently, the resonance decay is handled by NEXUS.

To compute the observables, NeXSPheRIO described here is run many times, corresponding to many different events or ICs. 
In the end, an average over final results is performed. Such a procedure is now commonly applied in the event-by-event hydrodynamics with fluctuating initial conditions.
See~\cite{hydro-vn-02, hydro-vn-03, hydro-v3-08, hydro-eve-04, hydro-eve-06, hydro-vn-04} and references therein.

Now that the tool is described, let us explain how we fix the parameters of the model and compute the observables of interest~\cite{sph-v2-fluct, sph-v2-fluct2}.
\begin{figure}[htb]
\begin{center}
\includegraphics*[width=8.5cm]{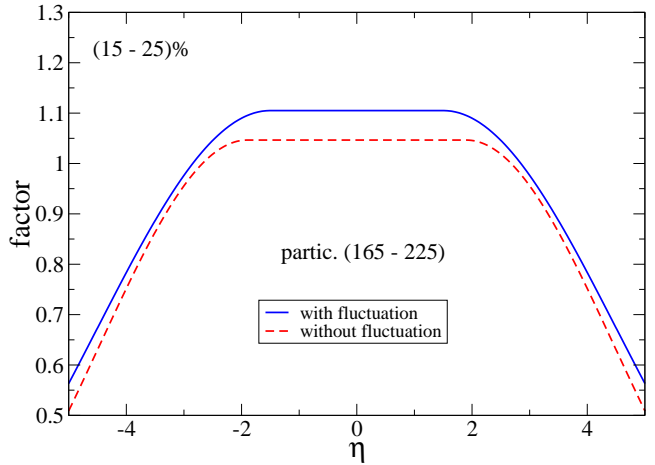}
\end{center}
\vspace*{-.2cm}
\caption{Example of $\eta$-dependent factor which was used to multiply the energy density given by NEXUS in the initial conditions for collisions at $\sqrt{s_{NN}}=200$ GeV. 
Remark that, once the centrality is chosen, the same factor is applied for all the fluctuating events.
}
\label{rf}
\end{figure}
First of all, one of the most important parameters to characterize an event of nucleus-nucleus collisions is the impact parameter or centrality. 
However, experimentally it is not possible to control this quantity a priori. 
Usually, the centrality bin of an event is determined probabilistically using the correlation between ZDC energy and multiplicity. 
On the other hand, in our code, it is possible to determine the number of participant nucleons in each event, which is closely related to the experimental centrality.
In specific cases, they may give rise to nontrivial effects~\cite{b-part} due to the fluctuations.
In the calculations presented below, however, we still utilize the impact parameter to determine the centrality class.

%First of all, since it is impossible to know the impact parameter in experiments, we use some quantity, which in principle can be experimentally determined to define the centrality. 
%For instance, in the code, it is possible to determine in each event the number of participant nucleons, which is intimately related to often used ZDC energy. 
%Although the participant number is closely connected to the impact parameter. 
%It is not the same~\cite{b-part} due to the fluctuation.
%\smallskip
\begin{figure}[htb]
\begin{center}
\includegraphics*[width=8.5cm]{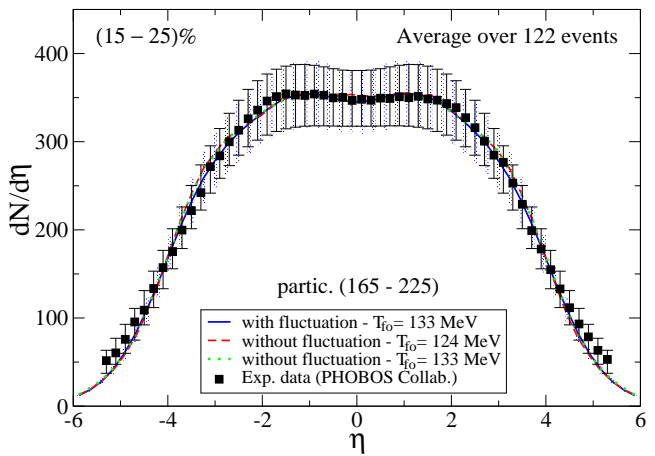}
\end{center}
\caption{Results of pseudo-rapidity distributions at 200 A GeV, calculated with the NEXUS initial conditions as explained in the text. 
PHOBOS data~\cite{Phobos-1} are shown for comparison.}
\label{dndeta3}
\vspace*{-.2cm}
\end{figure}
Now, certainly, any model to be considered as such should reproduce the most fundamental, global quantities involving the class of phenomena for which it is proposed. 
So, we begin by fixing the IC to properly reproduce the (pseudo-)rapidity distributions of charged particles in each centrality window. 
This is done by applying an $\eta$-dependent factor $\sim1$ to the initial energy density distribution of all the events of each centrality class, produced by NEXUS.
An example of such factors is shown in FIG.~\ref{rf} for the centrality (15 - 25)\% class both for fluctuating IC and the averaged IC. 
We show the resultant pseudo-rapidity distributions in FIG.~\ref{dndeta3}. 
Here, {\it without fluctuation} means that the computation has been done for one event whose IC are the average of the same 122 fluctuating IC, used in the other case, except for the normalization factor shown in FIG.~\ref{rf}. 
Results with averaged IC are being shown in comparison to exhibit clearly the effects of fluctuating initial conditions. 
Observe that, to obtain the same multiplicity as shown in FIG.~\ref{dndeta3}, we have to start with a smaller average energy density in the case of averaged IC, as implied by the normalization factor of FIG.~\ref{rf}. 
This is a manifestation of the effect already discussed in Ref.~\cite{topics}. 
Another observation concerning FIG.~\ref{dndeta3} is that the freezeout temperature, $T_{fo}\,$, gives a negligible influence on the (pseudo-)rapidity distributions.

\begin{figure}[t!hb]
\begin{center}
\includegraphics*[width=8.5cm]{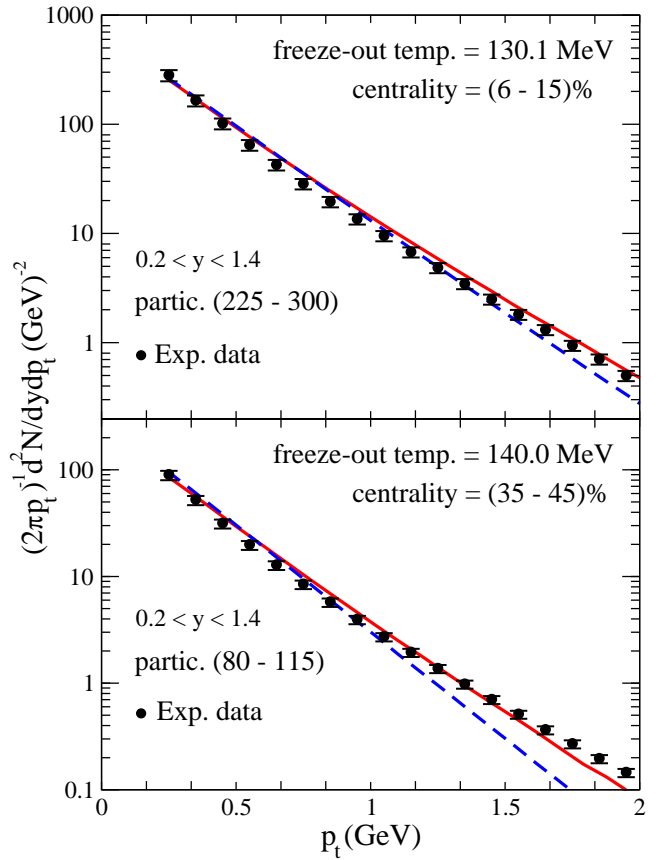}
\end{center}
\vspace*{-.5cm}
\caption{Charged-particle $p_T$ distributions at 200 A GeV, in two centrality windows, computed in two different ways as explained in the text. 
The red solid lines indicate results for event-by-event fluctuating IC, whereas the blue dashed lines the ones for the averaged IC. 
The data points are also plotted for comparison~\cite{Phobos-2}.}
\label{dndpt}
\vspace*{-.4cm}
\end{figure}

Next, we would like to correctly reproduce the transverse-momentum spectra of charged particles, which can be achieved by choosing an appropriate freezeout temperature, $T_{fo}\,$.
In FIG.~\ref{dndpt}, we show examples of choice with the corresponding spectra in two different centrality windows. 
One can see in this figure that the fluctuating IC make the transverse-momentum spectra more concave and closer to data.
The corresponding $\chi^2/\mathrm{nfs}$ values are found to decrease from 1.6 to 1.3 for the 6\% - 15\% centrality windows, and from 28 to 1.6 for the 35\% - 45\% centrality window.

\begin{figure}[h]
\begin{center}
\includegraphics*[width=7.5cm]{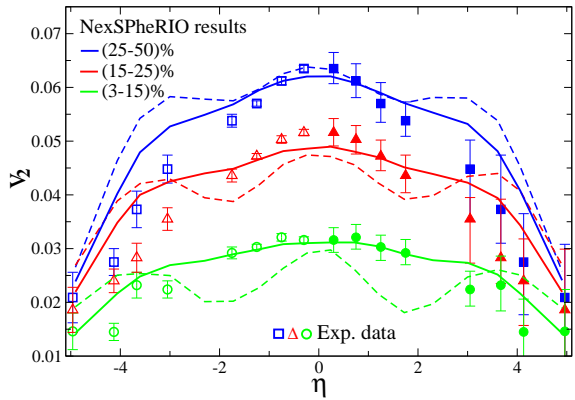}
\end{center}
\vspace*{-.5cm}
\caption{$\eta$ dependence of $\langle v_2\rangle$ for three centrality windows. 
The solid lines indicate results for event-by-event fluctuating IC, whereas the dotted lines the ones for the averaged IC. 
The data points~\cite{Phobos} are also plotted for comparison. 
$T_{fo}$ has been taken as indicated in FIG.~\ref{v2pt}.}
\label{v2eta}
\end{figure}

\begin{figure}[h]
\begin{center}
%\vspace*{-.2cm}
\includegraphics*[width=7.5cm]{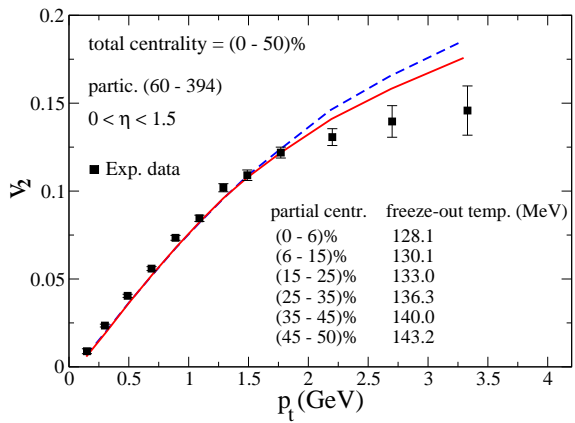}
\end{center}
\vspace*{-.5cm}
\caption{$p_T$ dependence of $\langle v_2\rangle$ in the centrality window and $\eta$ interval as indicated, compared with data~\cite{Phobos}. 
The solid line indicates result for fluctuating IC, whereas the dotted one that for the averaged IC. 
The curves are averages over PHOBOS centrality sub-intervals with freeze-out temperatures as indicated.
 }
\label{v2pt}
\end{figure}

Once all the parameters have been fixed now, let us see results on some other variables~\cite{sph-v2-fluct,sph-v2-fluct2,granular}.
In FIG.~\ref{v2eta}, we show the pseudo-rapidity distributions of $v_2$ for charged particles calculated in three centrality windows as indicated. 
It is seen that they reasonably reproduce the overall behavior of the existing data, both the centrality and the $\eta$ dependences.
For average IC, the obtained $v_2$ exhibits a pair of shoulders, in agreement with other hydrodynamical or transport model calculations~\cite{hydro-v2-hirano-04,hydro-v2-hirano-05,hydro-v2-nonaka-01}.
When fluctuating ICs are employed, the resulting curves become smoother, as compared to those with the averaged IC. 
The FIG.~\ref{v2pt} shows the transverse-momentum dependence of $v_2$ in the mid-rapidity region. 
Notice that the curve shows some additional bending at large-$p_T$ values, compared with the averaged IC case. 
These results will be discussed in more detail in the next section.
%\vspace*{-.5cm}

\begin{figure}[h!tb]
\begin{center}
\begin{overpic}[percent,width=15.cm]{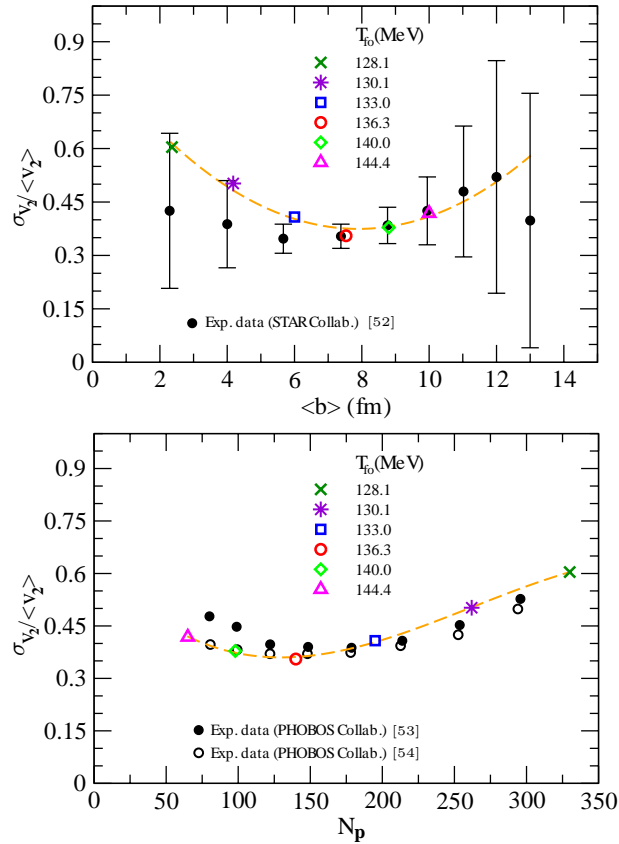}
%\put(5,18){\rotatebox{90}{$\sigma_{v_2}/\langle v_2\rangle$}}
%\put(5,54){\rotatebox{90}{$\sigma_{v_2}/\langle v_2\rangle$}}
\put(4.5,18){\rotatebox{90}{$\sigma$}}
\put(4.5,56){\rotatebox{90}{$\sigma$}}
\put(35.8,47.3){{\tiny \cite{v2-fluct1}}}
\put(38,11){{\tiny \cite{v2-fluct2}}}
\put(38,9){{\tiny \cite{v2-fluct3}}}
\end{overpic}
\end{center}
\caption{$\sigma_{v_2}/\langle v_2\rangle$ computed for $Au+Au$ collisions at $200\,A\,$GeV, compared with data.
In the upper panel, $\sigma_{v_2}/\langle v_2\rangle$ is given as function of the impact parameter $\langle b\rangle$ and compared with the STAR data~\cite{v2-fluct1}. 
The same results are expressed in the lower panel as a function of participant nucleon number $N_p$ and compared with the PHOBOS data~\cite{v2-fluct2,v2-fluct3}. 
The latter is obtained by the event-plane method~\cite{v2-fluct2} (filled circles) and a hybrid Glauber Monte-Carlo approach~\cite{v2-fluct3} (empty circles), respectively.
The size of the uncertainties of the PHOBOS data is mostly smaller than that of the symbols and thus not explicitly depicted in the lower panel.
}
\label{flv2}
\end{figure}

Let us turn to the results on $v_2$ fluctuation. 
In Refs.~\cite{spherio-v2-fluct,spherio-v2-fluct2}, for $Au+Au$ collisions at $130\,$GeV, we showed that fluctuations of $v_2$ are quite large.
When $\sigma_{v_2}/\langle v_2\rangle$ data were obtained in $200\,$GeV $Au+Au$ collisions~\cite{v2-fluct1,v2-fluct2,v2-fluct3}, they showed good agreement with the above results when QGP is included.
In Refs.~\cite{sph-v2-fluct,sph-v2-fluct2}, we further evaluate $\sigma_{v_2}/\langle v_2\rangle$ at the collision energy where the measurements were performed.
We show the results of this check in FIG.~\ref{flv2}. 
The freeze-out temperature has been chosen as explained above and increases with the impact parameter $\langle b\rangle$ (decreases with the participant nucleon number $N_p$).
In other words, the fluid decouples hotter and hotter as one goes from more central to more peripheral collisions as expected. 
Also, these temperature values have been used and then averaged over the partial windows in computing the $p_T$ distribution of $v_2$ shown in FIG.~\ref{v2pt}.
The curves of $v_2$ fluctuations, plotted in FIG.~\ref{flv2}, indicate that the NeXSPheRIO results remain in reasonable agreement with the data.

%\begin{figure}[h!tb]
%\vspace*{-.2cm}
%\begin{center}
%\hspace*{-.5cm}
%\includegraphics*[width=15.cm]{fig-sec2-flv2}
%\end{center}
%\vspace*{-.4cm}
%\caption{$\sigma_{v_2}/\langle v_2\rangle$ computed for $Au+Au$ collisions at $200\,A\,$GeV, compared with data.
%In the upper panel, $\sigma_{v_2}/\langle v_2\rangle$ is given as function of the impact parameter $\langle b\rangle$ and compared with the STAR data~\cite{v2-fluct1}. 
%In the lower panel, the same results are expressed as a function of participant nucleon number $N_p$ and compared with the PHOBOS data~\cite{v2-fluct2,v2-fluct3}. 
%The latter is obtained by the event-plane method~\cite{v2-fluct2} and a hybrid Glauber Monte-Carlo approach~\cite{v2-fluct3}, respectively.
%}
%\label{flv2old}
%\end{figure}

In the model, the fluctuations of the observables appear mostly because of the IC fluctuations introduced by the NEXUS  generator~\cite{NEXUS}, in addition to small effects due to the freeze-out procedure related to the Monte-Carlo method. 
The latter is, however, usually made negligible with increasing Monte-Carlo events at the freezeout.

Then, let us go further and try to learn more about the high-density matter using this prototype of the realistic fluctuating hydrodynamic model.

\section{Importance of the high-density tubular structure in the initial conditions}

In this Section, we will discuss some basic results~\cite{granular}, which revealed the importance of the characteristic feature of the NEXUS IC, namely the presence of fluctuating high-energy-density peripheral tubes, shown in the previous Section.
As a matter of fact, such results mentioned above already appeared in the previous Section, and what we do here is to try to capture the origin of such results, namely which is the connection between the remarkable characteristic of Event-Generator type IC and the hydrodynamic results which compare well with data.

\begin{figure}[htb]
%\vspace*{-.3cm}
\begin{center}
\includegraphics*[width=7.cm]{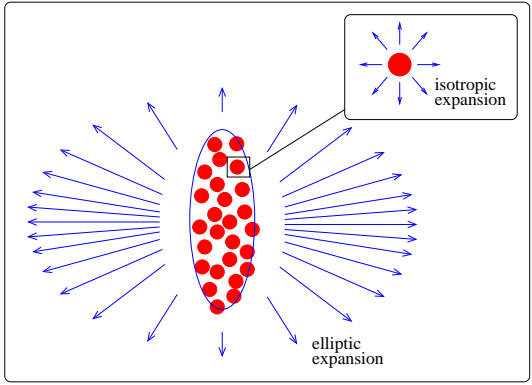}
\end{center}
\caption{Pictorial representation of the energy density distribution
in a fluctuating non-central IC.}
\label{blobs}
\end{figure}

In Ref.~\cite{granular}, we started by showing a pictorial representation of the energy-density distribution in some NEXUS IC, depicted in FIG.~\ref{blobs}, and asking {\it What is expected from the hot tubes} (or {\it blobs} in the transverse plane)? Upon answering this question, and then looking at the corresponding data to several observables, we could understand what they do and the importance of such high-energy-density peripheral tubes in NEXUS IC. Let us repeat this reasoning below.

Because of the high concentration of energy in {\it tube-like} regions, we imagine that each tube would initially suffer a violent explosion and, because of their small transverse size, isotropically in transverse directions.
If one such tube is deep inside the hot matter, its initial motion would be quickly absorbed by the surrounding medium, so it would not result in an appreciable observable effect. However, if such a tube is at the surface of the matter, the outgoing part of this initial acceleration would likely survive, producing high-$p_T$ particles, which would be isotropically distributed in the transverse plane.
\smallskip

{\bf Charged-particle spectra}--- Thus, first, we expect that the high-$p_T$ part of the $p_T$ spectra is enhanced when NEXUS IC are used in the computations, in comparison with the results with averaged (smooth) IC.
Such effects are clearly seen in FIG.~\ref{dndpt}$\,$ of the previous Section where, as one goes from the smooth averaged IC to lumpy fluctuating IC, the high-$p_T$ component of the spectra increases as expected, making them more concave and closer to the data~\cite{Phobos-2}.
\smallskip

{\bf$p_T$ dependence of $\langle v_2\rangle$}---In the second place, we would expect that the anisotropic-flow coefficient $v_2$ suffers reduction, as we go to high-$p_T$ region, due to the additional high-$p_T$ isotropic component.
FIG.~\ref{v2pt} of the previous Section exhibits this effect.
To be specific, as compared with the averaged IC case, the introduction of spiky IC makes the curve for the fluctuating IC more bending down at large-$p_T$ values, so closer to the data~\cite{Phobos}.
Here, since our purpose is to show clearly the effects of granular IC, we plot the results obtained with the same PHOBOS hit-based method for both curves, but without any correction for event-plane fluctuations.
The correction mentioned above shifts the curve upward,
corresponding to the fluctuating IC, but without modifying its shape.
The fact that the averaged smooth IC lead to rising $v_2(p_T)$ for large $p_T$ and not flattening is usually interpreted as a breakdown of the hydrodynamic model. We suggest it could be in part related to the granular structure of the IC.

{\bf $\eta$ dependence of $\langle v_2\rangle$}---As for the $\eta$ dependence of $v_2\,$, we know that the average matter density decreases as $\vert\eta\vert$ increases as reflected in the $\eta$ distribution of charged particles (see FIG.~\ref{dndeta3} of the previous Section).
In the case of NeXSPheRIO, the multiplication factor shown in FIG.~\ref{rf} further guarantees an appropriate matter density distribution in $\eta$.
When such a hot tube extends to the large-$\vert\eta\vert$ regions, its effects of reducing $v_2$ would appear more efficient there.
As a result, we would expect a considerable reduction of $v_2\,$ in those regions. FIG.~\ref{v2eta} of the previous Section indicates clearly this effect, making $v_2(\eta)$ smooth without peaks as exhibited in the computation with averaged IC.
\smallskip

It is encouraging to see that the description of the $p_T$ spectra, as well as $p_T$ and $eta$ dependence of $v_2$, improve when taking fluctuations of the IC into account.
\smallskip

As we proceeded the study of the relationship between several data and the characteristics of the ICs, we recognized the importance of the global shape and the emergence of high-energy tubular fluctuations.
In this stage, we studied only the direct effects of the flow produced by high-energy peripheral tubes. 
That is, all the issues mentioned above appear directly connected with the isotropic flow produced by high-energy tubes.
Besides, there is a substantial effect that peripheral tubes produce, which will be discussed in the next Section.

%\newpage

\section{Long-range two-particle correlation}

%In the previous Section, by using the NeXSPheRIO code, we explained how we studied several characteristics of the elliptic flow in comparison with the existing data obtained at RHIC. The code can reproduce the details of the existing data quite well, showing that the IC are perfectly realistic.
%In particular, we noticed that the existing lumpy structure, especially near the surface of the hot matter, is essential to get a good agreement with data.

In this section, we proceed to investigate the two-particle correlations in heavy-ion collisions. 
In practice, the two-particle correlations had often been studied in connection with jets, where particles appear to be correlated for small differences of $\phi$ and $\eta$. 
However, the present study mostly concerns the so-called long-range two-particle correlation, where the two chosen particles appeared separated far apart in pseudo-rapidity difference $\Delta\eta$. 
The above measurements were first reported by RHIC~\cite{Putschke, McCumber}, and many efforts have been devoted to the topic. 
The main feature of the data is as follows.
The {\it ridge} correlation appears on the near-side of the trigger, namely, $\Delta\phi\simeq0$ in the azimuthal direction, and it stretches for several units in $\Delta\eta$ direction.
At the same time, a symmetric double-ridge structure, often referred to as {\it shoulders}, is also observed on the away side from the trigger.
The present section is devoted to the numerical studies of the above long-range three-ridge correlation structure using the hydrodynamic approach and its possible interpretations.

\subsection{NeXSPheRIO results on long-range correlation}\label{4A}

The first step to study the long-range two-particle correlation was to compute it directly by using the NeXSPheRIO code~\cite{Jun-Nu}.
In this study, we generated on the order of 200 000 NeXSPheRIO events for $Au+Au$ collisions at $\sqrt{s_{nn}}=200\,$GeV, in two different centrality classes.
\begin{figure}[h!t]
\begin{center}
\includegraphics[width=8.0cm]{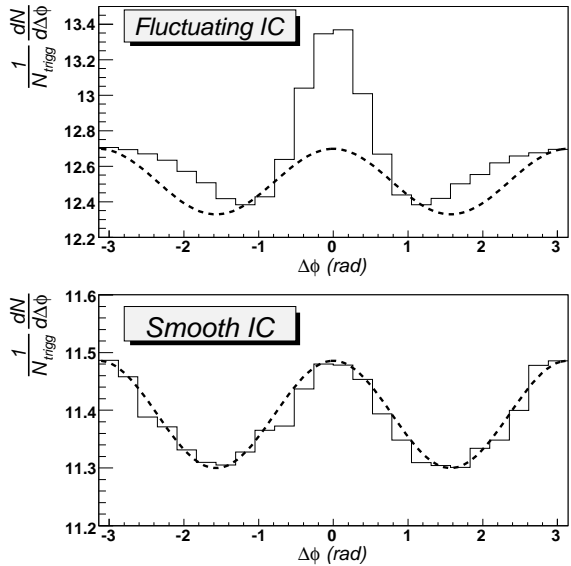}
\end{center}
\caption{Projection in $\Delta\phi\,$, for the $\Delta\eta$ range of $\pm0.5$.
The curve in the dashed line is the average flow contribution normalized using ZYAM~\cite{zyam-02} method.
This is from top 10\% $Au+Au$ collisions at $\sqrt{s_{NN}}=200\,$GeV.}
\label{4A1}
\end{figure}
The first set of events corresponded to the upper 10$\%$ of the total cross section and the second one to the fraction from 40$\%$ to 60$\%$.
Only the charged particles were considered in this analysis, in accordance with the available experimental data at RHIC. %\cite{star-ridge4}
We then computed the two-particle correlation with these NeXSPheRIO data by applying similar methods as used by the STAR experiment~\cite{RHIC-star-ridge-4}.
Accordingly, the trigger particles have been chosen with a threshold of $p_T>2.5\,$GeV, and associated particles, with a requirement of $p_T>1.0\,$GeV.

The average-flow subtraction has been done, first for $\Delta\eta$ direction, by using the event mixing technique. 
FIG.~\ref{4A1} shows the projection of the correlation, thus obtained, in the $\Delta\phi$ direction, integrating in the $\Delta\eta$ range of $\pm0.5$. 
The top plot corresponds to the projection histogram of the (0-10\%) central events, and the dashed curve represents the $v_2$ contribution normalized using the ZYAM (zero yield at minimum)~\cite{zyam-02} method. 
The $p_T$ dependence of $v_2$ was calculated based on the event-plane method as described in~\cite{event-plane-method-2}. 
The amplitude of the $v_2$ contribution is adjusted based on the mean $v_2$ value of the trigger, and the associated particles and the pedestal offset is equalizing the minimum yield of the correlation function with the averaged flow contribution.
It is clear from the comparison between the $\Delta\phi$ projection and the average elliptic flow contribution that there is an excess of the correlation yield suggesting that the correlation observed cannot be explained considering just the $v_2$ contribution. 
The same comparison was also made for events generated with the SPheRIO code using smooth IC (without fluctuations), and the equivalent result is shown in the bottom part of FIG.~\ref{4A1}. 
Here, the smooth IC were generated averaging over several NEXUS events, as shown in FIGS.~\ref{fic} and \ref{fic2}. 
Remark that the correlation function agrees with the $v_2$ curve, indicating that the anisotropy $v_2$ is the only contributor to the two-particle correlation function in this case.

\begin{figure}[h!t]
\vspace{-.3cm}
\begin{center}
\includegraphics[width=8.7cm]{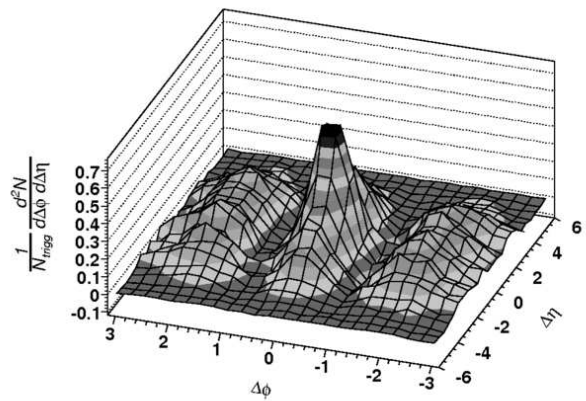}
\end{center}
\vspace{-.8cm}
\caption{Two-particle correlation in $\Delta\eta$ and $\Delta\phi\,$, computed with (0-10\%) central NeXSPheRIO data at $\sqrt{s_{nn}}=200\,$GeV, after subtracting the averaged flow contribution.}
\label{ridge}
\end{figure}

\begin{figure}[h!]
\begin{center}
\includegraphics[width=7.5cm]{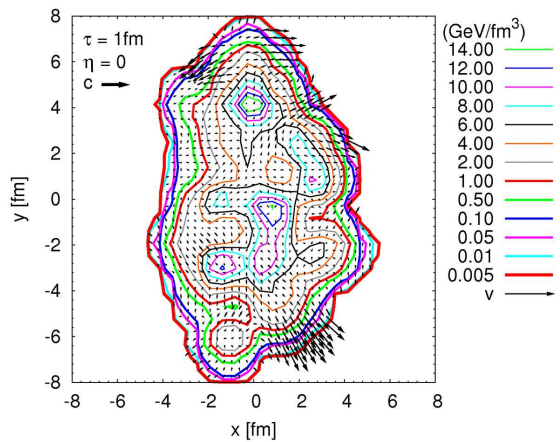}
\end{center}
\vspace{-.8cm}
\caption{NeXSPheRIO initial fluid velocity for a $Au+Au$ collision at  $\sqrt{s_{nn}}=200\,$GeV, in the 25-35\% centrality window.}
\label{jet}
\end{figure}
The subtraction of the average transverse flow, namely, the yield of particle pairs normalized by the number of triggers and $v_2$, has been done using ZYAM method, applied to each small interval of $\Delta\eta$.
The final result of the two-particle correlation is shown in FIG.~\ref{ridge}, 
where we can clearly see some excess of the correlation yield both in the near side of the trigger particle and also in the away side.
\begin{figure}[h!]
\begin{center}
\includegraphics[width=8.0cm]{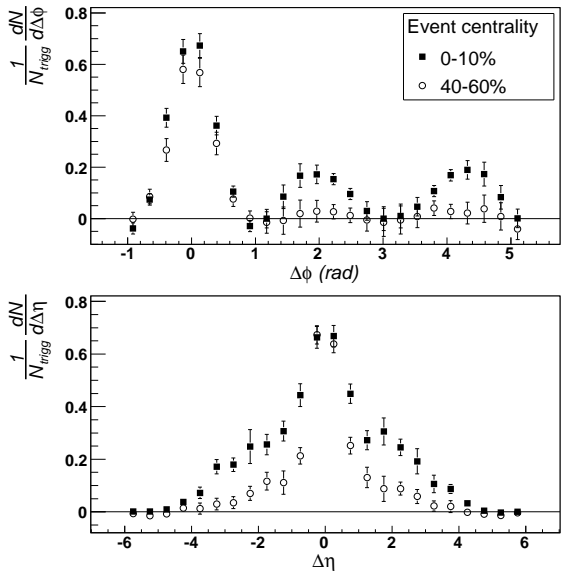}
\end{center}
\vspace{-.5cm}
\caption{$\Delta\phi\,$ and $\Delta\eta$ projections of the final NeXSPheRIO simulated data.
Solid squares represent the central collisions, whereas open circles the peripheral collisions. 
$\Delta\eta$ distributions are for near side ($-1.0<\Delta\phi<1.0$) only. 
The error bars include the statistical error of each bin and the error due to $v_2$ of the ZYAM method.}
\label{4A3}
\end{figure}
We note in FIG.~\ref{ridge} that the near side structure is composed of a ridge with a narrow width in $\Delta\phi\sim 0$ but extending over several units of pseudo-rapidity, and a peak in the middle, as seen experimentally.
On the away side, the correlation function also extends over several units of the rapidity with two symmetrical ridge-like structures close to  $\Delta\phi\sim\pm2$.
As for the peak at $\Delta\eta\simeq\Delta\phi\simeq0$, we can understand it in the following way.
This region is not correlated with a tube, so peak and ridge, appearing in FIG.~\ref{ridge}, are independent.
As explained in Ref.~\cite{sph-corr-02}, in this approach, the IC constructed by ``thermalizing" the NEXUS output~\cite{topics} do not explicitly involve jets. 
However, these are not totally forgotten in the IC as they leave some localized regions with higher transverse velocity, as shown in FIG.~\ref{jet}.

\begin{figure*}
\vspace{-1.5cm}
\includegraphics[angle=90,width=12.cm]{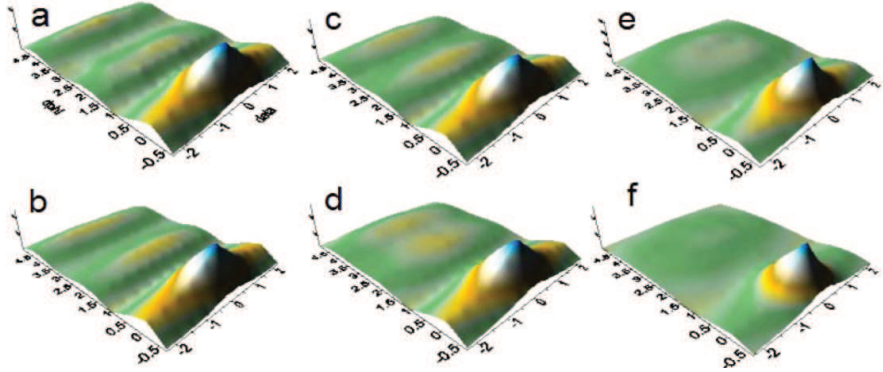}
\vspace{-1.8cm}
\caption{Two-particle correlation, as computed with the NeXSPheRIO code, for different centrality windows of Au+Au collisions at 200A GeV(a: 0-5\%, b: 5-10\%, c: 10-20\%, d: 20-30\%, e: 30-40\%,
f: 40-50\%).
The transverse momenta are chosen as $p_T^{trig}>2.5\,$GeV and $p_T^{ass}>1.5\,$GeV for triggers and associated particles, respectively.}
\vspace{-.3cm}
\label{CentDep}
\end{figure*}

We show in the top plot of FIG.~\ref{4A3}, the $\Delta\phi$ projection of the final two-particle correlation, integrated over $\Delta\eta$ range of $\pm1$, for events both of central (solid squares) and peripheral (open circles) collisions we considered. Both data sets show a clear narrow correlation peak (actually ridge in 3-dimensions) in the near side. On the away side, the central data show a double-peak (actually double-ridge) structure with a dip at $\Delta\phi=\pi$ that appears after the subtraction of the $v_2$ contribution, while the peripheral data show no clear correlation in the away side. Actually, this is strongly dependent on $v_2$ values.

In the second plot of FIG.~\ref{4A3}, we show the $\Delta\eta$ projection of the near side correlation for the same two different centrality classes shown above. 
The uncertainties in each point include each bin's statistical error and the uncertainty in the subtracted $v_2$ value. 
It is found that the residual correlation extends over several pseudo-rapidity units. 
This ridge structure in $\Delta\eta$ was also studied in the experimental data at RHIC by the STAR~\cite{Putschke} and PHENIX experiments.\cite{McCumber}. 
The observed feature extends up to two units of $\Delta\eta$ within the limits of the detector acceptance and has a sharp peak that corresponds to the near side jet correlation on top of the ridge structure.

In short, what we learned with this study, Ref.~\cite{Jun-Nu}, may be schematized as follows: 
{\it i}) NeXSPheRIO hydro code with fluctuating IC, containing high-density tubes, can produce ridge-like and shoulder-like structures in the two-particle correlations.
{\it ii}) Smooth IC (like average over many NEXUS events) $+$ SPheRIO do not.
{\it iii}) NEXUS only, without hydrodynamic evolution (SPheRIO), does not. So, briefly, {\it what produces these characteristic structures is the hydro expansion starting from fluctuating IC, containing hot tubes}.

\begin{figure}[h!]
\hspace*{-1.2cm}
\includegraphics[angle=90,width=10.5cm]{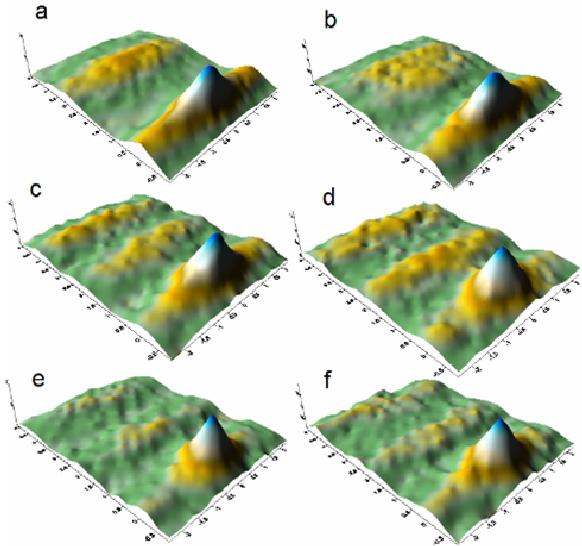}
%\vspace*{-.6cm}
\caption{\label{inout}Two-particle correlation for 20-30\%
centrality Au+Au collisions at 200A GeV, for different  %trigger-particle azimuthal angles
$\phi_t$
(a: $\phi_t=0^\circ-15^\circ$, b: $\phi_t=15^\circ-30^\circ$, c: $\phi_t=30^\circ-45^\circ$, d: $\phi_t=45^\circ-60^\circ$, e: $\phi_t=60^\circ-75^\circ$, f: $\phi_t=75^\circ-90^\circ$).
$p_T^{trig}>3.\,$GeV and $p_T^{ass}>1.\,$GeV.}
\vspace*{-.5cm}
\end{figure}

We mentioned above some questions about the centrality dependence of the away side ridge structure.
In order to observe more clearly such behavior, we have also computed these quantities, by using the NeXSPheRIO code with appropriate specification~\cite{sph-corr-03,sph-corr-02,sph-corr-ev-01,sph-corr-05}. 
We show the results in FIG.~\ref{CentDep} are in a qualitative agreement with the new PHENIX Collab. data~\cite{RHIC-phenix-ridge-5}.
In the same papers, we also calculated the trigger-angle ($\phi_t$) dependence, which has been measured by STAR Collab.~\cite{RHIC-star-plane-1} obtaining reasonable agreement (See FIG.~\ref{inout}). 
More quantitative studies, as well as discussions about possible mechanism of such behaviors have been achieved in~\cite{sph-corr-ev-02,sph-corr-ev-07,sph-corr-ev-04,sph-corr-ev-06,sph-corr-ev-08} and will be described in detail in Section V.

\subsection{How ridge+shoulders appear in NeXSPheRIO: Peripheral-Tube Model}

%In the preceding Subsection, the NeXSPheRIO code has shown to produce not only the ridge + shoulders structures~\cite{Jun-Nu}, but also some qualitative features as centrality dependence and trigger-angle effect~\cite{sph-corr-03,sph-corr-02,sph-corr-ev-01,sph-corr-05}.
%Now, the question is {\it how ridge + shoulders are produced}.
%In trying to answer this question we were lead in~\cite{sph-corr-03}, and in the following papers~\cite{sph-corr-02,sph-corr-05,sph-corr-04,sph-corr-06} to the central issue of this paper, namely the {\it peripheral-tube model}.

\begin{figure}[h!bt]
\begin{center}
%\hspace*{1.5cm}
\hspace{-0.5cm}
\includegraphics[width=8.5cm]{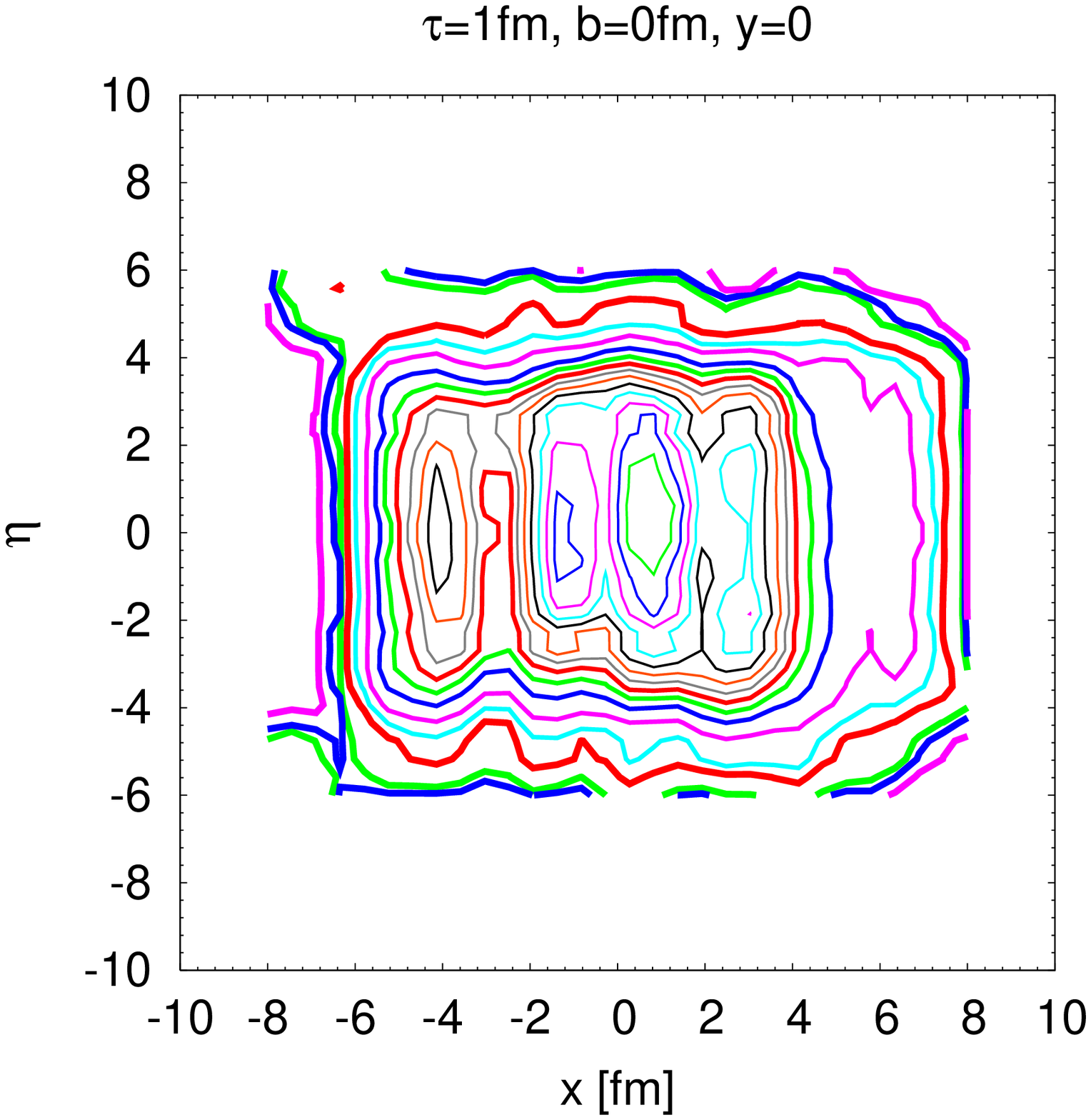}
\includegraphics[width=8.5cm]{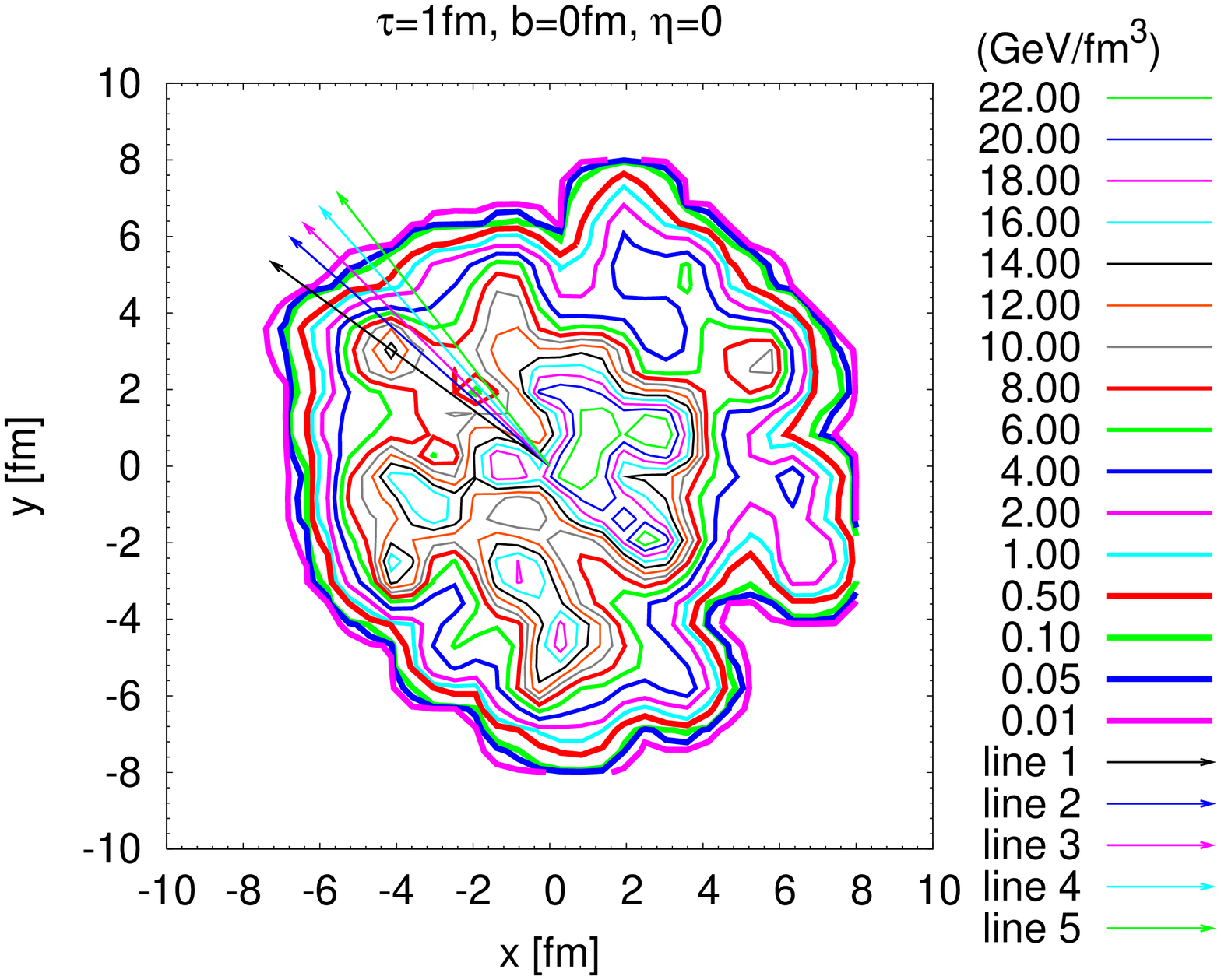}
%\vspace{.4cm}
\vspace{-.4cm}
\caption{\label{xyenergyic}Energy density distribution at $\tau=1$fm for a central Au+Au collision at $200A$ GeV, given by NEXUS generator, in the plane $y=0$ (up) and in the transverse plane $\eta=0$ (down).}
\end{center}
\end{figure}

As mentioned above, each event in the model is characterized by IC with many high-energy-density tubes (see FIG.~\ref{xyenergyic} and also FIGS.~\ref{fic},\ref{fic2} in Section II).
Also, as shown in subsection~\ref{4A}, expansion is necessary for producing the structures. Therefore one may naturally associate these tubes $+$ transverse expansion with the ridge structure. 
However, the phenomenon is not so trivial. 
Moreover, it is not clear {\it how} the away-side ridges (or shoulders) are generated. 
In order to clarify the basic mechanism of the formation of such structures, especially of the away-side one, we proceed to simplify the problem, avoiding unnecessary complications. 
So, to begin with, let us examine only the central collisions (see Section V, for peripheral collisions).

Consider, then, a typical event like the one depicted in FIG.~\ref{xyenergyic}, where we can see many tubes. 
In the previous Section, we discussed the importance of these tubes, especially when they are located close to the hot matter's surface. 
Let us focus, then, on one of such tubes, for instance, the one where the arrow 1 passes on in the lower plot. 
To carefully follow what happens in the neighborhood of this peripheral tube, one may replace the complex bulk of the hot matter, as shown there, by the average over many events, as shown in FIGS.~\ref{fic} and \ref{fic2}, right. 
This procedure seems reasonable because deep inside the hot matter, the initial motion would be quickly absorbed by the surrounding, making the matter more homogeneous. 
To be specific, we parametrize the energy density of an IC with a single peripheral ``tube" sitting top of the background and investigate the resultant hydrodynamical evolution, particularly in the vicinity of the peripheral tube.
Then, we used a 2D model with boost-invariant longitudinal expansion to make the computation easier. 
This procedure is natural if we observe the approximate translation invariance along $\eta$ of the IC.
\begin{figure}[h!]
%\vspace*{.6cm}
\includegraphics[width=6.5cm]{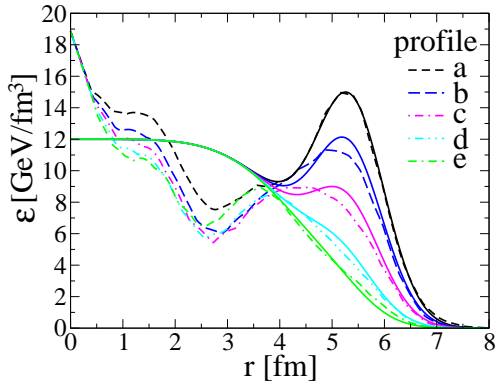}
\caption{Comparison of the parametrization given by Eq.~(\ref{par}) (solid lines) with the original energy density, as given in FIG.~\ref{xyenergyic} (dashed lines), along the lines 1-5.}
\label{lines}
\end{figure}

We, then, parametrized the energy density as
\begin{equation} 
\begin{aligned}
\epsilon=c_1\exp[-c_2 r^5] +c_3 \exp[-\frac{|{\bf r}-{\bf r}_0|^2}{c_4}]\ ,
  \label{par} 
\end{aligned}
\end{equation}
where the position of the peripheral tube $r_0=5.4\,$fm, $c_1=12\,$GeV/fm$^3$, $c_2=0.0004\,$fm$^{-5}$, $c_3\equiv \epsilon_t=12.8\,$GeV/fm$^3$, $c_4\equiv \Delta r=0.845\,$fm$^2$, and the initial velocity of the fluid as
\begin{equation}
  v_T=\tanh[4.57\exp(-27.2/r)]\ ,
  \label{parv}
\end{equation}
corresponding to the average NEXUS IC.
We show in FIG.~\ref{lines}, a comparison of this parametrization with the original energy-density distribution as given by the NEXUS event we are studying, namely, the one plotted in FIG.~\ref{xyenergyic}. 
Notice that, except for the inner region, the agreement is reasonable.

What did the high-energy-density tube produce, in conjunction with the background?
\begin{figure}[hbt]
\vspace{-.5cm}
\hspace{1.5cm}
\begin{center}
\includegraphics[width=6.3cm]{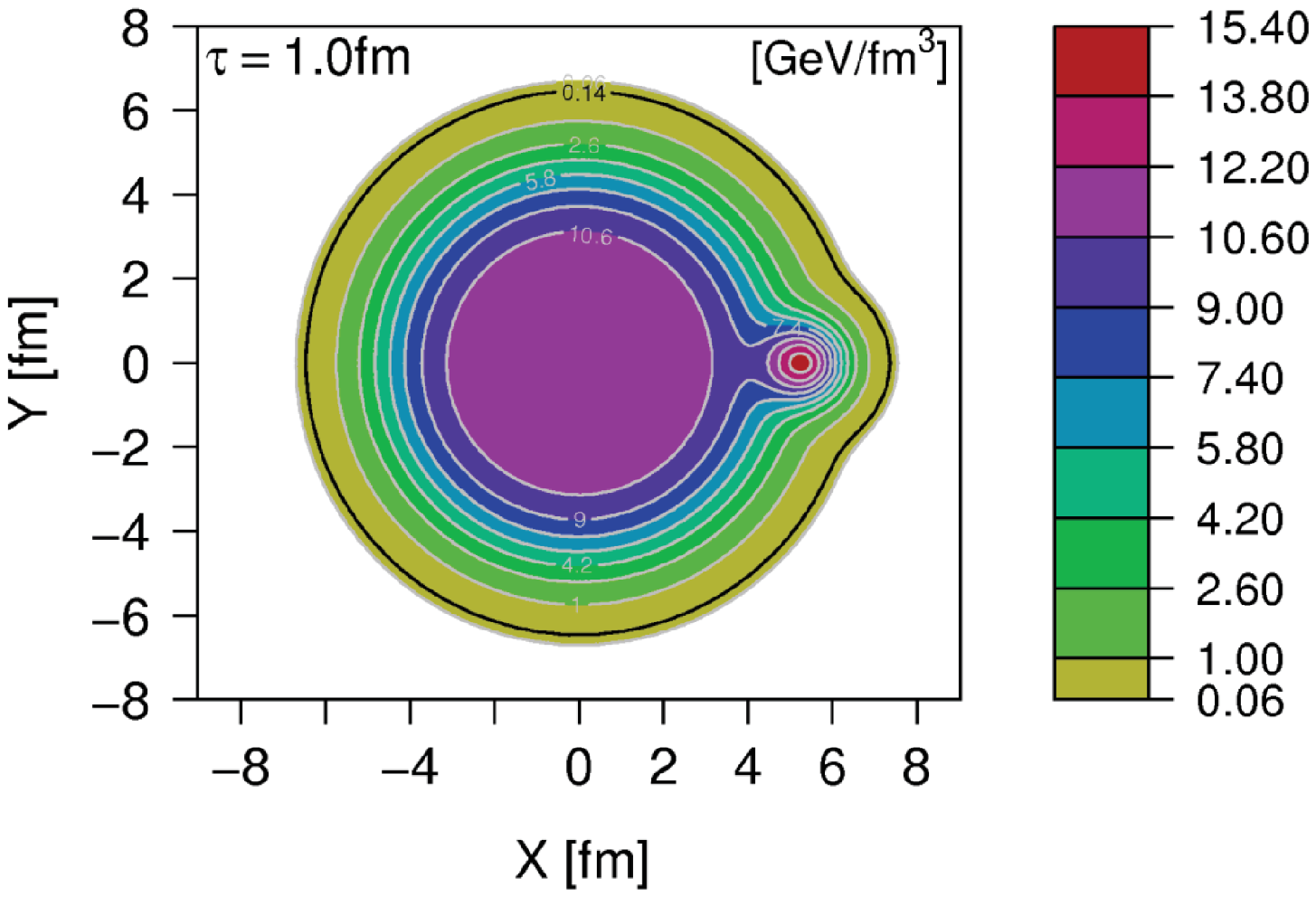}
\includegraphics[width=6.3cm]{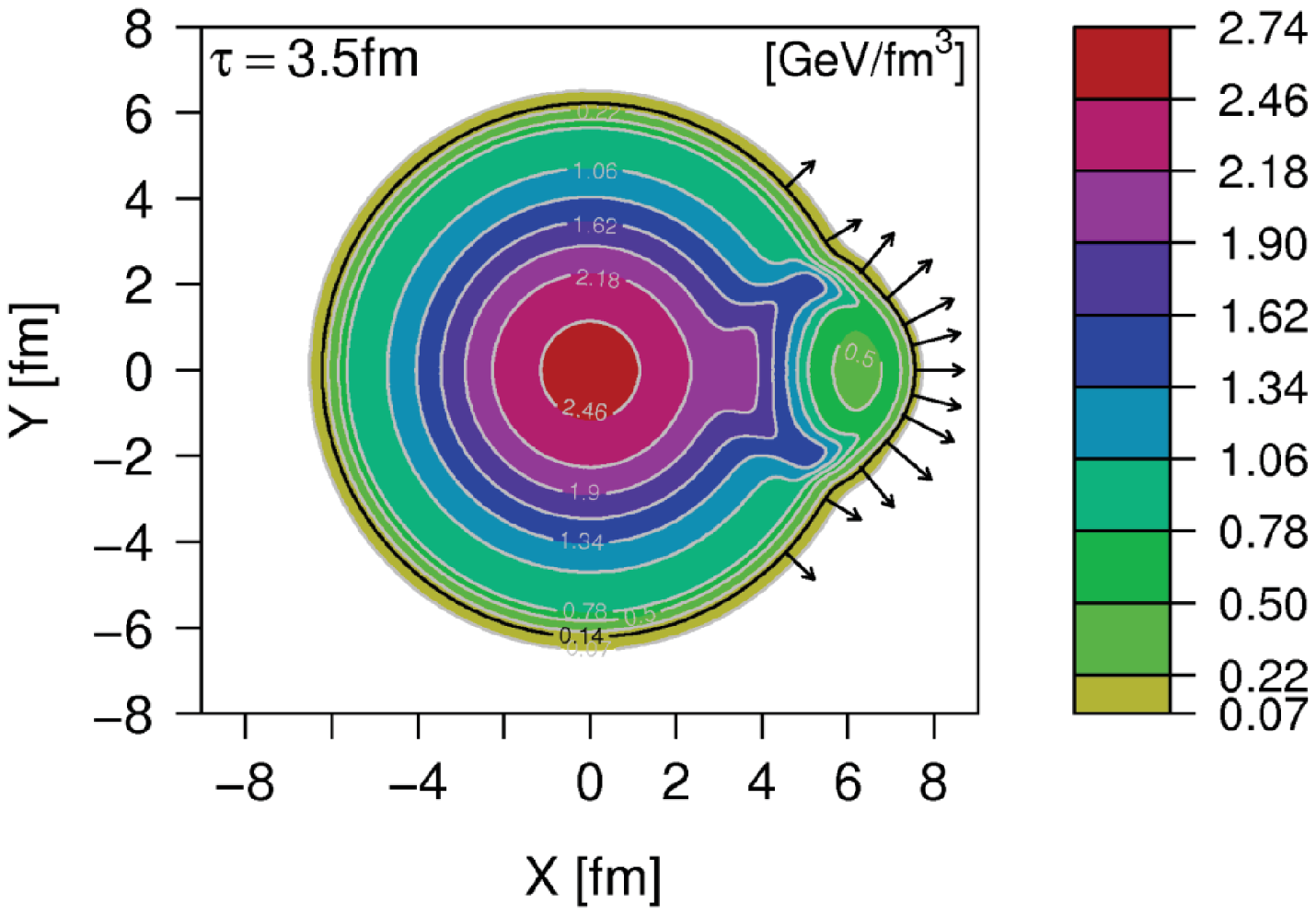}
\hspace{.5cm}
\includegraphics[width=6.3cm]{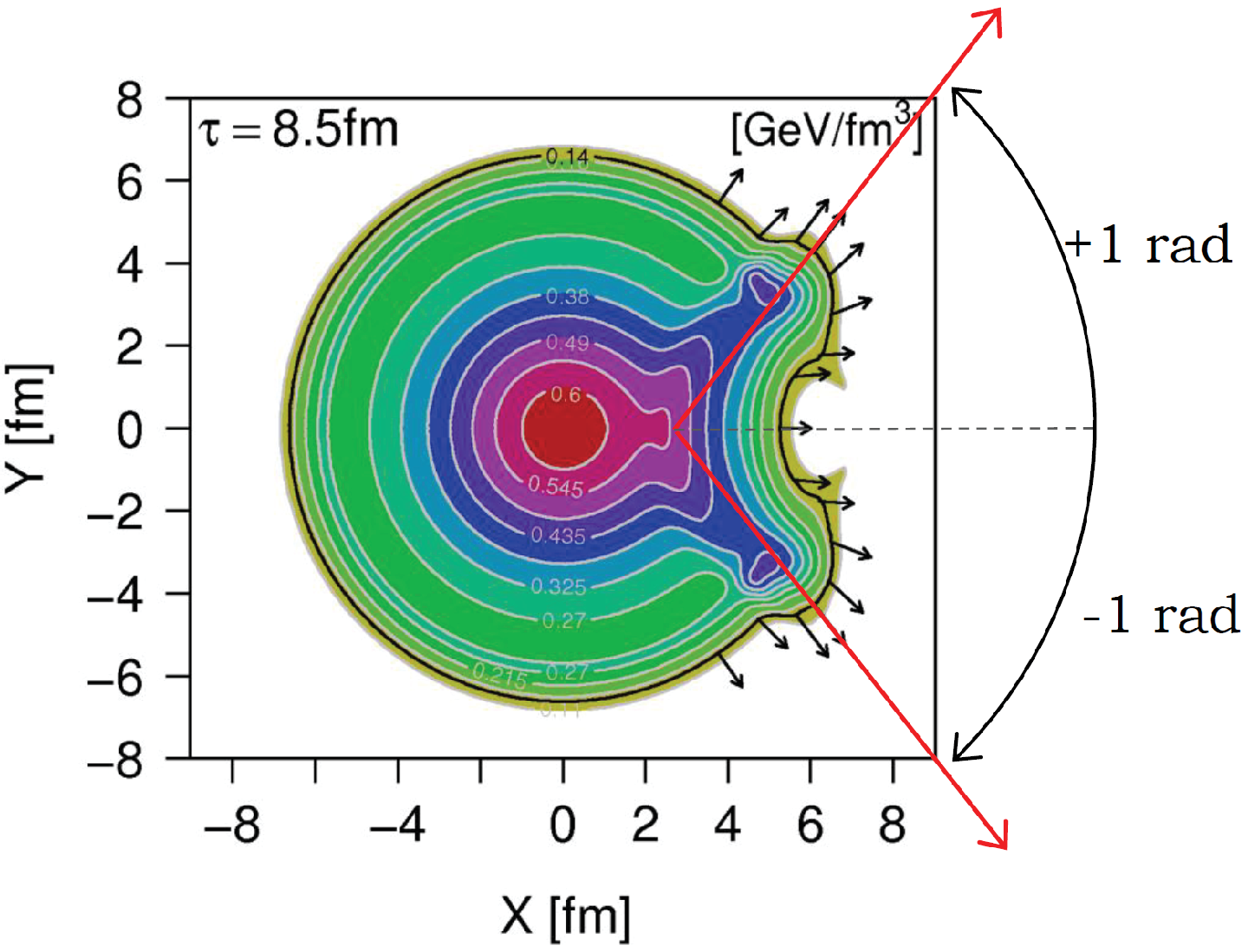}
\end{center}
\vspace{-.5cm}
 \caption{Temporal evolution of the energy density for the one-tube model (times: 1.0, 3.5 and 8.5 fm).
 Short arrows indicate fluid velocity on the freeze-out surface, the thicker curve labeled by the freeze-out temperature
 0.14 GeV. The longer arrows in the last plot indicate the maxima of the flow caused by the tube.}
 \label{flow}
% \vspace{-.4cm}
\end{figure}
FIG.~\ref{flow} shows the temporal evolution of the hot matter in this model. 
As seen, pressed by the violent expansion of the high-energy-density tube, the otherwise isotropic radial flow of the background was deflected and guided into two well-defined directions, symmetrical with respect to the initial tube position. Notice that the flow is clearly non-radial in this region. 
The outer part of the flow coming from the tube went quickly in isotropic expansion toward the vacuum, producing altogether a nearly circular hole with the center near the original position of the tube (as in~\cite{Shuryak-ridge}).

\begin{figure}[thb]
 \includegraphics[width=6.3cm]{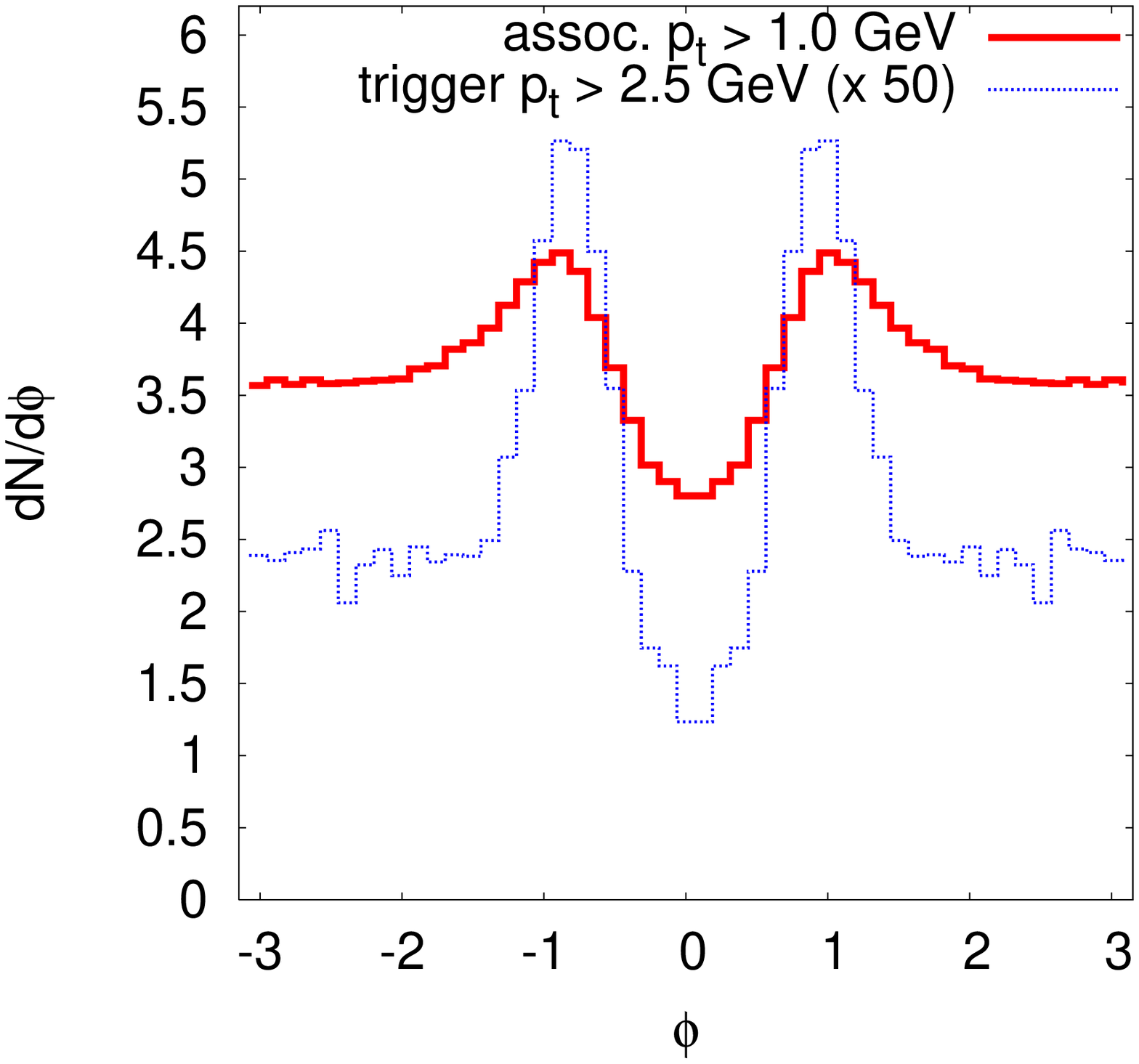}
 \includegraphics[width=6.7cm]{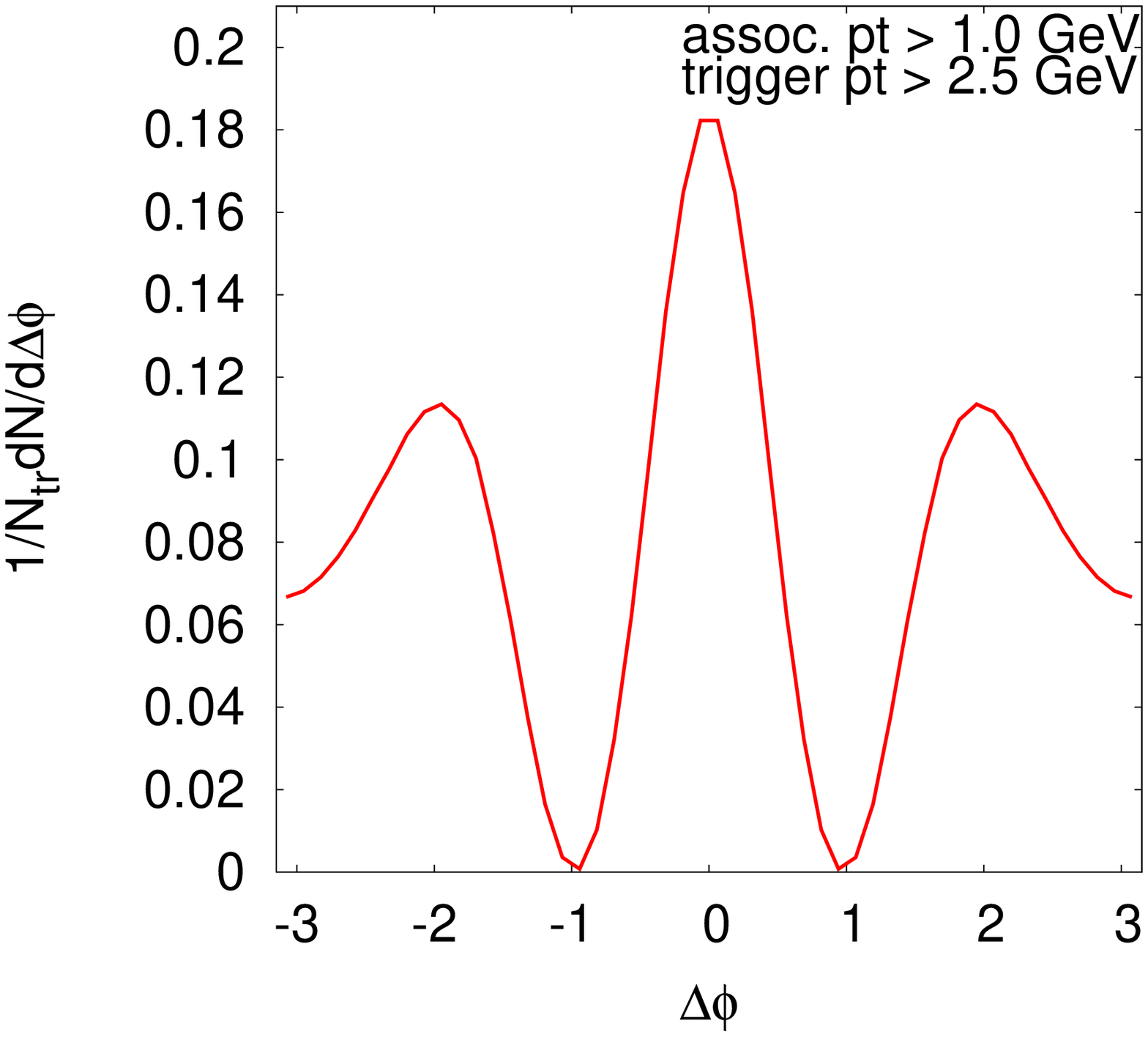}
 \vspace{.3cm}
 \caption{Angular distributions of particles in two different $p_T$ intervals (top) and resulting two-particle correlations (bottom) in peripheral-tube model.}
\label{angular}
\end{figure}

The resultant single-particle angular distributions for two different
$p_T$-intervals are plotted in FIG.~\ref{angular} (top), with respect to the tube position $\phi_{tube}$ set $=0$. 
As expected, they show a symmetrical two-peak structure.
From this plot, we could easily guess how the two-particle angular
correlation would be. 
The trigger particle would be more likely to be in one of the two peaks. 
We first chose the left-hand side peak. 
The associated particle would more likely to be also in the peak at the same position, {\it i.e.}, $\Delta\phi=0$ or in the right-hand peak with $\Delta\phi\sim+2$.
If we chose the trigger particle in the right-hand side peak, the associated particle would be likely also in the peak at this position, {\it i.e.}, with $\Delta\phi=0$ or in the left-hand side peak with $\Delta\phi\sim-2$.
So the final two-particle angular correlation, which actually is obtained by integrating over the trigger momentum and angle $\phi_t$, should have a large central peak at $\Delta\phi=0$ and two smaller peaks respectively at $\Delta\phi\sim\pm 2$. 
FIG.~\ref{angular} (bottom), which has been computed in the trigger- and associated-particle-$p_T$ regions indicated, shows that this is indeed the case. 
The peak at $\Delta\phi=0$ corresponds to the near-side ridge and the peaks at $\Delta\phi\sim\pm 2$ form the double-hump away-side ridge (shoulders).
Moreover, the above discussions imply that the height of the near-side ridge is naturally about twice in comparison to those of the away-side double hump.
This result is quite similar to the data (see FIG.~\ref{correlations} of Subsection VI.D, corresponding to the most central collisions we are considering here), not only the three-ridge structure but also their relative heights.
Often, one interprets the three-ridge structure of the two-particle correlation due to the so-called triangular flow, $v_3\,$. 
However, in order to get the correct shape of this structure, inclusively the relative heights as mentioned above, one needs also $v_2$, whose magnitude is correlated with that of $v_3$. 
How this correlation appears? 
The presence of the triangular flow can be attributed to the IC fluctuations. 
Nevertheless, the apparent correlation in magnitude between $v_2$ and $v_3$ is less straightforward. 
In this context, we find that {\it peripheral tube} we discuss in this paper provides a natural explanation to such a correlation.

\begin{figure}[htb]
 \includegraphics[width=5.5cm]{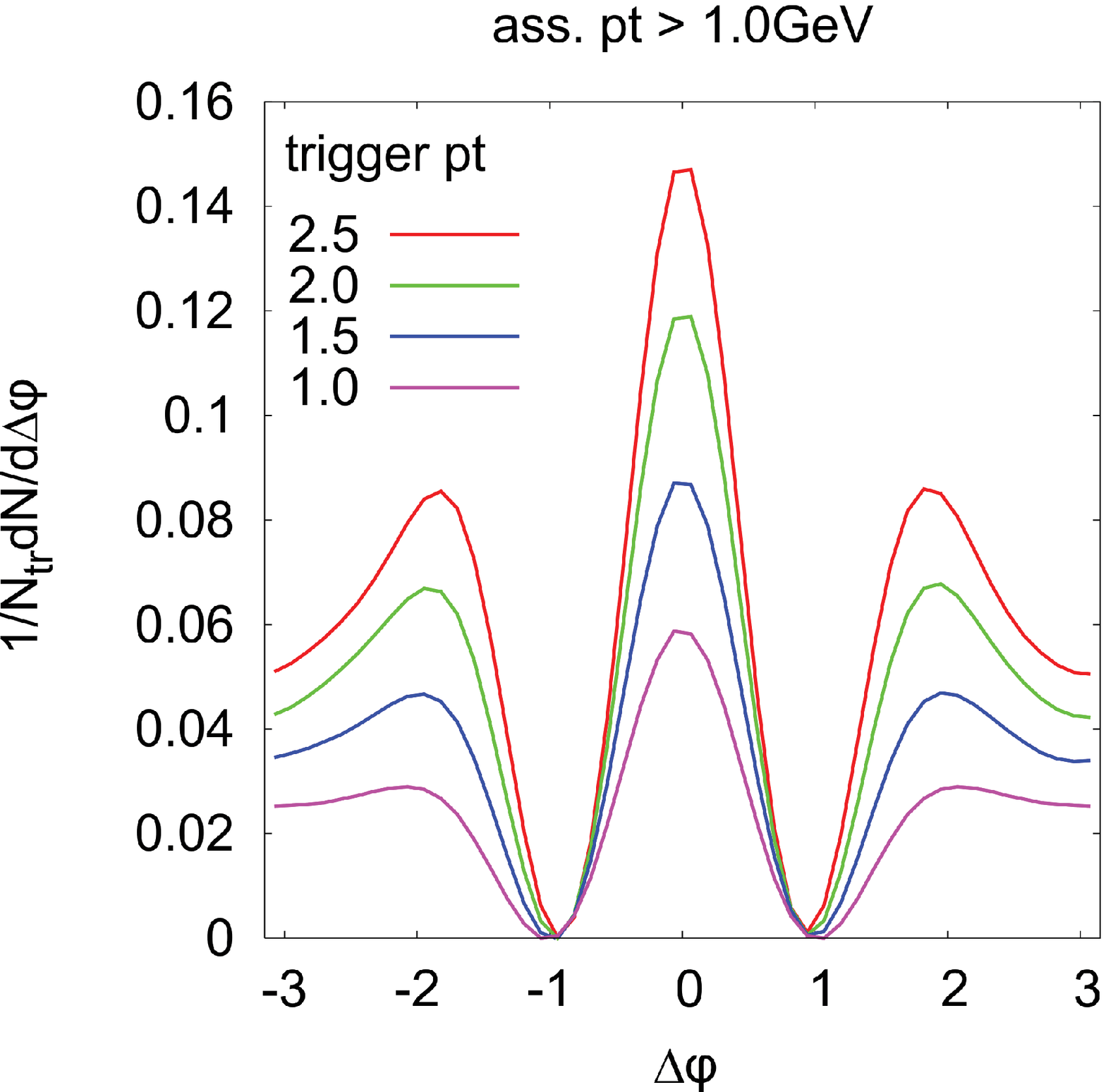}
 \vspace{-.3cm}
 \hspace{1.3cm}
 \includegraphics[width=6.1cm]{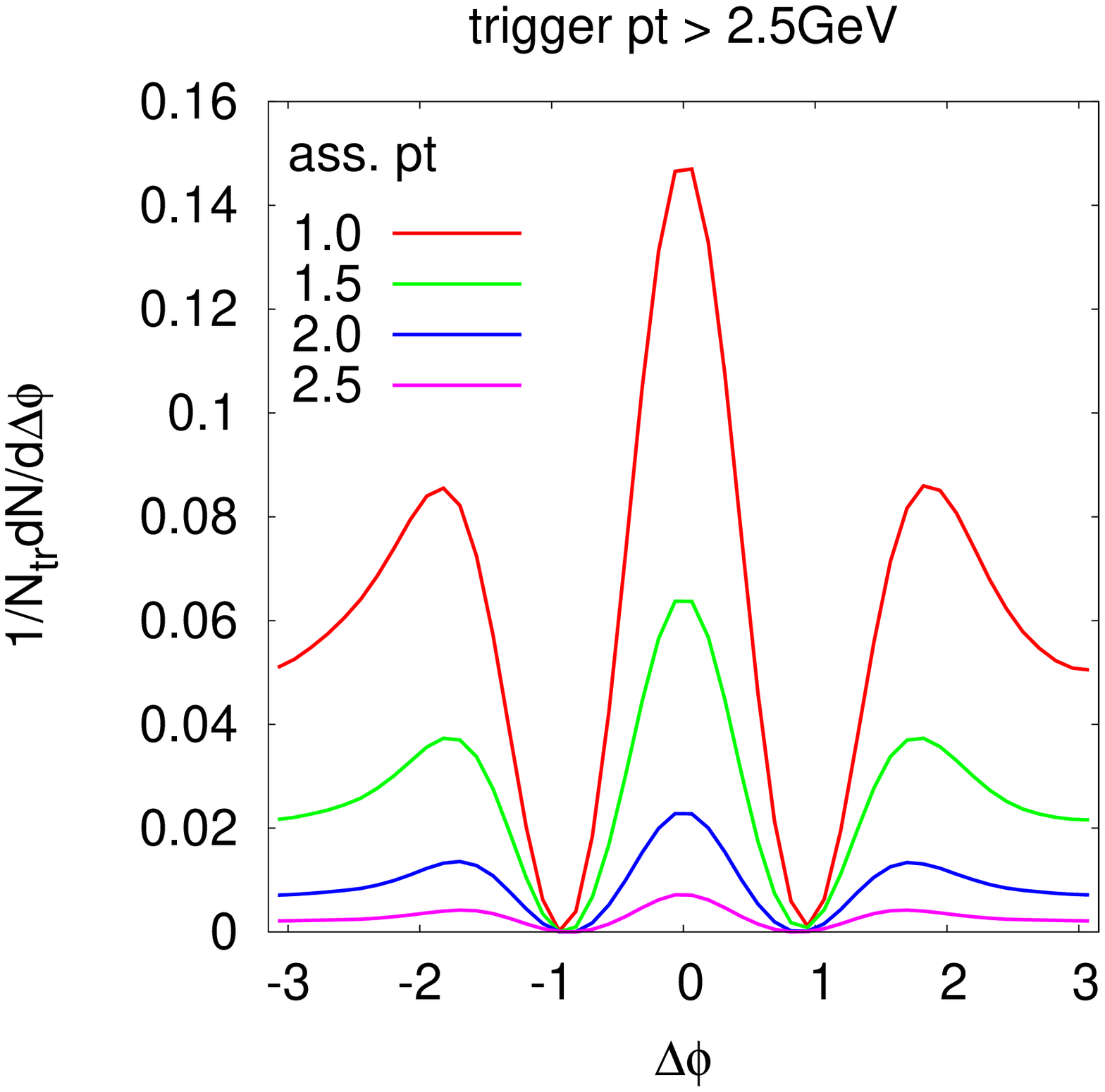}
 \vspace*{.5cm}
 \caption{\label{corr_p}Trigger-particle (top), and associate-particle $p_T$ dependence of two-particle correlations in peripheral-tube model (bottom).}
\end{figure}

We got some more detailed results, which are shown in FIG.~\ref{corr_p} (see also~\cite{sph-corr-05} for more $p_T$ dependence). 
We have checked that this structure is robust by studying the effect of several parameters, such as the height and shape of the background, initial velocity, height, radius, and location of the tube~\cite{sph-corr-03} (see also~\cite{sph-corr-02}).
This will be the topic of the next section.
%We will recaptulate it below.

It is also noted that the peripheral tube produced the ridge+shoulders {\it locally}.
To be specific, the observed three-peak structure in correlation is attributed to a causally connected local structure, rather than any global feature in terms of flow harmonics.
This point will be reinforced in Subsection IV-D.

\subsection{Parameter dependence}

\begin{figure}[bh]
 \vspace*{-.3cm}
 \begin{center}
 \includegraphics[width=5.5cm]{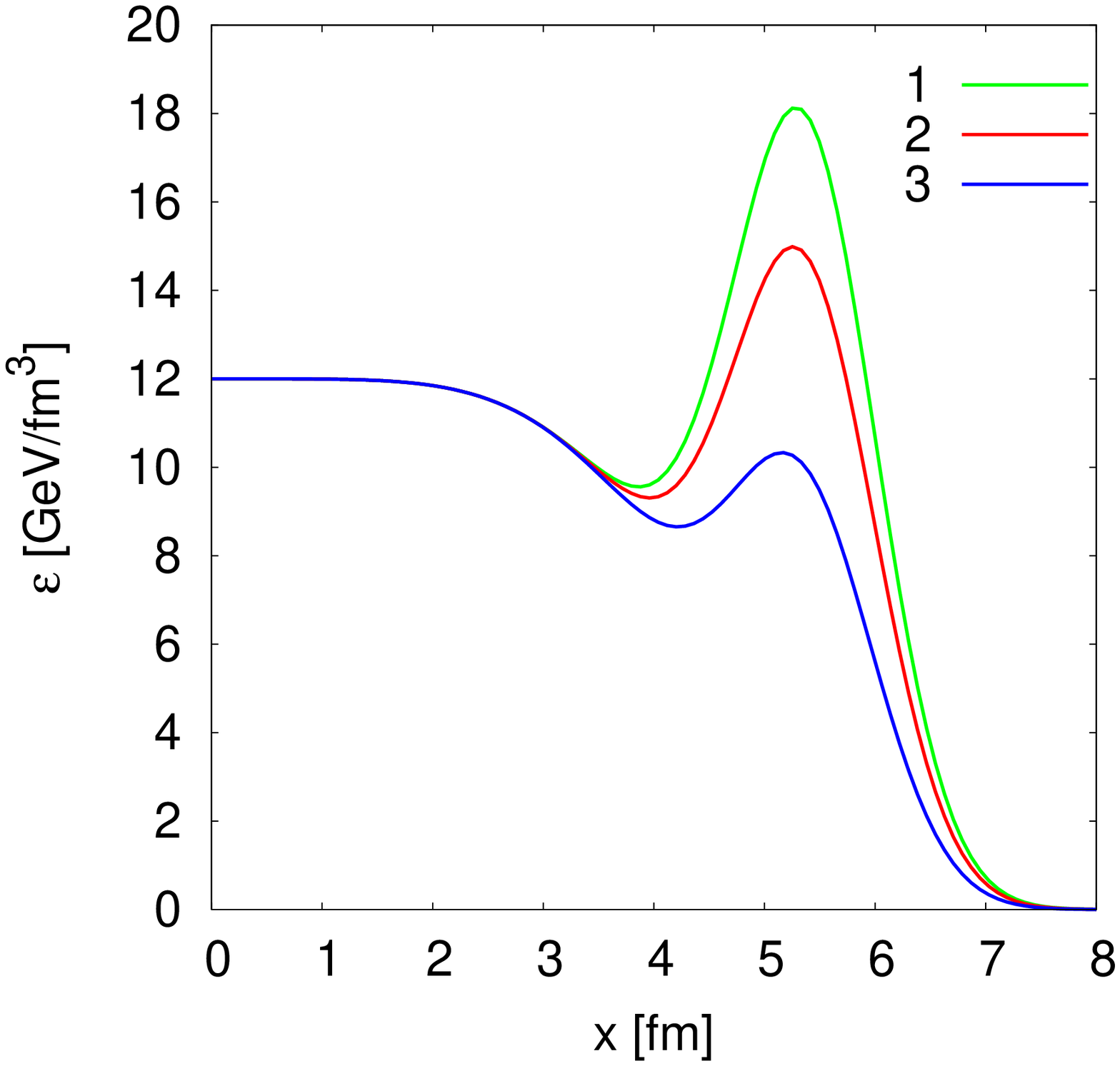}\\
 \includegraphics[width=5.5cm]{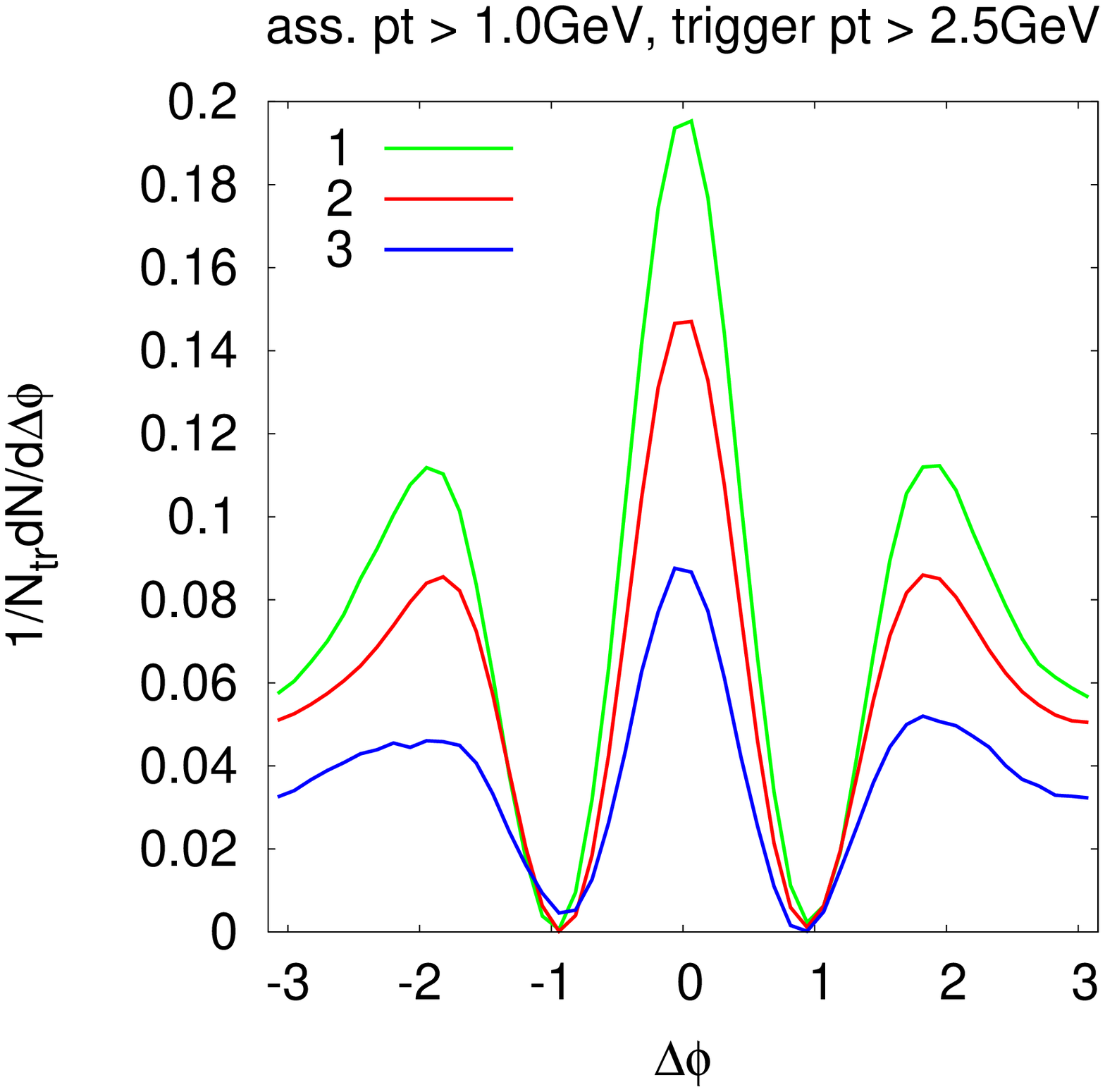}
 \end{center}
  \caption{Profile of smooth background $+$ peripheral tube of three different heights $\epsilon_t$ (top) and the corresponding two-particle correlations (bottom) in the peripheral one-tube model.}
  \vspace*{-.2cm}
  \label{h_dependence}
\end{figure}
In the above calculations, we used a tube extracted from a typical NEXUS initial conditions shown in FIG.~\ref{xyenergyic}, 
which has a radius $\Delta r$ of order 0.9 fm and (maximum) energy density $\epsilon_t$ of order 12 GeV $fm^{-3}$ (at proper time 1 fm), 
located at a distance of $r_0\simeq5.4\,$fm from the axis of the hot matter, as can be seen from Eq.~(\ref{par}) or from FIG.~\ref{xyenergyic}. 
As for the initial transverse velocity, we took the average NEXUS IC as given by Eq.~(\ref{parv}). 
The background energy density, which replaced the complex distribution as it appears in the original event, FIG.~\ref{xyenergyic}, we parametrized approximating the average distribution by a simple appropriate function.
Let us see how the correlation changes if we change these parameters. 
We will see that some parameters affect much the correlation, whereas this is almost insensitive to others. 
However, one aspect which calls attention is that the shape of the correlation in $\Delta\phi$ is very robust.

Let us begin with the height of the tube, {\it i.e.}, energy density, in the middle of the tube, maintaining all the other parameters constants.
To be more specific, the three curves, denoted by ``1" to ``3" in the top plot of FIG.~\ref{h_dependence}, are generated by merely varying $\epsilon_t=c_3$ in Eq.~\eqref{par}.
The resultant two-particle correlations are presented by those labeled by 1 to 3 in the bottom plot of FIG.~\ref{h_dependence}.
As one can see, the height of the correlation is more or less proportional to the tube's height above the background.
Unless specified otherwise, here and in all this subsection, we investigate three different variations of model parameters.
The model with the original parameter given in Eq.~(\ref{par}) is drawn as the middle curve, usually depicted in red and denoted by ``2" in the plot.

\begin{figure}[tbh]
% \vspace*{-1.cm}
 \begin{center}
  \includegraphics[width=5.5cm]{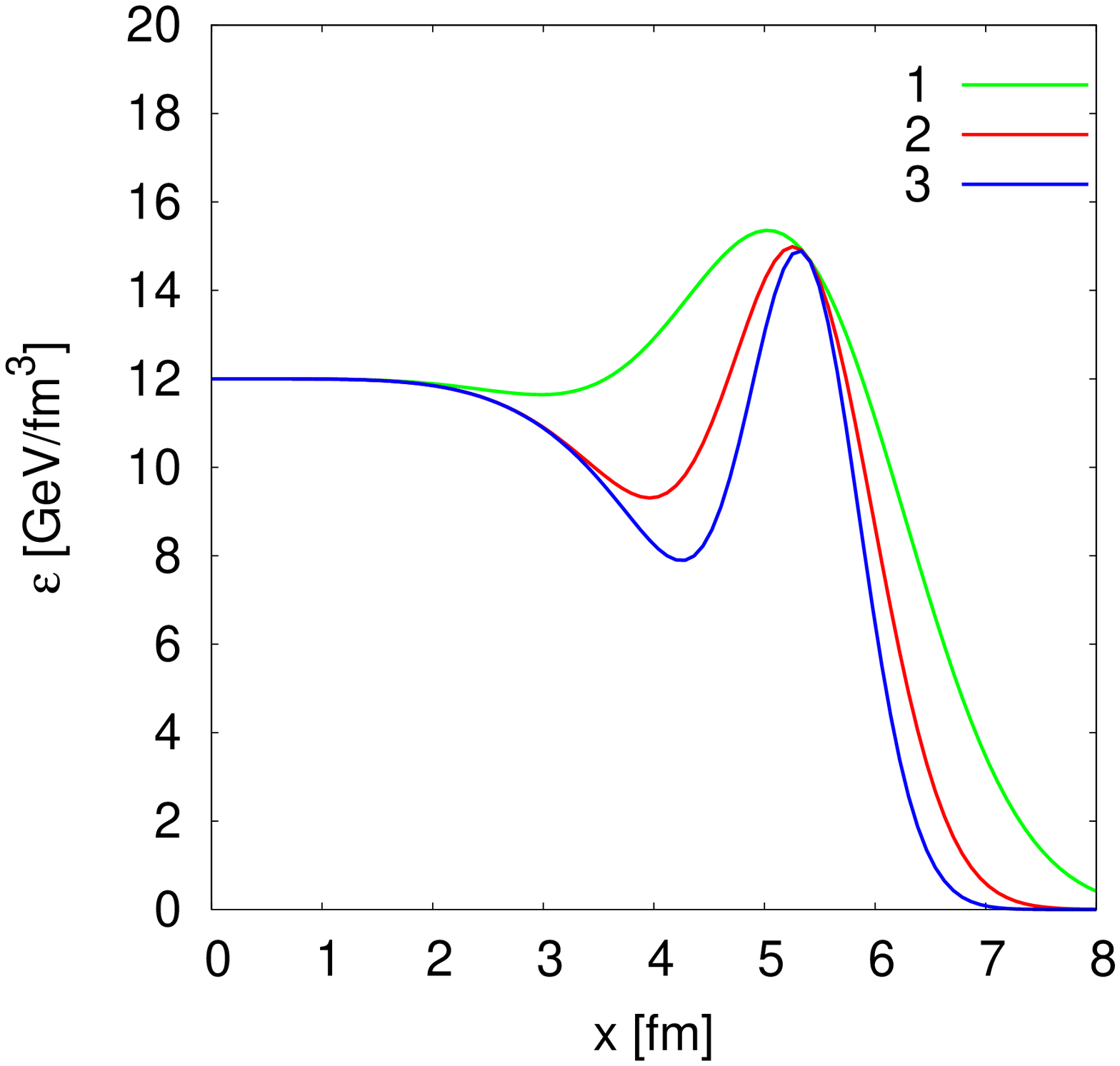}\\
  \includegraphics[width=5.5cm]{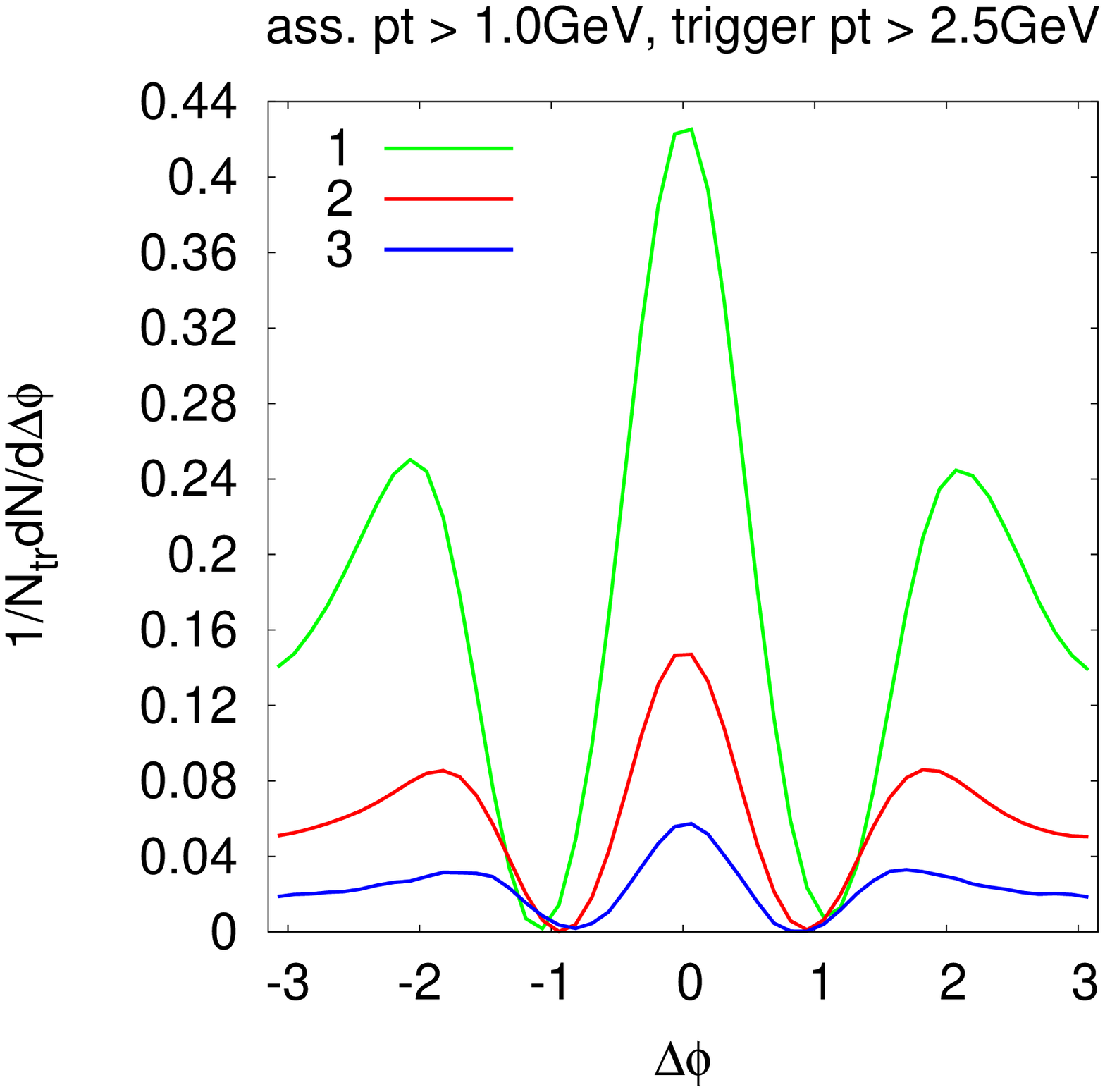}
 \end{center}
%  \vspace*{-.5cm}
  \caption{\label{d_dependence} Profile of smooth background $+$ peripheral tube of three different radii $\Delta r$ (top) and the corresponding two-particle correlations (bottom) in the peripheral one-tube model.}
  \vspace*{-.2cm}
\end{figure}
\begin{figure}[h]
 \vspace*{-.3cm}
 \begin{center}
  \includegraphics[width=5.7cm]{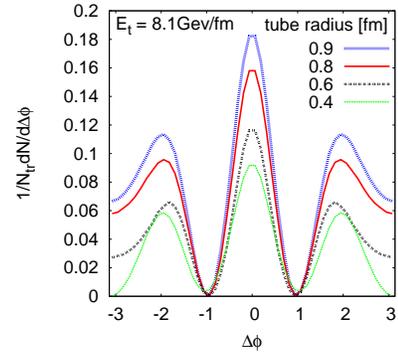}
 \end{center}
  \vspace*{-.3cm}
  \caption{\label{Econst} Two-particle correlations with similar energy $E_t$ in the peripheral tube, with variable radius.}
%  \vspace*{-.2cm}
\end{figure}

Next, let us vary the radius ($\Delta r=c_4$ in Eq.~\eqref{par}) of the tube, and study the resultant correlations for three different ICs in FIG.~\ref{d_dependence}.
Here, one sees that the height of the correlation increases when the radius increases, and it raises more quickly than the radius of the tube.
Moreover, the spacing in the peaks is also affected, as shown. 
By examing these two figures, FIGS.~\ref{h_dependence} and~\ref{d_dependence}, it is found that what imports in the phenomenon is neither $\epsilon_t$ nor $\Delta r$, but the energy content of the tube ($E_T$). 
In order to verify it, we made a plot, FIG.~\ref{Econst}, by fixing $E_T$ and varying the tube radius $\Delta r$. 
It is observed that the two-particle correlation maintains its overall shape. Namely, the heights of the peaks and the angle between them are almost unchanged.
We should mention here that, although for simplicity, we parametrized the shape of the tube as Gaussian, what is really relevant here is the energy content $E_T$.

\begin{figure}[h]
 \vspace*{-.3cm}
 \begin{center}
  \includegraphics[width=5.5cm]{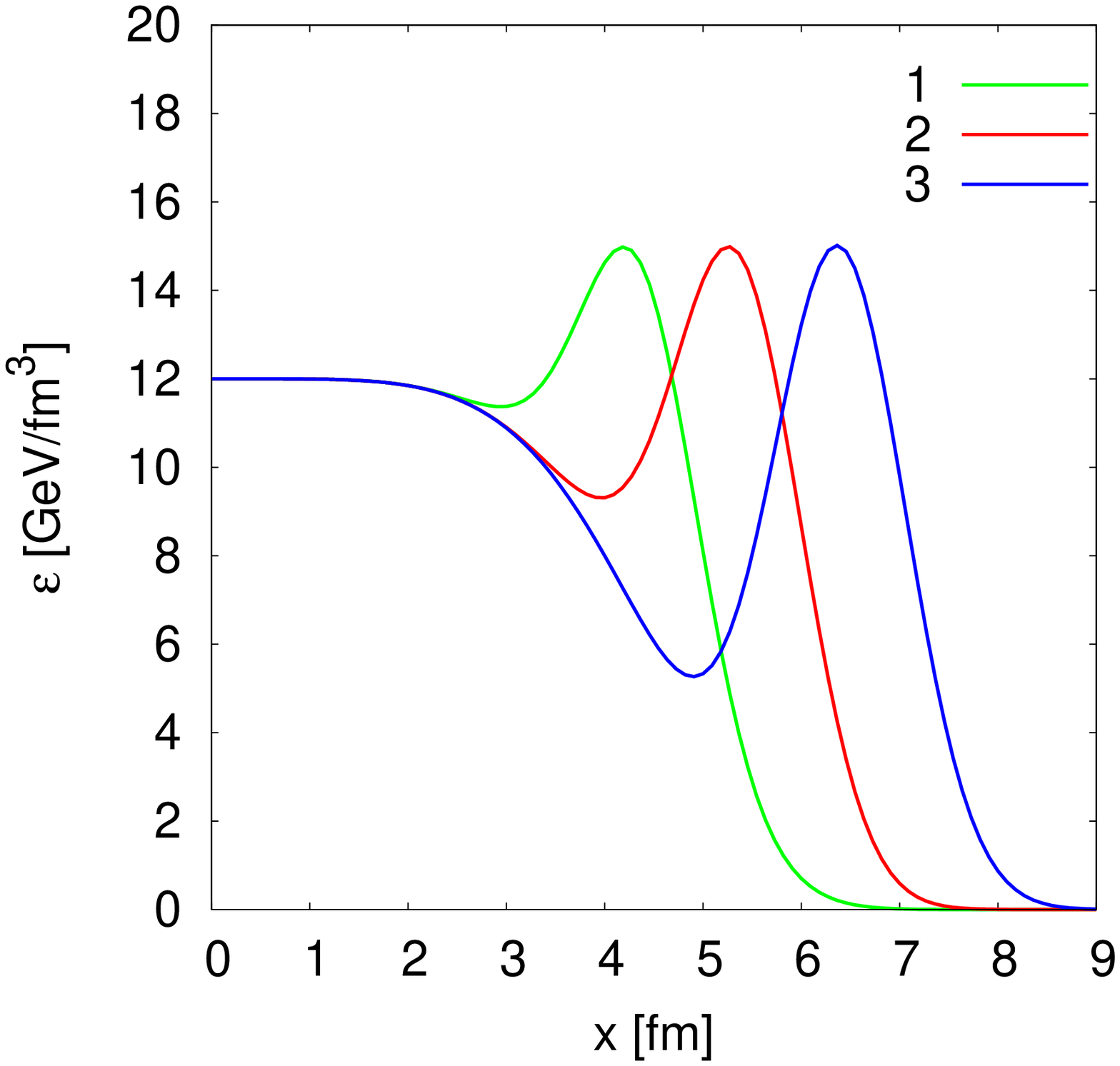}\\
  \includegraphics[width=5.5cm]{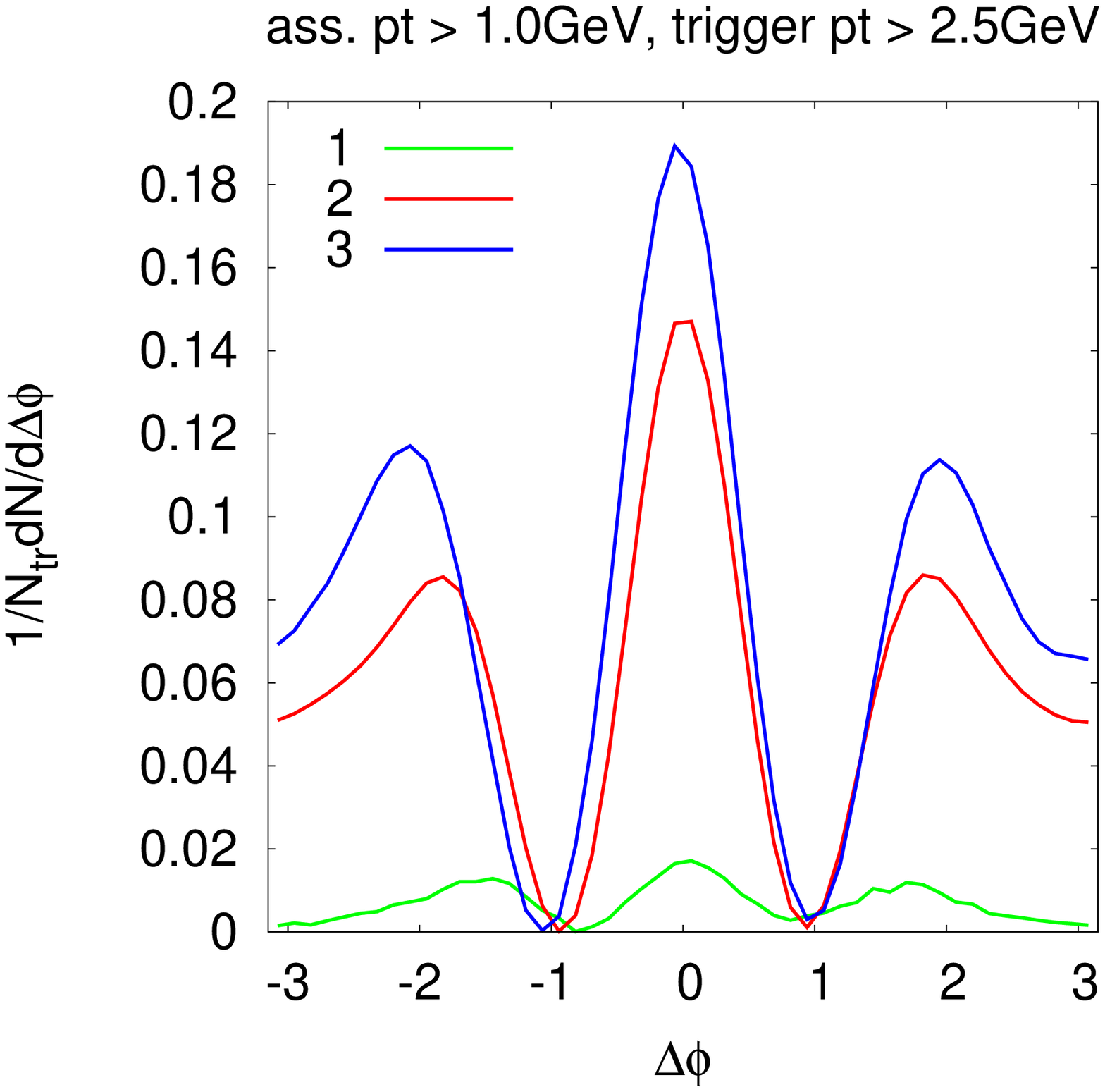}
 \end{center}
%  \vspace*{-.8cm}
  \caption{\label{r_dependence} Profile of smooth background $+$ peripheral tube of three different positions $r_0$ (top) and the corresponding two-particle correlations (bottom) in the peripheral one-tube model.}
  \vspace*{-.4cm}
%  \vspace*{-.8cm}
\end{figure}
Now, let us see what happens if we move the position of the tube ($r_0$). 
In FIG.~\ref{r_dependence} (top), the middle one corresponds to the original peripheral tube, and we moved the same distance (1 fm), one inward and the other one outward the original position. 
The difference in correlation they provoked (below) is remarkable. 
While the one moved outward produced more or less the same correlation (slightly higher), the one moved inward became ineffective with regard to the correlation issue.
It is true that in the comparison, we maintained constant the total height of the tube ($\epsilon_t$), including the background and not the tube energy itself. 
Even though, the size of the inner tube itself is more or less only half, not smaller than, that of the original middle one.
\begin{figure}[thb]
 \begin{center}
%\vspace{-.7cm}
\includegraphics[width=5.5cm]{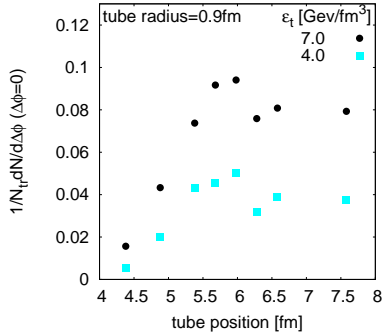}
\vspace*{-.2cm}
\caption{\label{rlimit} Height of the two-particle correlation at $\Delta \phi=0$ as function of the tube position with respect to the axis.}
\end{center}
\end{figure}

This result suggested to us how to define the peripherality of a tube, by considering its power of producing the correlation structure we are studying.
In FIG.~\ref{rlimit}, we plot the central maximum of the correlation as a function of the distance of the tube from the axis of the hot matter.
As seen in FIG.~\ref{lines}, for Au+Au collisions we are studying here, 
the average energy distribution presents a plateau at $r\leq3\,$fm and decreases quickly to zero at the border ($r\simeq7\,$fm).
\begin{figure}[t!h]
 \vspace*{-.3cm}
 \begin{center}
  \includegraphics[width=5.5cm]{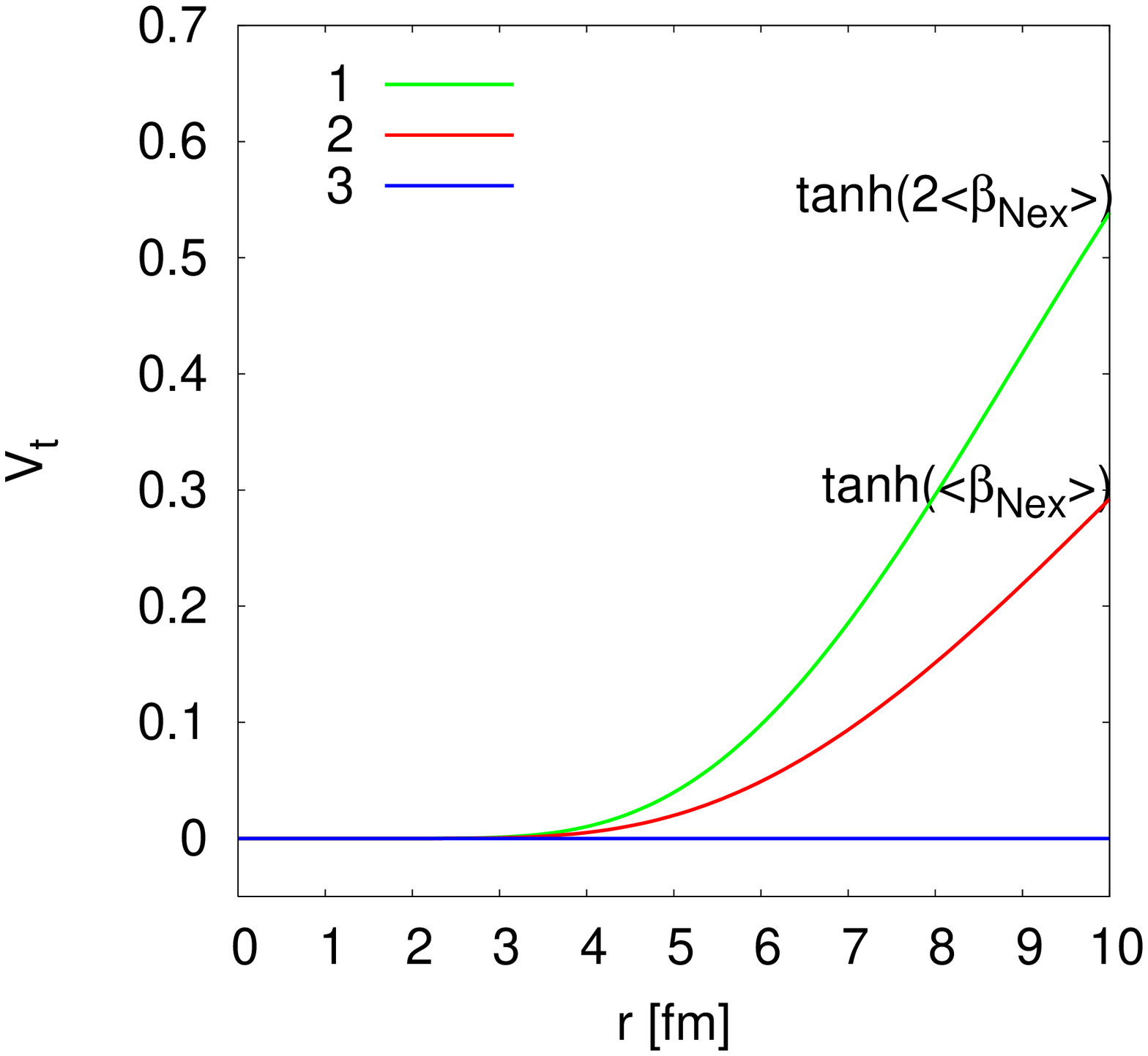}\\
  \includegraphics[width=5.5cm]{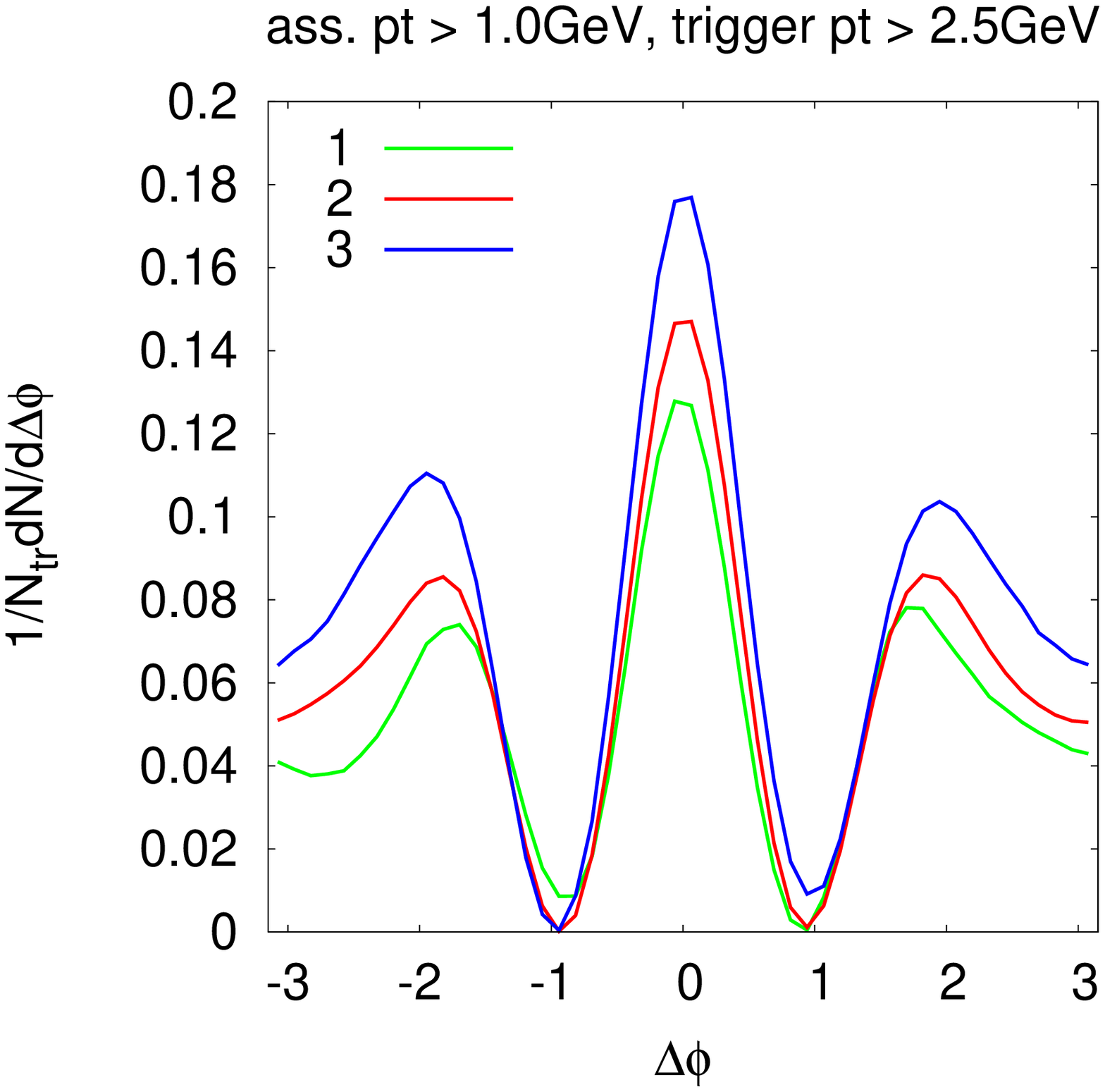}
 \end{center}
%  \vspace*{-.8cm}
  \caption{\label{v-dependence} Three different velocity distributions (top) and the corresponding two-particle correlations (bottom).}
  \vspace*{-.2cm}
%  \vspace*{-.8cm}
\end{figure}
So, FIG.~\ref{rlimit} shows that, independent of the size, on the plateau and its immediate vicinity, the resultant hydrodynamical expansion does not produce correlation.
Then, as the location of the tube goes to the border, the related correlation increases linearly, reaching the maximum value at $r\simeq5.5$fm.
The radius related to maximal correlation is where the selected typical tube is located, and we will maintain it as a constant. 
To be specific, according to the above discussions, {\it peripheral tubes} that can produce two-particle correlation are located at $r\geq4.0\,$fm.
See also FIG.~3 of Ref.~\cite{sph-vn-02} and related discussions.

Now, what is the effect of the initial (transverse) velocity? 
As understandable, FIG.~\ref{v-dependence} shows that the correlation decreases as the initial velocity increases. 
However, as seen, the influence of the initial velocity is almost negligible.

Finally, let us see what the influence of the background is. 
We will see that for specific changes, this is entirely negligible. 
However, for other kinds of changes, one will see that this is a crucial factor in the present model.
First, let us see what happens if we conserve the short-tailed shape of background
represented by $\sim\exp[-.0004r^5]$ in Eq.~(\ref{par}), as we did in the whole present subsection, IV.C, as given by averaging NEXUS fluctuating events, and by changing only its height or energy density.
\begin{figure}[t!h]
% \vspace*{-.3cm}
 \vspace*{-.4cm}
 \begin{center}
  \includegraphics[width=5.9cm]{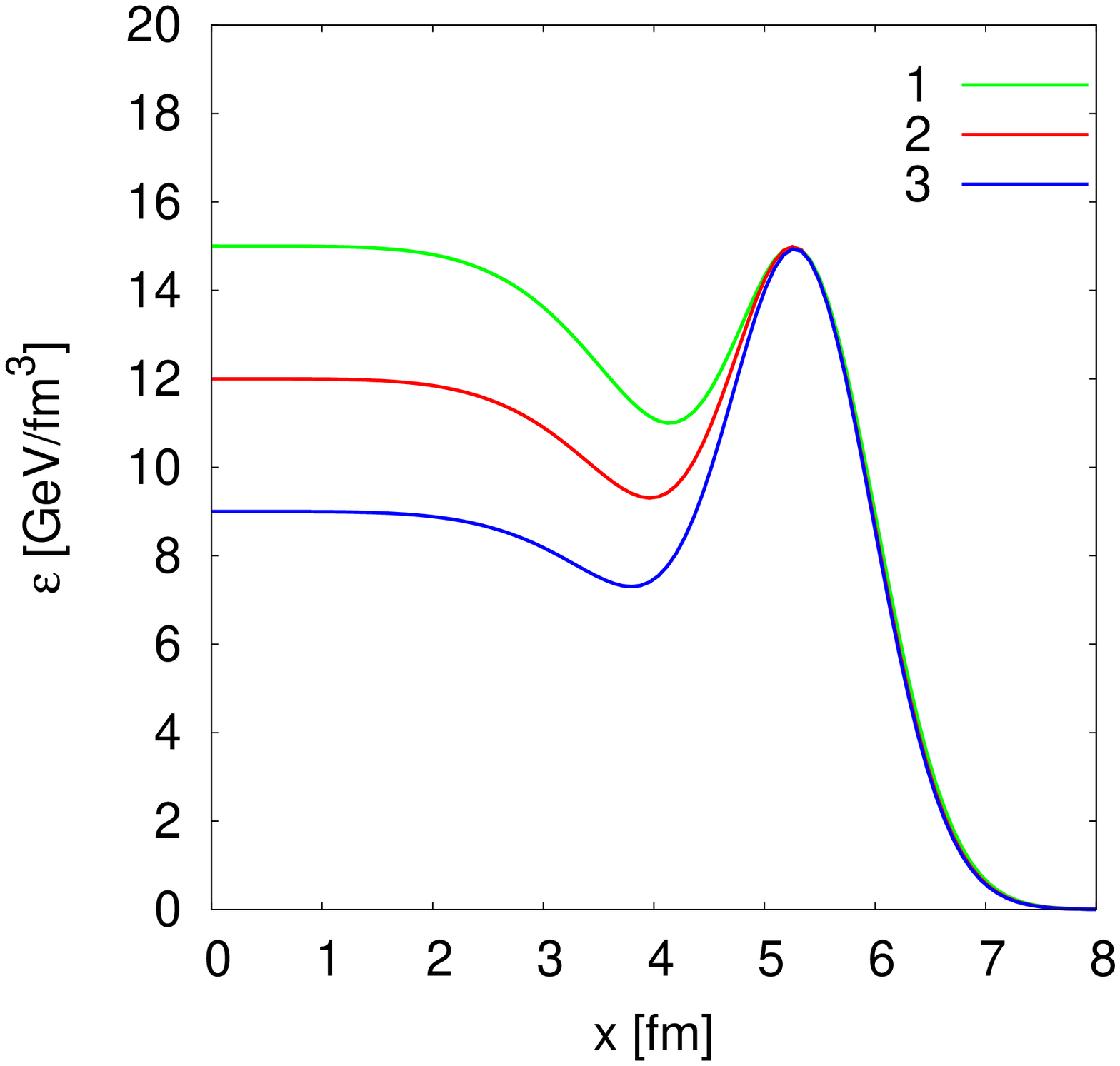}\\
  \includegraphics[width=6.2cm]{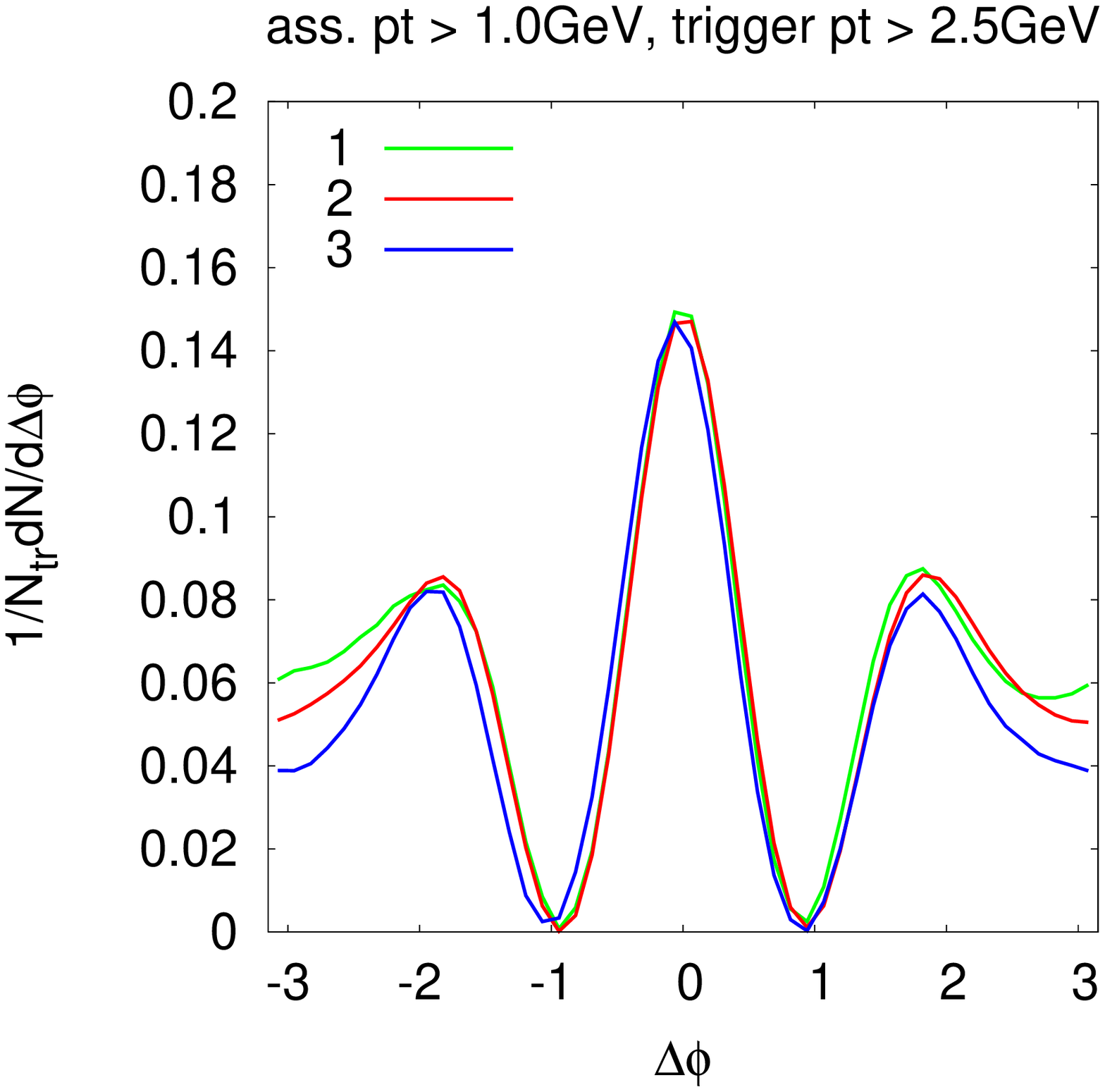}
 \end{center}
%  \vspace*{-.8cm}
  \caption{\label{bg-size} Backgrounds with three different energies, with the shape given by the first term of Eq.~\eqref{par} and the selected tube (top). The corresponding two-particle correlations are shown in the (bottom plot).}
  \vspace*{-.2cm}
%  \vspace*{-.8cm}
\end{figure}
The middle curve in FIG.~\ref{bg-size} (top) corresponds to the present energy density. As seen, the correlations obtained are identical among themselves. 
Evidently, this does not mean that the correlation is energy independent, because usually, the energy content of the tube itself follows the background energy.

\begin{figure}[t!h]
 \vspace*{-.5cm}
 \begin{center}
  \hspace{.5cm}
  \includegraphics[width=6.2cm]{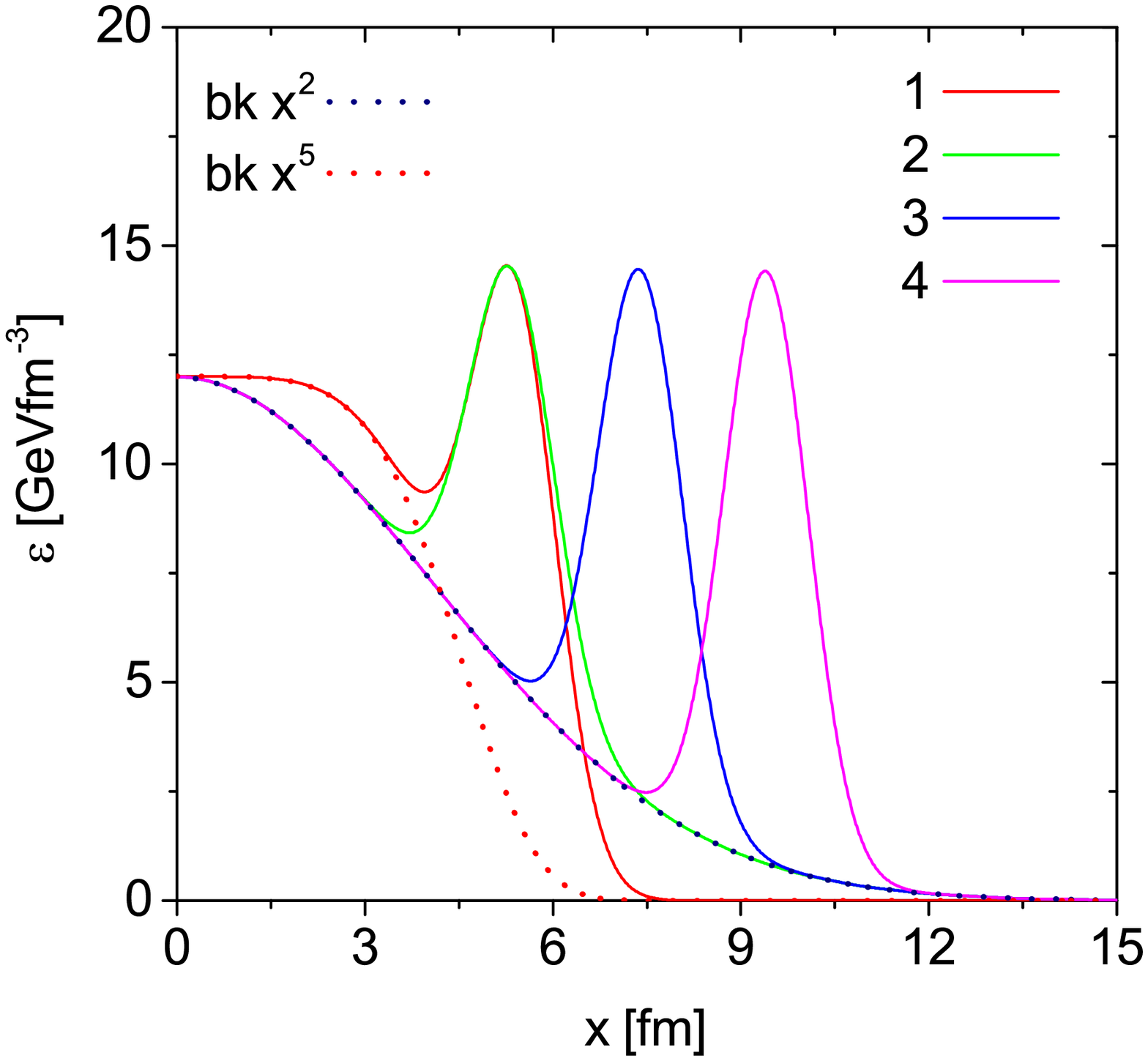}\\
  \vspace{-2.4cm}
  \hspace{.7cm}
  \includegraphics[width=6.2cm]{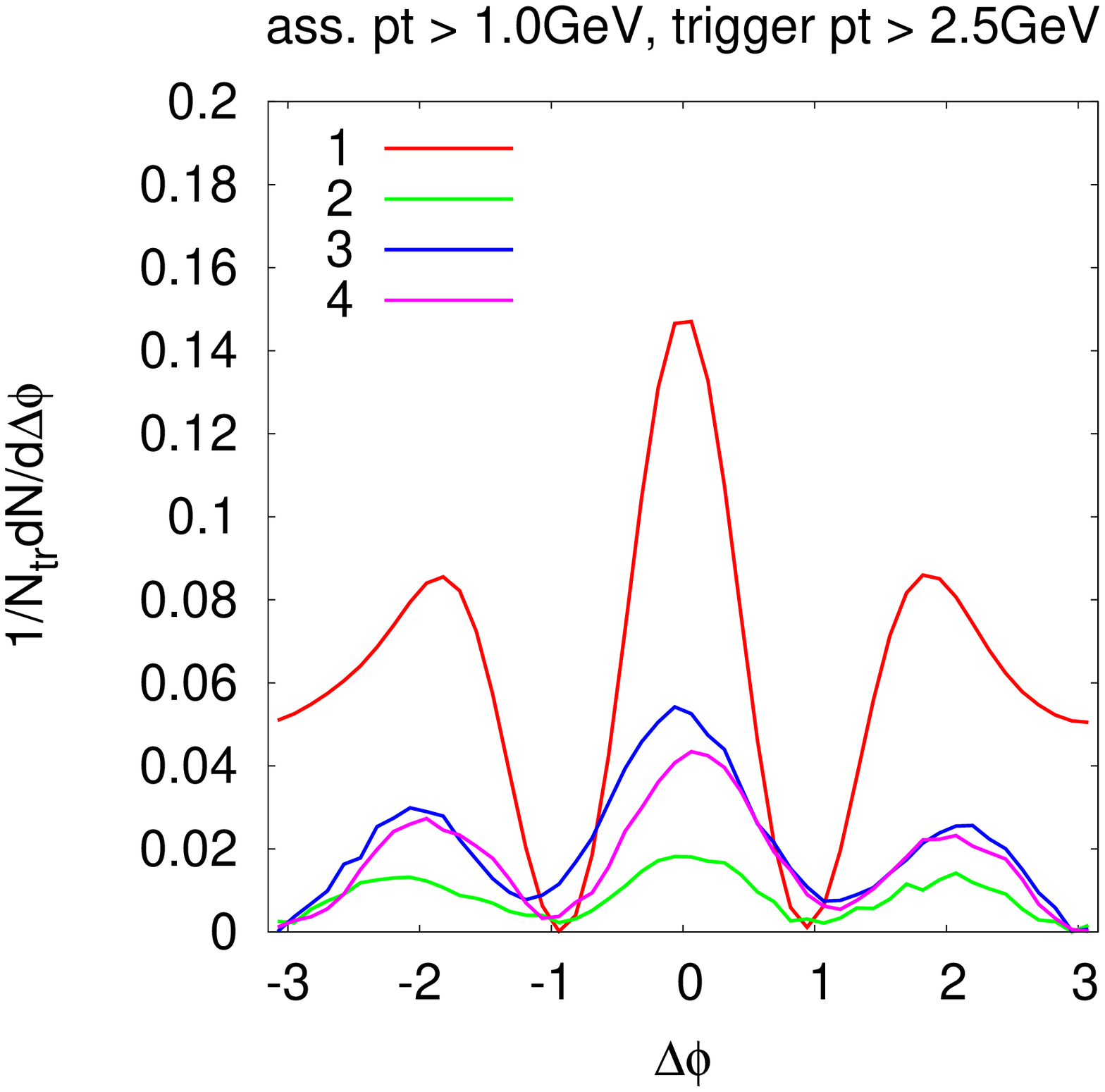}
 \end{center}
%  \vspace*{-.2cm}
  \caption{\label{bg-shape} Three tubes, similar to the selected one, placed in three different positions and Gaussian background (wider one). The original background (narrower one) is also shown in the top plot. 
The corresponding two-particle correlations are shown in the bottom plot, where the taller one is the one with the original edgy background.}
\end{figure}

Now, let us see what happens if we change the shape of the background to a long-tailed Gaussian. 
In practice, the Gaussian shape is sometimes used to represent a typical average energy distribution in nucleus-nucleus collisions. 
Here, the calculations are carried out for two different types of backgrounds. 
One defines in Eq.~(\ref{par}) with a short tail (exponential $r^5$) and the other with a broader tail (exponential $r^5$ replaced by $r^2$). 
We devise our ICs by adding a tube, located at different positions, going from inside to outside and labeled by ``2" to ``4", on top of the broader background. 
As a comparison, the original IC defined in Eq.~(\ref{par}) is also computed here and shown labeled by ``1". 
We remark that here the parameters have been chosen so that both backgrounds have the same energy content. 
The resultant correlations shown in FIG.~\ref{bg-shape} demonstrate a significant difference regarding the two types of background. 
Although the correlations do show characteristic three peaks, those with the Gaussian background are much smaller than the original edgy case, unable to compare with data.

\subsection{Extension to Peripheral Multi-Tube Model}

In Subsection IV-B, we made a simplified model of IC in order to clearly observe the dynamics of the ridge formation in two-particle correlations.
This will be referred to as {\it peripheral single-tube model}.
In this approach, we left just one original high-density tube located close to the surface and replaced the rest of the more realistic NEXUS lumpy matter distribution by a smooth background matter distribution. 
If we take into account what we discussed about FIG.~\ref{rlimit}, it is relevant to consider only tubes in the {\it peripheral region}, defined there as being out of the central cylinder of radius $r\simeq4\,$fm for Au+Au collisions.
In the case of the event shown in FIG.~\ref{xyenergyic}, the prominent peripheral tube is indeed the one we studied, being the others very small in comparison. 
However, as shown in FIG.~\ref{fic}, there may exist events with more than one comparable peripheral tubes.
What happens in such a case?

\begin{figure}[ht]
 \begin{center}
  \includegraphics[width=6.cm]{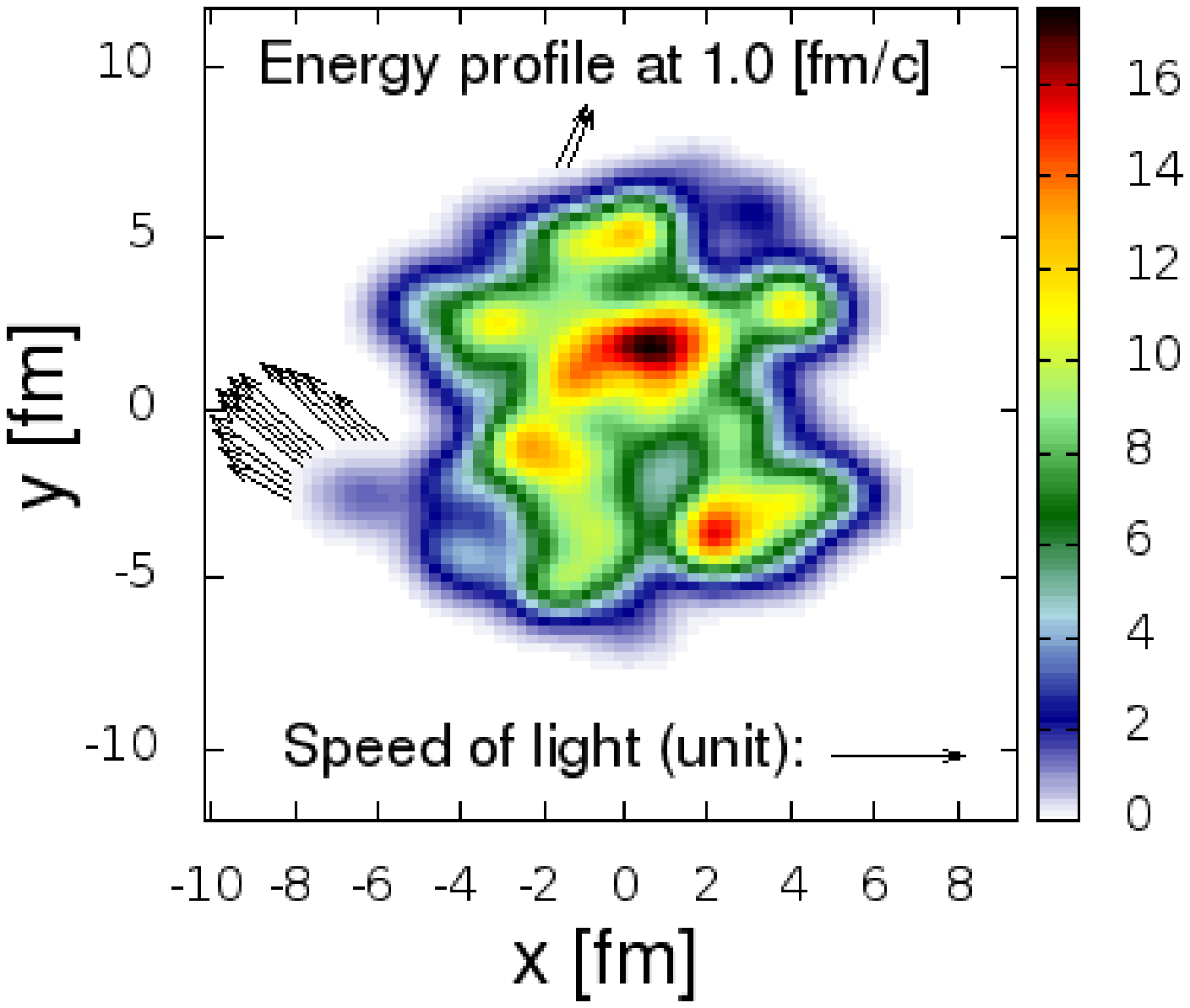}
 \vspace*{-.9cm}
  \includegraphics[width=6.cm]{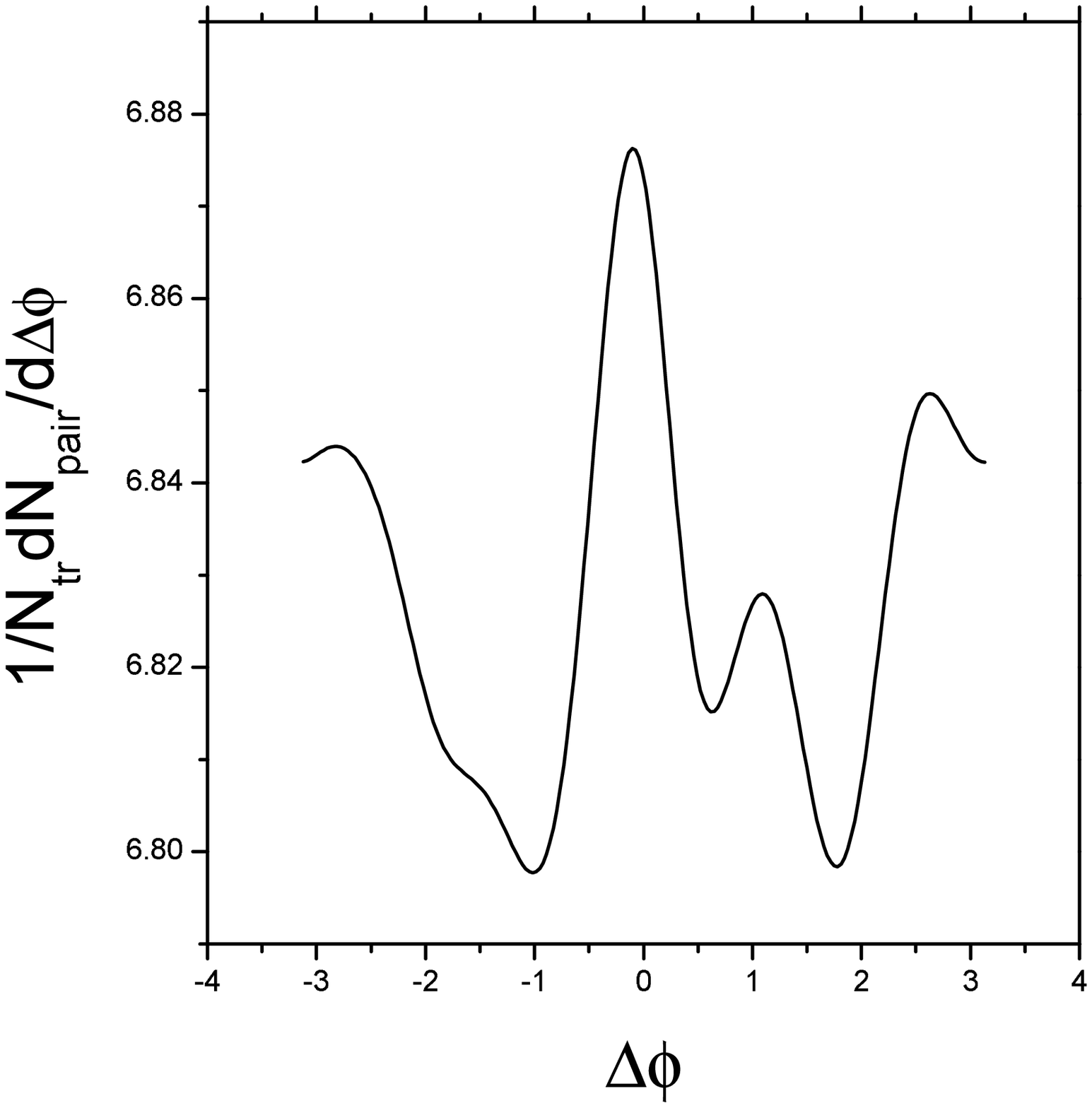}
  \vspace*{-2.1cm}
  \caption{\label{mtube-event} A NEXUS event with 5 peripheral tubes is plotted (top plot), with corresponding two-particle correlation (bottom plot).}
 \end{center}
\end{figure}
Before going into a more systematic study of this question, let us first show an example of NEXUS IC with many peripheral tubes and the correlation it produces. 
FIG.~\ref{mtube-event} shows such an example, where we can pinpoint five peripheral tubes. The very high-density and fat tube approximately in the middle of this energy distribution produces the more or less radial flow, which is deflected by these peripheral tubes, casting a complex shadow to it.
In the previous Subsection (IV-C), we saw that the shape of the resultant two-particle correlations from a single tube (particularly both the peak spacing and the relative heights) is almost independent of its features.
Therefore, the various tubes in the NEXUS event under consideration should contribute with somewhat similar two-peak emission pattern at various angles in the single-particle angular distribution. 
As a consequence, the two-particle correlation, computed by using just this event, is expected to have a well-defined main structure similar to that of the one-tube model (FIG.~\ref{angular}) surrounded by several other peaks and depressions due to the trigger and associated particles coming from different tubes. 
This is indeed the case, as shown in FIG.~\ref{mtube-event}, bottom plot.
However, one can already visualize some trace of the three-peak structure that one tube causes due to the constructive interference they produced. 
Evidently, they also introduce some destructive interference, which makes the resultant correlation asymmetric and more complex.

Recalling these qualitative aspects we have just pointed out, we went one step further in Ref.~\cite{sph-corr-07} and systematically studied what happens if we have more than one peripheral tube, randomly distributed in azimuth. 
Our results were that, although the azimuthal flow parameters $v_n$ and directions $\Psi_n$ were random, the final two-particle correlations were more or less constants, and similar to the one produced by just one peripheral tube.

Let us show how we studied this question. 
Instead of only one as before, we considered 2, 3, or 4 peripheral tubes in this study, all equal in shape, size and radial position as one we studied in Ref.~\cite{sph-corr-03}, namely, Gaussian energy distribution, with radius $d\simeq.92\,$fm and distance $r_0=5.4\,$fm from the axis. 
They are distributed randomly in azimuth on top of the same isotropic background as before. 
Explicitly, the energy distribution is parameterized, while compared with Eq.~(\ref{par}), as
\begin{equation}
  \epsilon=c_1\exp[-c_2 r^5]+\sum_{i=1}^n c_3 \exp[-\frac{|{\bf r}-{\bf r}_i|^2}{c_4}]\ ,
  \label{parmulti}
\end{equation}
where $n=1, 2, 3, 4$ is the number of peripheral tubes, $\vert{\bf r}_i\vert=5.4\,$fm, and their azimuths are chosen randomly.
As discussed in the previous subsection, since the characteristics of these tubes we have chosen here are of a typical peripheral tube, and because the essential features of two-particle correlation such tubes produce individually are robust, we find that fixing some of the parameters does not avoid to get a good vision of the effects.
In what follows, we present preliminary results computed with only 50 random events in each case.

\bigskip
\noindent
{\bf Fourier components of $dN/d\phi: v_n$}
\medskip

First, we computed the one-particle azimuthal distribution for each number of peripheral tubes we chose. 
Then, decomposed them into Fourier components.
\begin{figure}[t!hb]
\vspace*{-.3cm}
\begin{center}
\includegraphics[width=5.6cm]{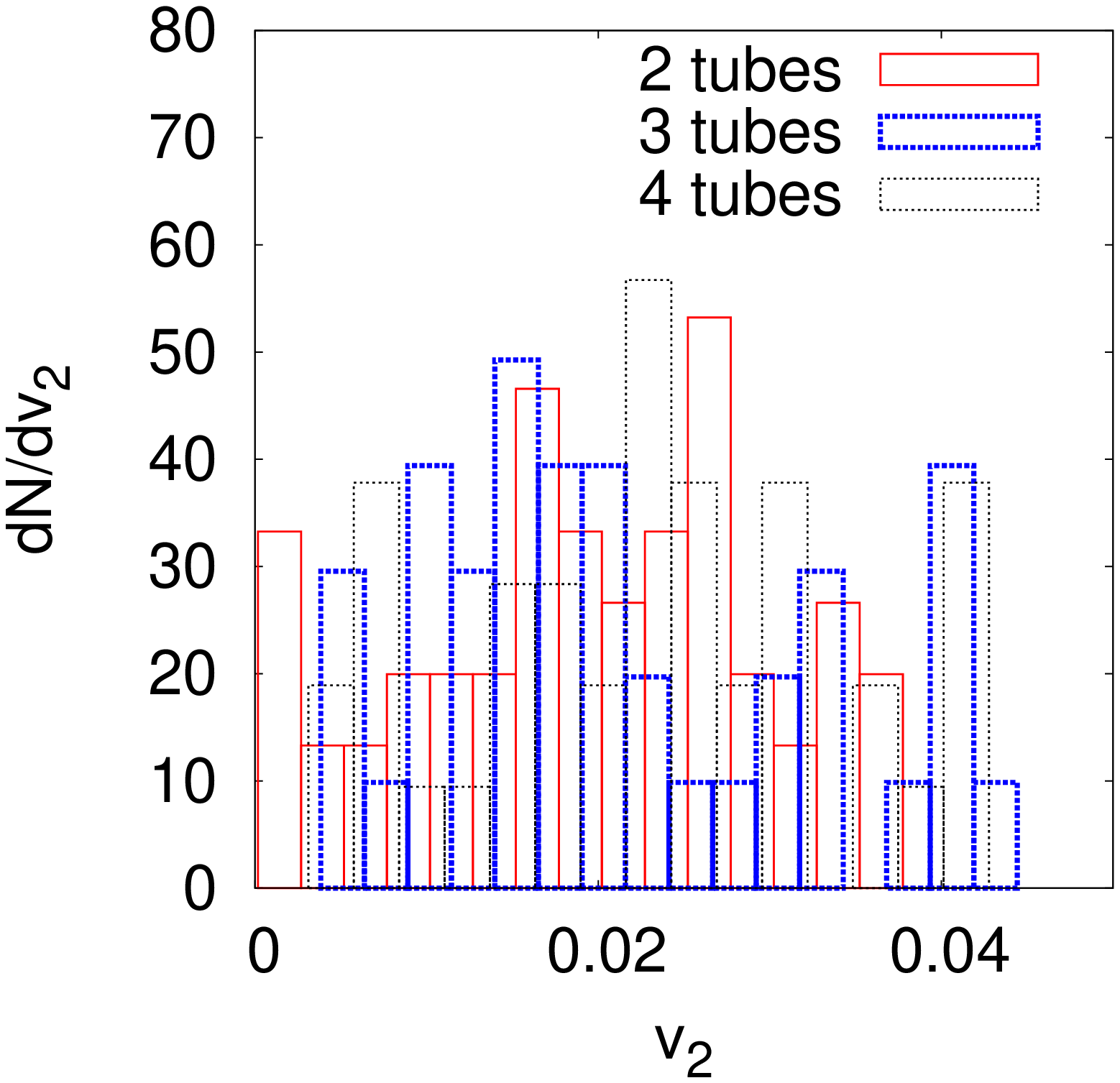}
\includegraphics[width=5.6cm]{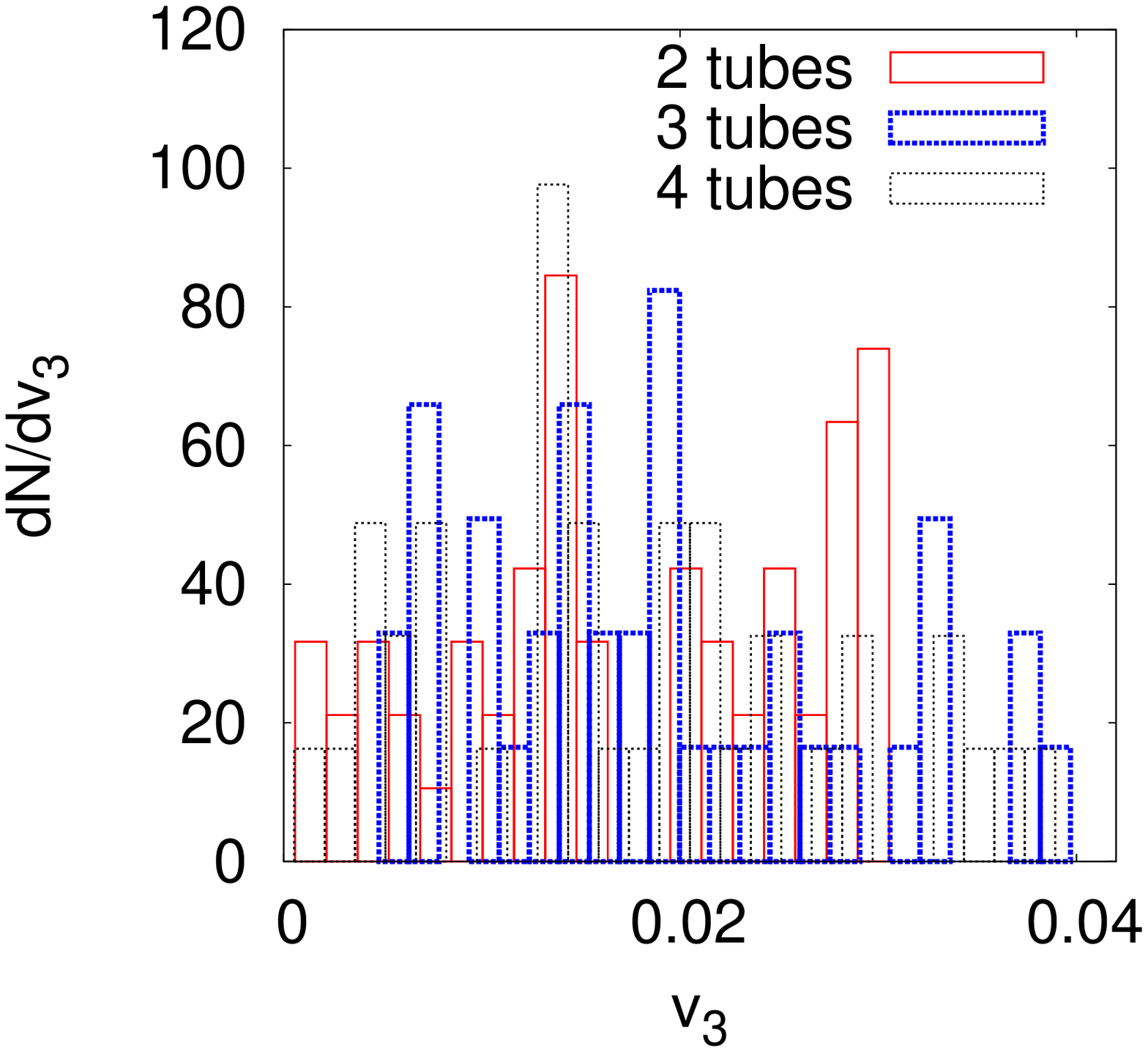}
%\vspace*{-.5cm}
\includegraphics[width=5.6cm]{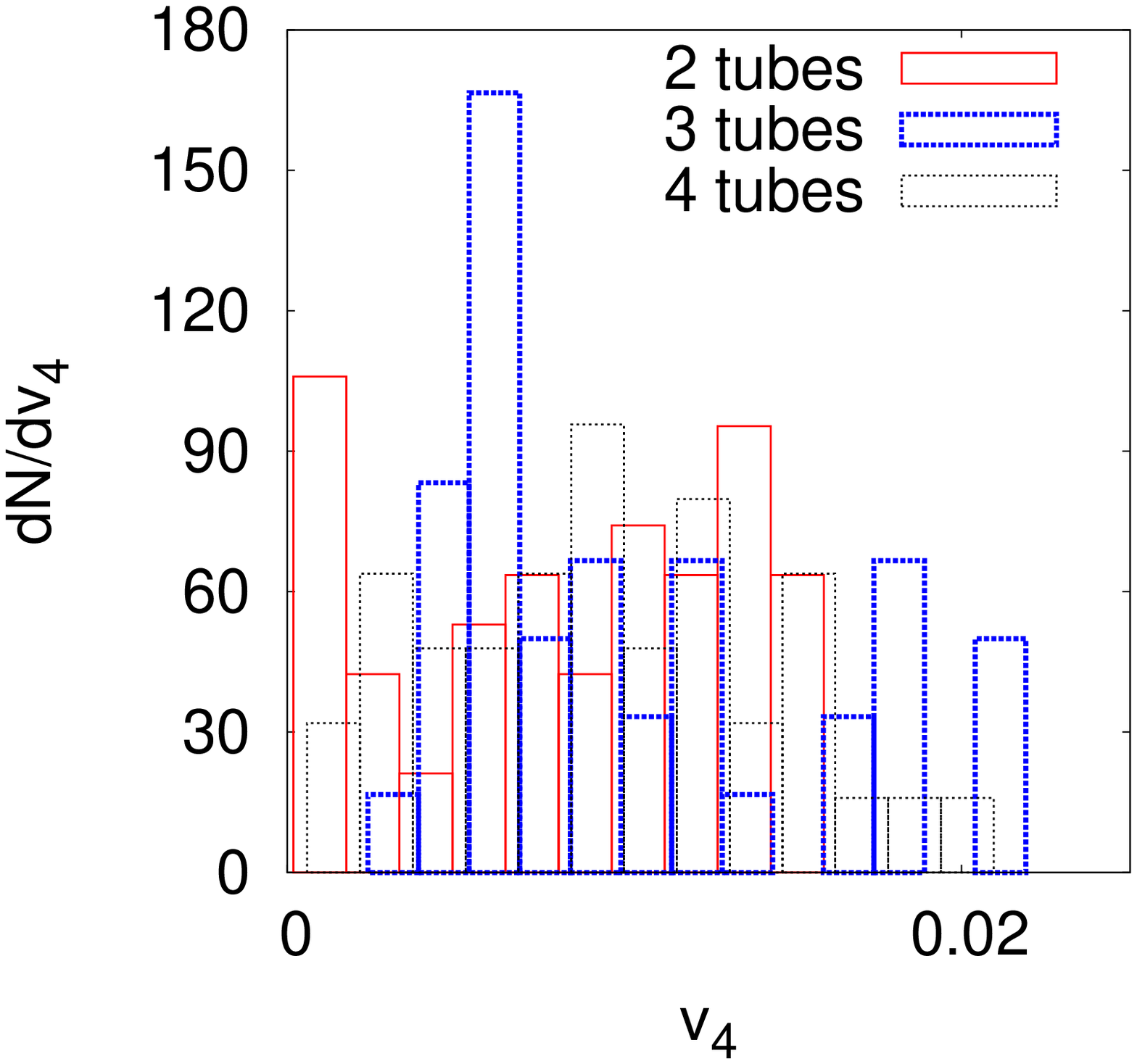}
\includegraphics[width=5.6cm]{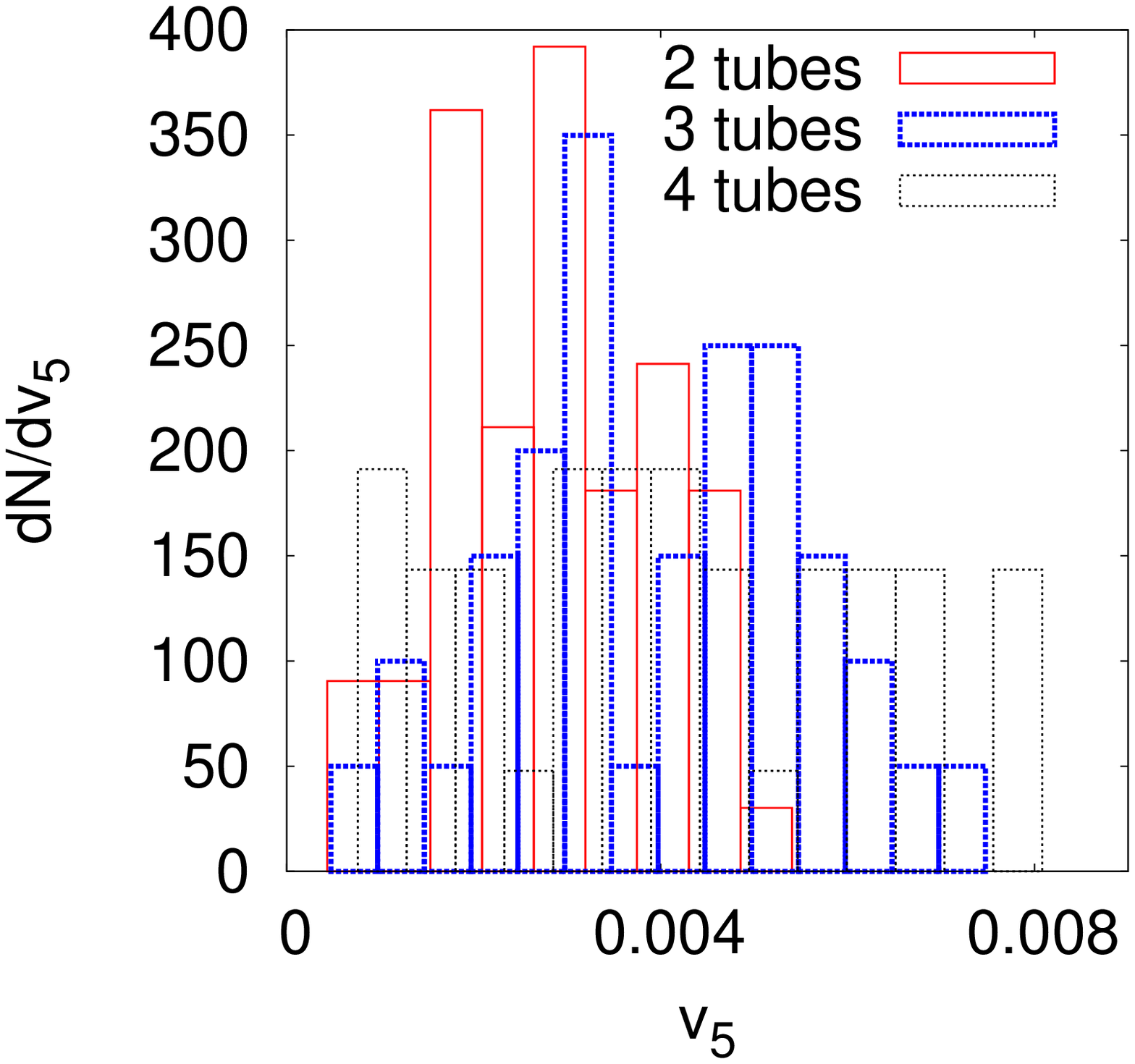}
\end{center}
%\vspace*{-.5cm}
\caption{Distributions of some of the Fourier components $v_n$ of the single-particle azimuthal distributions $dN/d\phi$, produced by 2-, 3- and 4-tube models.
For comparison, the corresponding values for the one-tube model are, respectively, $v_2=0.0569, v_3=0.0740, v_4=0.0479$ and  $v_5=0.0160$.}
\label{vn}
\end{figure}
In FIG.~\ref{vn}, we show how some of the Fourier coefficients $v_n$ are distributed in 2-, 3- and 4-tube cases. As expected, they are widely spread and also show smaller values as
compared to the one-tube case, where evidently they are sharply defined.

\bigskip
\noindent
{\bf Correlations among $\Psi_n$}
\medskip

Next, we proceed to discuss the correlation between the event planes $\Psi_n$ for harmonic components.
According to the picture of the peripheral tube model, a small fraction of elliptic flow is correlated to the triangular flow, as both of them are associated with the temporal evolution of a given tube.
We relegate the results and relevant discussions to section V.D.
There, it is shown that the correlations between event planes in our model are consistent with random distribution as the number of events becomes significant.

\bigskip
\noindent
{\bf Two-particle correlations in $\Delta\phi$}
\medskip

\begin{figure}[thb]
%\vspace*{-.5cm}
\begin{center}
\includegraphics[width=7.cm]{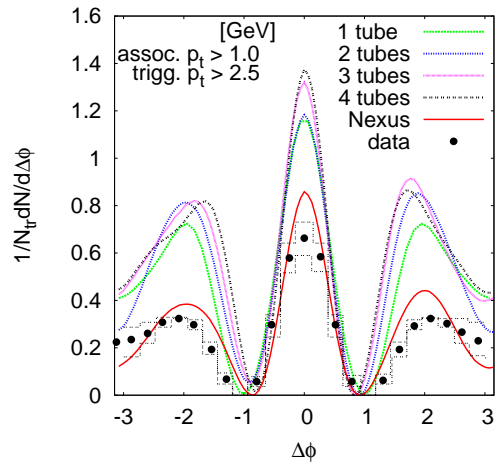}
\end{center}
%\vspace*{-2.8cm}
\caption{Two-particle correlations as functions of $\Delta\phi$, computed with 1-, 2-, 3- and 4-tube models. 
The results with NEXUS initial conditions and the STAR data for the 0-12\% centrality window~\cite{horner-ridge} are shown for comparison.
It should be mentioned that in NEXUS, both the tube size and their radial positions fluctuate and not maintained constant as in Eq.(\ref{parmulti}).
}
\label{correlations}
\end{figure}

Finally, we went on to compute the two-particle correlations, predicted by each $n$-tube model. 
FIG.~\ref{correlations} shows the  results. 
As expected, the two-particle correlation is almost independent of the number of peripheral tubes, showing that the interference arising from different tubes are already completely canceled with only 50 random events.

While in one-tube case, because there is just one event, $dN/d\phi\,$ is uniquely defined, in the multi-tube cases the IC are fluctuating from event to event, as given by Eq.~(\ref{parmulti}), so the one-particle azimuthal distribution $dN/d\phi$ varies from event to event.
Also, because of such fluctuations, the two-particle correlation due to one such event is similar to the one shown in FIG.~\ref{mtube-event}, bottom.
However, when averaged over the totality of such events, they give almost the same results as for the one-tube case. 
We understand that this almost coincidence indicates that {\it what determines the several structures of two-particle correlation is what each peripheral tube produces during the expansion of the bulk matter and has nothing to do with the global distribution of the matter at the initial time}.
Evidently, the causal connection we mentioned at the end of Subsection IV.B has to do, in the multi-tube case,  with the structure coming from each peripheral tube, which started from the same point.

Now, care should be exercised about this conclusion, that we should not forget that hydrodynamics {\it is not linear} in principle, so the overlapping of the effects coming from the multiple tubes producing constructive
and destructive interferences are only approximate. 

In the examples above, FIG.~\ref{correlations} tells us that this approximation is quite good, for $n=2,3,4$ tubes. 
Obviously, however, if the number of tubes increases further so that they themselves begin to overlap, the results could not hold anymore.
A rough estimate of such a number is about 15 in azimuth for the parameters we used in Eq.~(\ref{parmulti}), which is too big to happen in practice.

Another point FIG.~\ref{correlations} calls attention is, besides the almost independence of the tube multiplicity, the close similarity of these multi-tube results when compared with NEXUS correlation, that is computed directly from the IC, without using the tube model. 
Evidently, this similarity means that, as far as long-range two-particle correlation is concerned, the tube-model catches the essence of NeXSPheRIO, which by itself reproduces quite well the data, as shown. 
In order to get a better numerical agreement with data, we can adjust some of the parameters, as shown in Subsection IV-C.

%\newpage

\subsection{Some comparisons with eccentricities$\rightarrow$flow-harmonics interpretation of ridge correlation}

Several analyses have been performed in an effort to understand the relation between flow harmonics and the initial eccentricities~\cite{hydro-eve-06,hydro-v3-04},  pointing out a close relation between the elliptic flow $v_2$  and the ellipticity of the IC $\epsilon_2$, and also between triangular flow $v_3$ and the triangularity of the IC $\epsilon_3$. 
Nevertheless, although they did use fluctuating IC, in all these studies, fairly well-behaved (Gaussian-like) IC were used.

In Ref.~\cite{sph-vn-04}, we made a systematic analysis, trying to see, among others, which is the relation between individual cumulant and the produced flow harmonics, and by using NEXUS IC featured by many high-energy tubes. 
Among several interesting results, perhaps the one produced by the IC plotted below in FIG.~\ref{4tube-IC} is a typical example to elucidate the differences between the peripheral-tube model and the popular triangularity$\rightarrow$triangular flow interpretation of the long-range two-particle correlation.

\begin{figure}[ht]
%\vspace{-1.8cm}
 \vspace{-0.8cm}
 \begin{center}
  \includegraphics[width=10.cm]{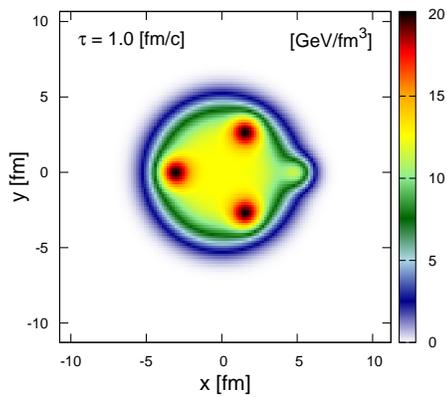}
% \vspace{0.5cm}
  \vspace{-1.0cm}
  \caption{\label{4tube-IC} Energy distribution of four tubes placed on top of the averaged NEXUS IC, with zero $\epsilon_3$.}
 \end{center}
 \vspace{-.5cm}
\end{figure}

\begin{figure}[ht]
 \begin{center}
  \includegraphics[width=8.cm]{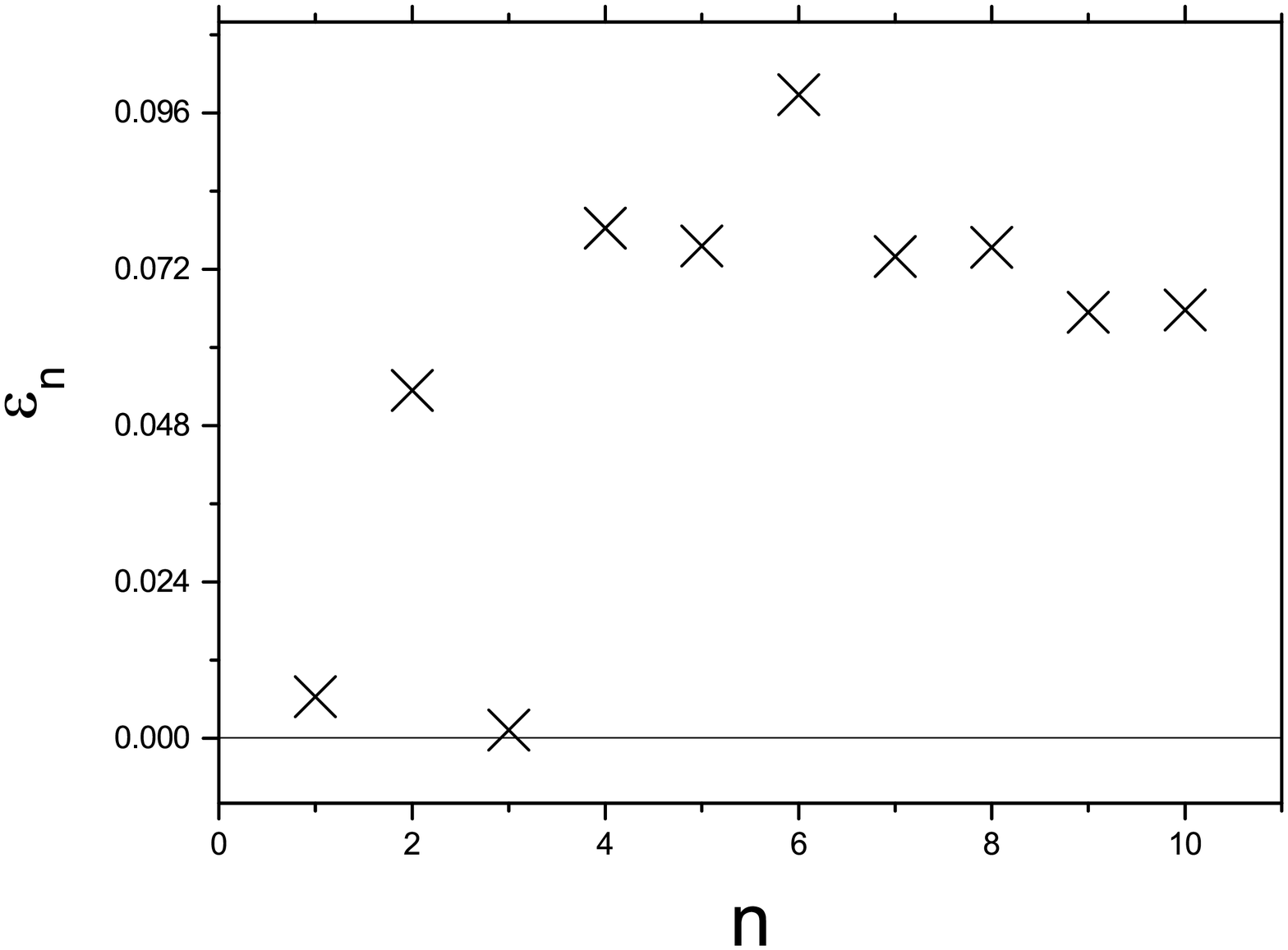}
  \vspace*{-.3cm}
  \includegraphics[width=8.cm]{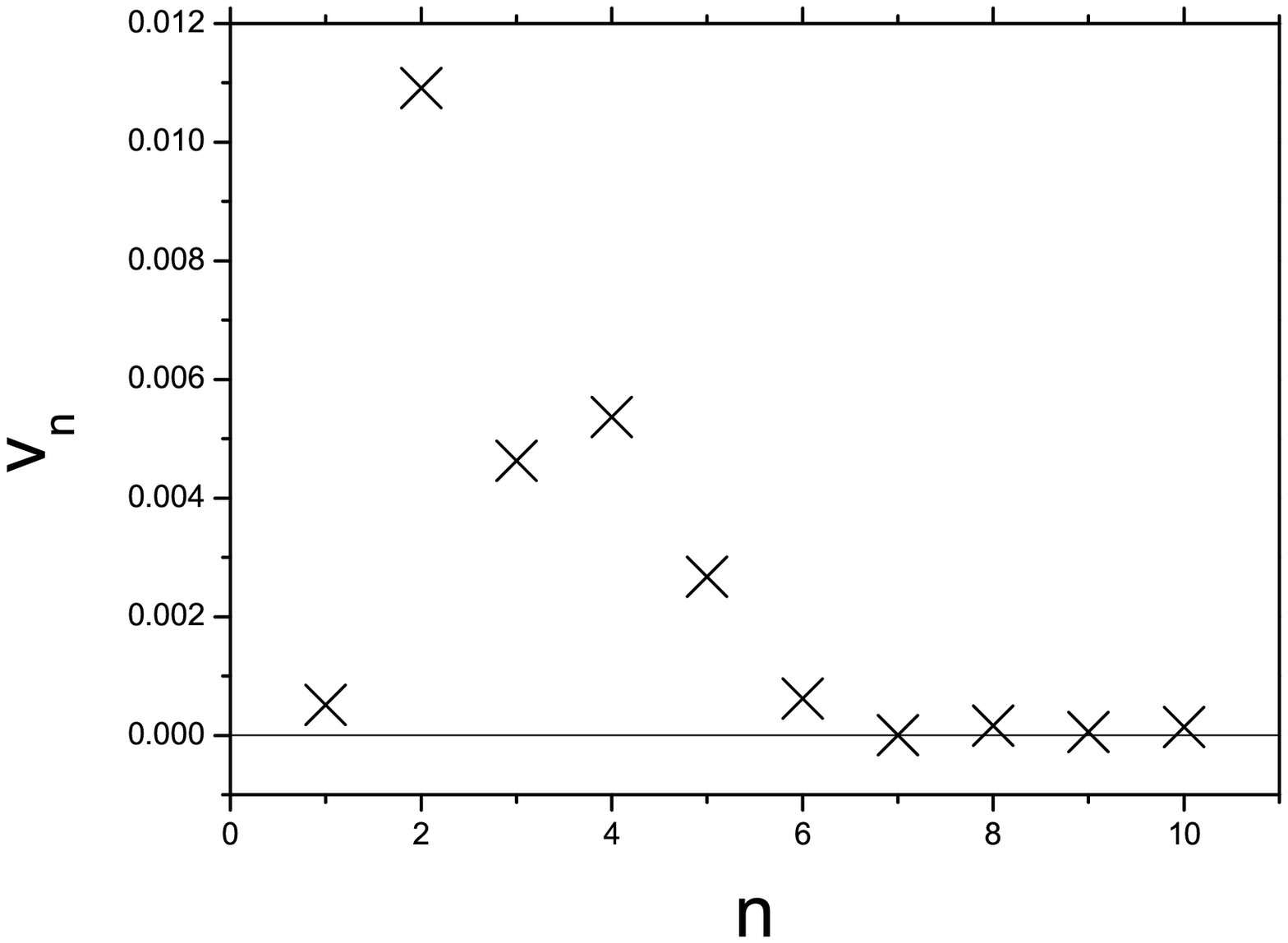}
%  \vspace*{-2.5cm}
  \caption{\label{4tube-event} Initial eccentricities and calculated flow harmonics of the IC in FIG.~\ref{4tube-IC}.}
 \end{center}
\end{figure}

This distribution has been made by using the average NEXUS $200\,$GeV central $Au+Au$ IC as background, and arranged a number of tubes so that $\epsilon_3=0$ (see also Ref.~\cite{sph-vn-02}). 
Thus, if flow harmonics are more or less linearly connected to the eccentricities of the energy distribution, $v_3$ would be =0, and no two-particle long-range correlation would appear. 
However, if we explicitly compute the flow harmonics, we do find, as shown in FIG.~\ref{4tube-event}, non-zero $v_3$, so, together with non-zero $v_2$, it should produce ridge and shoulders. 
How does one explain physically the appearance of $v_3$ and ridges in this interpretation?

Now, let us change the interpretation and make use of the peripheral-tube model. 
In this case, as shown by FIG.~\ref{rlimit}, the three biggest tubes are not peripheral enough, so do not give any relevant effect to the shadowing. 
Only the smallest, but the most peripheral, tube produces the relevant effect. 

This tube does provoke shadowing to the background matter, producing two peaks as in FIGS.~\ref{flow},\ref{angular}, and constructing the ridge + shoulders structure in two-particle correlation.
In terms of flow harmonics, the two-peak distribution contains both $v_2$ and $v_3$ with the correct symmetry angle between them.
So, we can physically understand the ridge and shoulder formation phenomenon here.

One might argue that the IC shown in FIG.~\ref{rlimit} is merely a particular case.
It can be shown, however, that on an event-by-event basis, ICs with vanishing $\epsilon_3$ may systematically produce nonvanishing $v_3$.
This is achieved by devising a collection of ICs by manually removing the $\epsilon_3$ component, and study the resultant collective flow on an event-by-event basis.
Although $\epsilon_3$ is identically vanishing, hydrodynamic evolution gives rise to sizable $v_3$, as shown in FIG.~7 of Ref.~\cite{sph-vn-04}.

Regarding the comparisons made here, we should not forget the results we discussed in the previous subsection. In IV,  D, we have shown that if many tubes are randomly distributed on top of a smooth averaged NEXUS energy distribution, not only the eccentricities $\epsilon_n$ but also the flow harmonics $v_n$ are entirely random, event-by-event. 
Nevertheless, after the averaging carried out, the final two-particle correlation is always the same with three peaks, independent of the number of tubes. 
As we understand, what really matters in two-particle correlation is what each peripheral tube does locally and not the overall shape of the initial state.

\bigskip

We can summarize this Section as follows:

\begin{itemize}
\item
Hydrodynamic approach with fluctuating IC, with tube-like structures, produces ridge structures in two-particle correlations at low and intermediate $p_T$.
\item
The peripheral-tube model (now extended to multi-tube configurations) is a possible production mechanism for the three-ridge structure seen in heavy-ion collisions.
\item
It gives a unified description of the ridge structures, both near-side and away-side ones (shoulders).
\item
The mechanism of ridge production is local: what is important is each peripheral tube, and what it does separately, not the global structure of the IC.
\item
Because these structures are produced by each of the peripheral tubes, they are causally related. 
The causal connection of the ridge~\footnote{In fact, most models that use an (approximately) boost-invariant IC plus hydrodynamics (or even transport theory) can reproduce similar longitudinal correlations.
In general terms, the relevant mechanism is that the initial interaction zone's shape is turned into a longitudinal correlation structure in the IC when the nuclei pass through each other. 
Moreover, hydrodynamics maintains the above correlations during the course of temporal evolution, as they are eventually manifested in those of final state hadrons.}, with respect to the longitudinal distribution, has been discussed elsewhere~\cite{dumitru}.
Regarding the azimuthal direction, the resultant correlation structures, including the near-side and away-side ones, are entirely connected by causality.
\end{itemize}

\newpage

\bigskip
%{\it These results seem to show that, in terms of flow components, what is important to %producing the observed ridge structure is not so much the global combination of the flow %components but the flow components due to each peripheral tube.}

%\fi
%\newpage
\section{Some observed properties and relation with the peripheral tubes}
%*5. Further results on the peripheral-tube model for ridge correlation, Yogiro Hama, Rone P.G. Andrade, Frederique Grassi, Jorge Noronha (Sao Paulo U. & UFOP, Ouro Preto), Wei-Liang Qian (UFOP, Ouro Preto & Sao Paulo U.). Dec 2012. 6 pp. Published in Acta Phys.Polon.Supp. 6 (2013) 513-518

This section focuses on some specific features of the two-particle correlations closely associated with the observations.
In the past few years, measurements of di-hadron azimuthal correlations were reported by various collaborations from RHIC and LHC, for different collision systems, centralities, as well as momentum intervals.
Many exciting features were observed.
For instance, it is found that the away-side correlation evolves from double- to a single-peak structure when one goes from most central to peripheral collisions.
In fact, such copious data provides an intriguing challenge to all existing theoretical models.
On the theoretical side, extensive studies of event-by-event based hydrodynamics indicate that the underlying physics of observed two-particle correlations at low and intermediate transverse momenta is mostly associated with the collective flow~\cite{hydro-v3-01,hydro-v3-02,hydro-corr-ph-01,hydro-v3-08,hydro-vn-02}.
To be specific, the form of the correlations has been given in terms of harmonic coefficients, $v_{n}$, and in particular, the triangular flow $v_3$ plays a vital role.
Moreover, many quantitative studies~\cite{hydro-v3-01,hydro-v3-02, sph-vn-04} regarding the relation between $v_{n}$ and $\epsilon_n$ have been carried out.
The peripheral tube model~\cite{sph-corr-02,sph-corr-03,sph-corr-04,sph-vn-04,sph-corr-07}, on the other hand, provides an alternative picture for the generation of the triangular flow and consequently the di-hadron correlations. 

In the framework of event-by-event hydrodynamics, the fluctuations in initial energy density distributions are separated into two parts.
They correspond to the background and characteristic localized fluctuations.
The latter is represented by a peripheral high-energy tube (hot spots), while the former is replaced by a smooth elliptic distribution determined by the collision geometry. 
The peripheral tube carries the role of irregular event-by-event fluctuations.
On the other hand, those properties that are less significant or common between different events are viewed as the background, and thus the remaining of the IC is replaced by an average. 
Subsequently, in this picture, we ignore those global fluctuations whose wavelength is comparable to the system size.  
The inhomogeneity in IC due to event-by-event fluctuations is captured by the localized peripheral hot tubes. 
As a result, for instance, the evolution of the shape of the away-side structure in the two-particle correlations from central to peripheral collisions is not attributed to the centrality dependence of harmonic coefficients.
As will be discussed below, it is viewed mostly due to the centrality dependence of the multiplicity fluctuations and that of the background elliptic flow.
Such an explanation also naturally describes some of the features observed in data analysis, which is less intuitive otherwise. 
The remainder of this section is dedicated to study the trigger-angle dependence~\cite{sph-corr-ev-02,sph-corr-ev-04,sph-corr-ev-07,sph-corr-ev-08}, centrality dependence~\cite{sph-corr-ev-06} of di-hadron correlation in Au+Au collisions, as well as the extracted model paremeters~\cite{sph-corr-08} in comparison with the data.

\subsection{Trigger-angle dependence}

%*7. On the Origin of the Trigger-Angle Dependence of the Ridge Structure, Yogiro Hama, Rone P.G. Andrade, Frederique Grassi (Sao Paulo U.), Wei-Liang Qian (UFOP, Ouro Preto). Nov 2011. 4 pp. Published in Prog.Theor.Phys.Suppl. 193 (2012) 167-171
%*6. Origin of the trigger-angle dependence of di-hadron correlations, Wei-Liang Qian (UFOP, Ouro Preto), Rone Andrade, Fernando Gardim, Frédérique Grassi, Yogiro Hama (Sao Paulo U.). Jul 2012. 8 pp. Published in Phys.Rev. C87 (2013) no.1, 014904
%*11. Trying to understand the in-plane/out-of-plane effect in long-range correlations, Yogiro Hama, Rone P.G. de Andrade, Frederique Grassi (Sao Paulo U.), Wei-Liang Qian (UFOP, Ouro Preto). 2011. 6 pp. Published in PoS WPCF2011 (2011) 048

\begin{figure*}[ht]
  \centerline{\includegraphics[width=500pt]{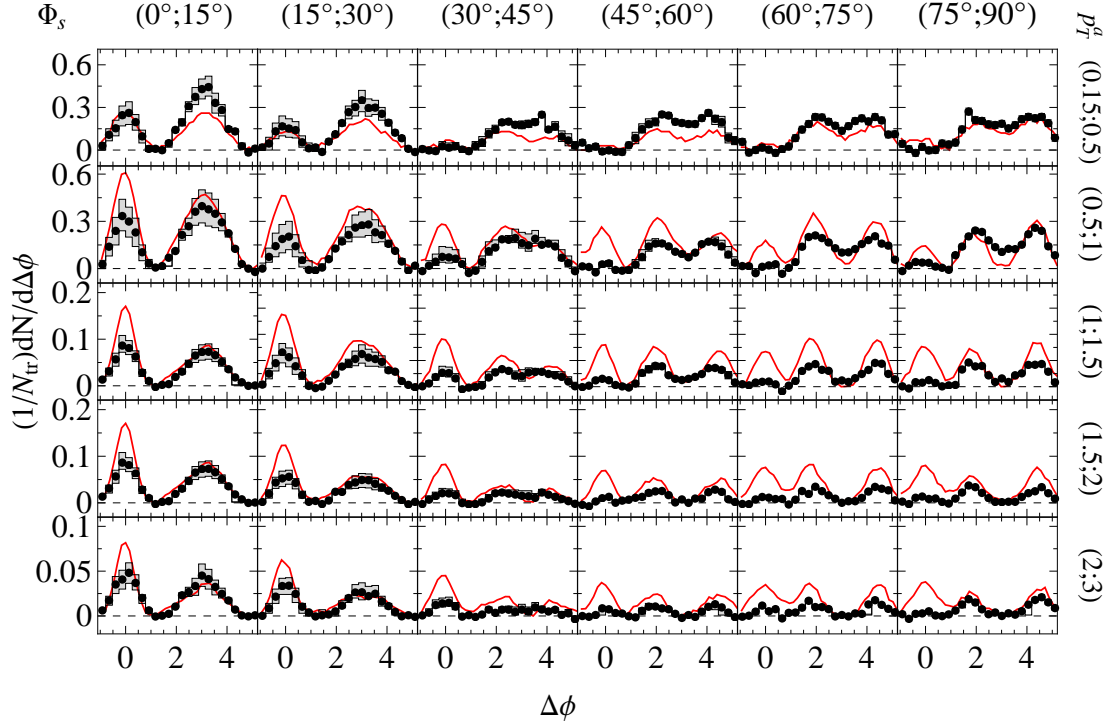}}
  \caption{
    The subtracted di-hadron correlations as a function of $\Delta\phi$ for different $\phi_s=\phi_{trig}-\phi_{EP}$ and $p_T^{a}$ with $3 < p_T^{\mathrm{trig}} < 4 GeV$ and 
   $|\Delta\eta| > 0.7$ in 20 - 60\% Au+Au collisions. 
   The $\phi_s$ range increases from 0-15$^{\circ}$ (left column) to 75-90$^{\circ}$ (right column); the $p_T^{a}$ range increases from 0.15-0.5 GeV (top row) to 2-3 GeV (bottom row). 
   NeXSPheRIO results in solid curves are compared with STAR data in filled circles~\cite{RHIC-star-plane-2}. 
   The shaded area between two histograms indicate the systematic uncertainties from flow subtraction.
   }
  \label{sec5-1-fig2}
\end{figure*}

In this subsection, we study the trigger-angle dependence of di-hadron correlations. 
STAR Collaboration reported measurements~\cite{RHIC-star-plane-1} of di-hadron azimuthal correlations as a function of the trigger particle's azimuthal angle relative to the event plane, $\phi_s=|\phi_{tr}-\Psi_{EP}|$ at the different trigger and associated transverse momenta $p_T$. 
The data are for 20-60\% Au+Au collisions at 200 A$\,$GeV. 
In a follow-up study~\cite{RHIC-star-plane-2}, the correlation was further separated into ``jet'' and ``ridge'', where the ridge yields are obtained by considering hadron pairs with large $|\Delta\eta|$. In this procedure, one assumes that the ridge is uniform in $\Delta\eta$ while jet yields are not. 
Such an assumption is quite reasonable when one considers the measured correlation at low $p_T$ without a trigger particle~\cite{RHIC-star-ridge-5}. 
In their work, all correlated particles at $|\Delta\eta| > 0.7$ are considered part of the ridge. 
After subtracting the contributions from the collective flow components, $v_2$ and $v_4$ using the ZYAM method, it was observed that the correlations vary with $\phi_s$ in both the near and away side of the trigger particle. 
As shown in FIG.~\ref{sec5-1-fig2}, the ``ridge'', indicated by the peak at $\Delta\phi\sim 0$, drops when the trigger particle goes from in-plane to out-of-plane, which corresponds to successive plots from left to right on a given row.
Moreover, with increasing $\phi_s$, the correlations in the away-side evolve from a single peak, as can be observed at $\Delta\phi\sim \pi/2$ in the plots on the two leftmost columns of FIG.~\ref{sec5-1-fig2}, to a double peak structure, as presented in the last three columns on the right of the same figure. 

We understand that it is quite probable that both experimental results on particle correlations, namely, Refs.~\cite{RHIC-star-plane-1, RHIC-star-plane-2}, primarily reflect the properties of the medium.
Also, the background's subtraction was carried out in terms of the {\it average flow} harmonics, from the proper correlations, including the information on {\it event-by-event fluctuating flow}. 
It is not clear that the remaining correlation associated with the collectivity of the system is insignificant.
In this context, it is worthwhile to carry out an explicit calculation to quantitatively verify to what extent the data can be reproduced by a hydrodynamic approach.
As the NeXSPheRIO code provides a good description of observed data in addition to the structures in di-hadron long-range correlations, it is therefore interesting to continue employing the model. 
In what follows, we first carry out a hydrodynamic study on the trigger-angle dependence of di-hadron correlations by using the NeXSPheRIO code. 
The calculations are done also with the pseudo-rapidity cut $|\Delta\eta| > 0.7\,$.  
To evaluate di-hadron correlations, we generate 1200 NEXUS events in the 20-60\% centrality window for 200 A$\,$GeV Au-Au collisions. 
At the end of each event, the Monte-Carlo generator is invoked for hadron production.
For the most central collisions, it is invoked 300 times, and the number increases as one goes to more peripheral centrality windows.
In the end, the generator is invoked 500 times for the most peripheral collisions.
Here we emphasize that there is no free parameter in the present simulation since the few existing ones have been fixed in earlier studies of $\eta$ and $p_T$ distributions~\cite{sph-v2-fluct}.   
Numerical results are presented in FIG.~\ref{sec5-1-fig2} in solid lines for different angles of trigger particles and at different associated-particle transverse momentum.
They are compared with the STAR data~\cite{RHIC-star-plane-2} in filled circles and flow systematic uncertainties in histograms. 
As the effect of the $\Delta\eta$ cut is implemented in the study, the calculated results shown in FIG.~\ref{sec5-1-fig2} are adequate for comparison with the STAR data. 
From FIG.~\ref{sec5-1-fig2}, one sees that the NeXSPheRIO code reasonably reproduces the main features of the data. 
The calculated correlations decrease, especially on the near side, when $\phi_s$ increases.
This feature has been interpreted to be related to the path length that the parton traverses~\cite{RHIC-star-plane-2}.
Here, however, we see that it is reproduced surprisingly well by a hydrodynamic approach. 
Nonetheless, some deviations are also observed.
For $\phi_s > 45^{\circ}$ degrees, the near side peak is more significant in the calculated correlation. 
Also, the associated $p_T$ dependence of the magnitude of the near-side peak is not produced properly. 
It is expected that the results will be refined if the statistics are further improved.
As expected, NeXSPheRIO results fit better for low momentum, and the deviations increase at higher momentum, as the correlations obtained by hydrodynamical simulations start to overestimate those of STAR measurements.

\begin{figure*}[ht]
\begin{tabular}{ccc}
\begin{minipage}{160pt}
\centerline{\includegraphics[width=180pt]{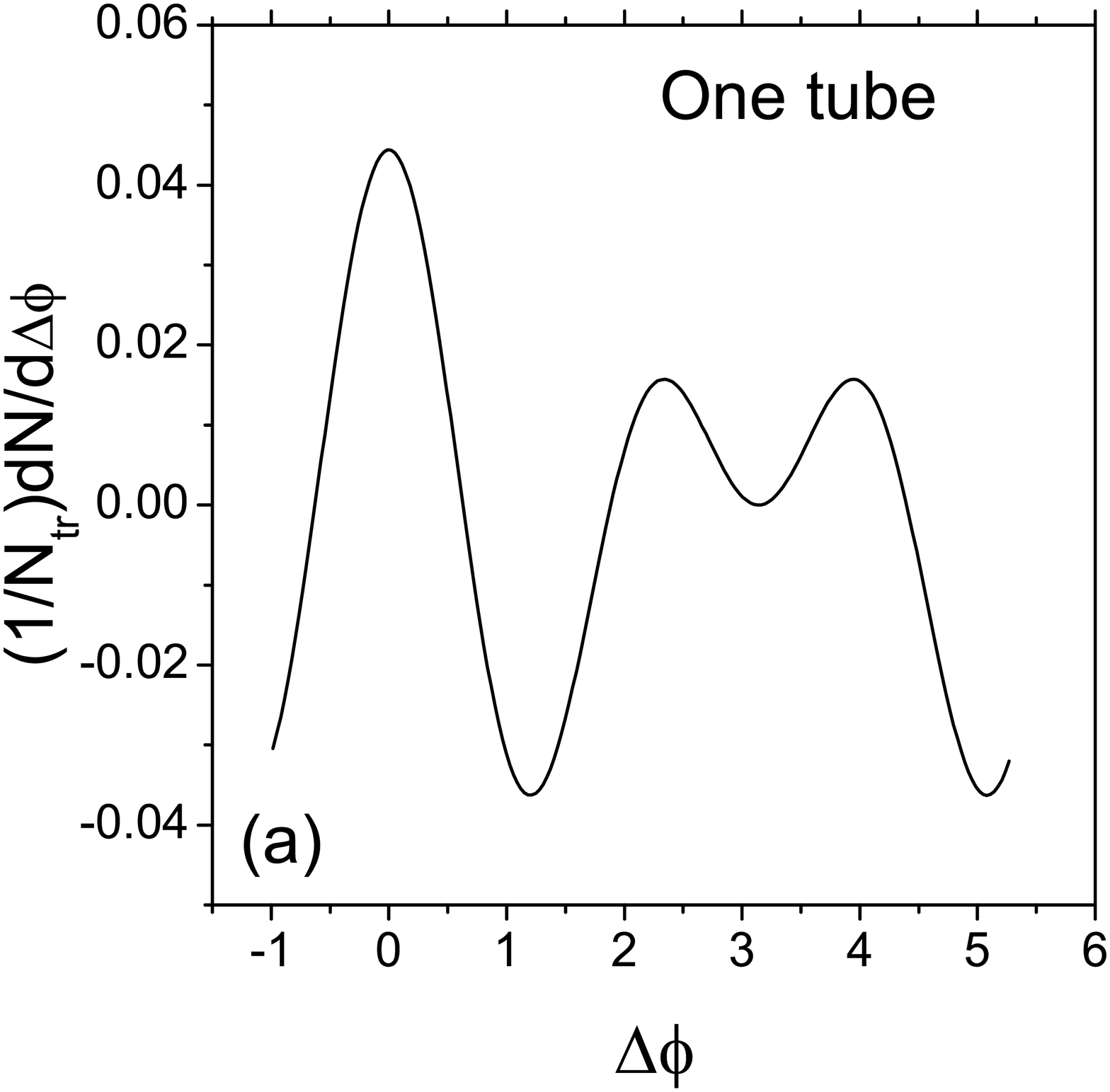}} 
\end{minipage}
&
\begin{minipage}{160pt}
\centerline{\includegraphics[width=180pt]{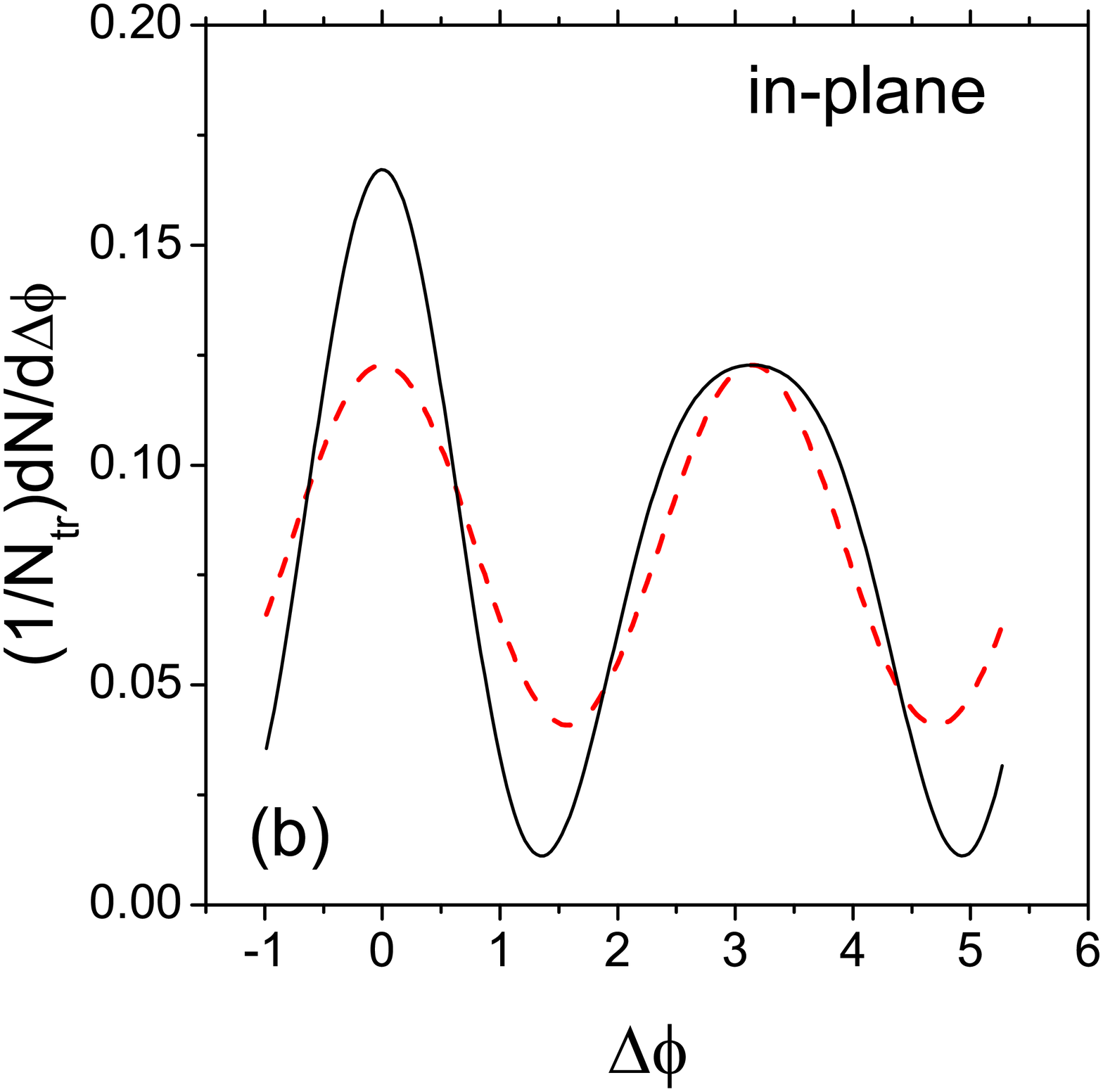}}
\end{minipage}
&
\end{tabular}
\begin{minipage}{160pt}
\centerline{\includegraphics[width=180pt]{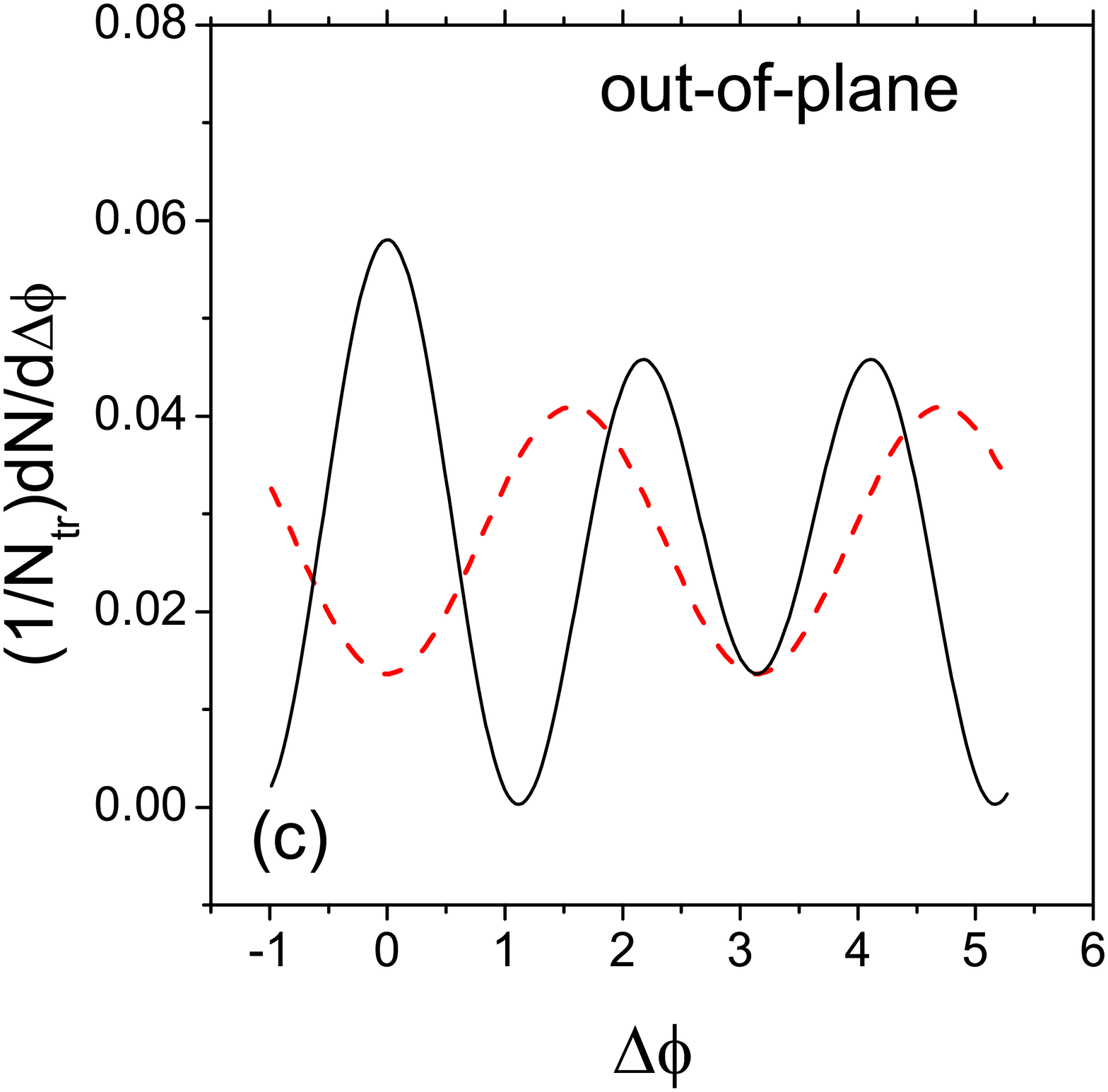}} 
\end{minipage}
 \caption{ Plots of di-hadron correlations calculated by cumulant 
 method: (a) the peripheral-tube contribution; 
 (b) the one from the background (dashed line) and 
 the resultant correlation (solid line) for in-plane triggers, as 
 given by Eq.~(\ref{eq-sec5-1-9}); and (c) the corresponding ones for the 
 out-of-plane triggers, Eq.~(\ref{eq-sec5-1-10}).} 
 \label{sec5-1-fig3}           
\end{figure*} 

Now we have shown that hydrodynamic simulations can reproduce the main features observed experimentally. 
Then, how is this effect produced?
In previous sections, we tried to clarify the origin of the effect by using the tube model. 
Here, we will further show that the model can be employed to interpret the main features of the data by using an approximate analytical approach. 
The derivation relies on the following three hypotheses: 
\begin{itemize}
\item The collective flow consists of contributions from the background radial + elliptic flow and those induced by a peripheral tube with a double-peak structure.
\item The latter is generated due to the interaction between the background and the peripheral tube and, 
therefore, the flow produced in this process is correlated to the tube. 
\item Event by event multiplicity fluctuations are further considered as a correction that sits atop of the above collective flow of the system. 
\end{itemize} 
Let us briefly comment on these hypotheses. 
As discussed in previous sections, according to the idea of the one-tube model, a small portion of the background flow is deflected by the peripheral tube, as extra Fourier components of the flow are generated by this process. 
Though these components are essential to the present study, their magnitudes are small, so we treat it perturbatively.
Also, we are considering just one tube, because as shown in subsection IV-D, the correlation due to random tubes is independent of the number of tubes. See FIG.~\ref{correlations}.
There are many possible sources of fluctuations, such as flow fluctuations, multiplicity fluctuations, etc., here, we will only consider multiplicity fluctuations in this simple model as assumed in the third hypothesis. 
As will be shown below, this turns out to be enough to derive the observed feature in di-hadron correlations.

Using the hypotheses stated above, we write down the one-particle distribution as a sum of two terms: the distribution of the background and that of the tube. 
\begin{eqnarray} 
  \frac{dN}{d\phi}(\phi,\phi_t) =\frac{dN_{\mathrm{bgd}}}{d\phi}(\phi) +\frac{dN_{\mathrm{tube}}}{d\phi}(\phi,\phi_t), 
 \label{eq-sec5-1-1}  
\end{eqnarray} 
where 
\begin{eqnarray} 
 \frac{dN_{\mathrm{bgd}}}{d\phi}(\phi)&=&\frac{N_b}{2\pi}(1+2v_2^b\cos(2\phi)),    \label{eq-sec5-1-2}\\  
 \frac{dN_{\mathrm{tube}}}{d\phi}(\phi,\phi_t)&=&\frac{N_t}{2\pi}\sum_{n=2,3}2v_n^t\cos(n[\phi-\phi_t])   \label{eq-sec5-1-3}     
\end{eqnarray} 
As for the background flow, in Eq.~(\ref{eq-sec5-1-2}) we consider the most simple case, by parametrizing it with the elliptic flow parameter $v_2^b$ and the overall multiplicity, denoted by $N_b$. 
The contributions from the tube are assumed to be independent of its angular position $\phi_t$, and we take into account the minimal number of Fourier components to reproduce the two-particle correlation due to the sole existence of a peripheral tube in an isotropic background. 
Therefore, only two essential components $v_2^t$ and $v_3^t$ are retained in Eq.~(\ref{eq-sec5-1-3}).
We note here that the overall triangular flow in the present approach is generated only by the tube, {\it i.e.}, $v_3=v_3^t$, and so its symmetry axis is correlated to the tube location $\phi_t$. 
The azimuthal angle $\phi$ of the emitted hadron and the position of the tube $\phi_t$ are measured with respect to the event plane $\Psi_2$ of the system. 
We note that the overall elliptic flow of the event is dominated by that of the background, as most of the yield comes from the background, not from the tube.
Subsequently, $\Psi_2$ is mostly determined by the elliptic flow of the background $v_2^b$. 

Following the methods used by the STAR experiment\cite{RHIC-star-plane-1,RHIC-star-plane-2}, the subtracted di-hadron correlation is given by
\begin{eqnarray} 
 \left<\frac{dN_{\mathrm{pair}}}{d\Delta\phi}(\phi_s)\right>   =\left<\frac{dN_{\mathrm{pair}}}{d\Delta\phi}(\phi_s)\right>^{\mathrm{prop}} -\left<\frac{dN_{\mathrm{pair}}}{d\Delta\phi}(\phi_s)\right>^{\mathrm{mix}} , \nonumber \\
 \label{eq-sec5-1-4}  
\end{eqnarray} 
where $\phi_s$ is the trigger angle ($\phi_s=0$ for in-plane and $\phi_s=\pi/2$ for out-of-plane trigger). 
In one-tube model, 
\begin{eqnarray} 
\left<\frac{dN_{\mathrm{pair}}}{d\Delta\phi}\right>^{\mathrm{prop}}  =
 \int\frac{d\phi_t}{2\pi}f(\phi_t)
 \frac{dN}{d\phi}(\phi_s,\phi_t)  
\frac{dN}{d\phi} (\phi_s+\Delta\phi,\phi_t),   \nonumber \\ \label{eq-sec5-1-corr-proper}
\end{eqnarray}
where $f(\phi_t)$ is the distribution function of the tube. 
We will take $f(\phi_t)=1$, for simplicity. 

The combinatorial background 
$\left<{dN_{\mathrm{pair}}}/{d\Delta\phi}\right>^{\mathrm{mix}}$
%$\left<\frac{dN_{\mathrm{pair}}}{d\Delta\phi}\right>^{\mathrm{mix}}$ 
can be calculated by using either cumulant or ZYAM method~\cite{zyam-01}. 
In fact, both methods lend very similar conclusions in our model. 
Here, we first carry out the calculation using cumulant, it can be shown~\cite{sph-corr-ev-04}
\begin{equation} 
\begin{aligned}
&\left<\frac{dN_{\mathrm{pair}}}{d\Delta\phi} \right>_{\mathrm{in-plane}} ^{{\mathrm{(cmlt)}}}\\
&=\frac{<N_b^2>-<N_b>^2}{(2\pi)^2} 
     (1+2v_2^b)(1+2v_2^b\cos(2\Delta\phi))\\ 
&+(\frac{N_t}{2\pi})^2 
      \sum_{n=2,3}2(v_n^t)^2\cos(n\Delta\phi) ,
 \label{eq-sec5-1-9} 
\end{aligned}
\end{equation} 
and
\begin{equation} 
\begin{aligned}
 &\left<\frac{dN_{\mathrm{pair}}}{d\Delta\phi}\right>_{\mathrm{out-of-plane}}^{{\mathrm{(cmlt)}}}\\
  &=\frac{<N_b^2>-<N_b>^2}{(2\pi)^2}
     (1-2v_2^b)
     (1-2v_2^b\cos(2\Delta\phi))\\ 
  &+(\frac{N_t}{2\pi})^2 
      \sum_{n=2,3}2(v_n^t)^2\cos(n\Delta\phi).   
 \label{eq-sec5-1-10} 
\end{aligned}
\end{equation} 
One sees that, the cosine dependence of the background contribution gives an opposite sign in the out-of-plane correlation, as compared to that in the in-plane correlation. 
This negative sign leads to the following consequences. 
Firstly, there is a reduction in the amplitude of out-of-plane correlation both on the near side and the away side. 
More importantly, it naturally gives rise to the observed double peak structure on the away side.
Therefore, despite its simplicity, the above analytic model reproduces the main characteristics of the observed data.
The overall correlation is found to decrease. Meanwhile, away-side correlation evolves from a broad single peak to a double peak as $\phi_s$ increases. 
The correlations in FIG.~\ref{sec5-1-fig3} is plotted by an appropriate choice of parameters.
It is noted that very similar results will be again obtained if one evaluates the combinatorial mixed event contribution using the ZYAM method.
This can be shown by straightforward calculations~\cite{sph-corr-ev-04}. 

In our approach, the trigger-angle dependence of the di-hadron correlation is understood as due to the interplay between the elliptic flow caused by the initial almond deformation of the whole system and flow produced by fluctuations. 
The contributions due to fluctuations are expressed in terms of a high-energy-density tube and the flow deflected by it. 
However, the generic correlation due to the tube is preserved even after the background subtraction\cite{sph-corr-04}, by this simple model, it is shown explicitly that the result does not depend on either the cumulant or ZYAM method.
This is because, in either case, the form of combinatorial background is solely determined by the average flow harmonics.
From in-plane to out-of-plane direction, the background modulation is shifted, changing the phase.
As a result, the summation of the contributions of the background and that of the tube give rise to the desired trigger-angle dependence.

It is interesting to compare the present study with those using the Fourier expansion in~\cite{hydro-corr-ph-01,hydro-corr-ph-05}
\begin{equation} 
\begin{aligned}  
 &\left<\frac{dN_{\mathrm{pair}}}{d\Delta\phi}(\phi_s)\right>^{\mathrm{prop}} \\
 &=\frac{N^2}{(2\pi)^2}(1+2V_{2\Delta}\cos(2\Delta\phi)+2V_{3\Delta}\cos(3\Delta\phi))+\cdots
\end{aligned}
\end{equation}
For comparison, we rewrite proper correlation in terms of $V_{n\Delta}$ as follows
\begin{eqnarray}
V_{2\Delta}&=&\frac{N_t^2}{\left<N_b^2\right>\left(1+ 2v_2^b\cos(2\phi_s)\right)}\left(v_2^t\right)^2+ \cos(2\phi_s)v_2^b \label{V2d} \nonumber\\
\\
V_{3\Delta}&=&\frac{N_t^2}{\left<N_b^2\right>\left(1+ 2v_2^b\cos(2\phi_s)\right)}\left(v_3^t\right)^2 \label{V3d}\, .
\end{eqnarray}
One sees that the background elliptic flow $v_2^b$ dominates $V_{2\Delta}$ for both in-plane and out-of-plane directions, 
while $V_{3\Delta}$ is determined by the triangular flow $v_3^t$ produced by the tube.
Due to the factor $\cos(2\phi_s)$, the second term of Eq.~(\ref{V2d}) changes sign when the trigger angle goes from $\phi_s=0$ to $\phi_s=\pi/2$.
Dominated by this term, $V_{2\Delta}$ decreases with $\phi_s$, and it intersects $V_{2\Delta}=0$ at around $\phi_s=\pi/4$.
Since the first term in Eq.~(\ref{V2d}) is positive definite, the integral of $V_{2\Delta}$ with respect to $\phi_s$ is positive.
These features are in good agreement with the data analysis (see FIG.~1 of ref.\cite{hydro-corr-ph-01}).
On the other hand, the axis of triangularity is determined by the tube's position $\Psi_3=\phi_t$.
Since we have assumed a uniform distribution in the calculations, the event plane of triangularity is uncorrelated with the event plane $\Psi_2$, as generally understood~\cite{hydro-v3-01,glauber-en-3}, and consequently, the contribution from triangular flow should not depend much on the event-plane angle.
This is indeed shown in the above expression Eq(\ref{V3d}).
$V_{3\Delta}$ barely depends on $\phi_s$, if anything, it slightly increases with increasing $\phi_s$.
This characteristic is also found in the data~\cite{hydro-corr-ph-01}.

%==========================%

\begin{figure*}[ht]
\begin{center}
\includegraphics[width=500pt]{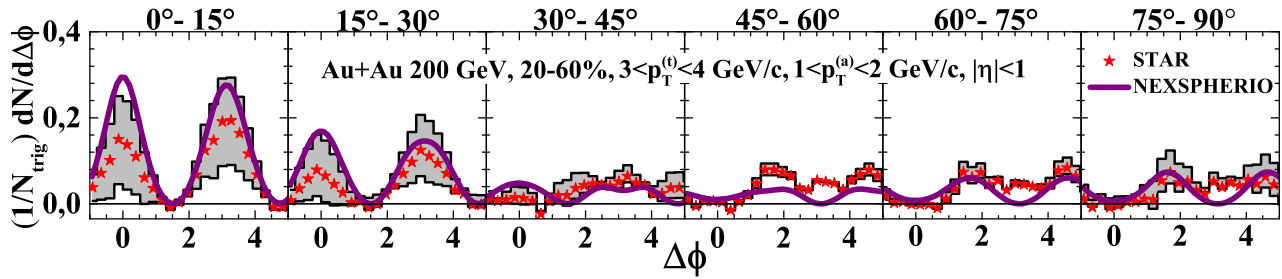}
\end{center}
\caption{The azimuthal di-hadron $\Delta \phi$ correlations for different values of $\phi_{s}$ = $| \phi_{t}-\psi_{2} |$.
The proper correlations are obtained by applying a cut on the pseudo-rapidity difference $| \Delta \eta | <$ 0.7 between the trigger and associated particles.
The resultant correlation is obtained using the ZYAM method, with $v_2$ and $v_3$ subtracted. 
The NeXSPheRIO results are shown by the solid purple curves, and the STAR data~\cite{RHIC-star-plane-3} are represented by the red stars whereas the gray area between the solid lines indicates the uncertainties.}
\label{corep}
\end{figure*}

The above STAR data~\cite{RHIC-star-plane-1, RHIC-star-plane-2} was analyzed by Luzum, who subsequently pointed out~\cite{hydro-corr-ph-01} that the observed two-particle correlations are consistent with being entirely generated by the collective flow.
Among others, the author argued that the flow background used in ZYAM subtraction did not consider the contribution from the triangular flow, $v_3$, which in turn leads to intrinsically different conclusions.
Answering the criticism, later, an improved study was carried out by the STAR collaboration~\cite{RHIC-star-plane-3}.
In particular, the ZYAM subtraction carefully includes the background due to the elliptic, triangular, and quadrangular flow.
Therefore, it is somewhat surprising to find that resultant correlation yields still maintain the same feature on the away side: it shows one peak in the in-plane trigger direction and double peaks in the out-of-plane trigger direction.
As the STAR experiment claims, it may be attributed to the pathlength-dependent jet-quenching.

Recently, in accordance with the experimental data~\cite{RHIC-star-plane-3}, we also carried out calculations which implement the same flow background due to the elliptic, triangular, and quadrangular flows in accordance with the STAR collaboration.
The calculated correlations~\cite{sph-corr-ev-08} are then compared with the data.
The simulations are carried out for the 20 - 60\% centrality window.
The transverse momentum range of the trigger particles is chosen to be $3 \le p_{T}^{t} \le 4\,$GeV, and that of the associated particles is $1 \le p_{T}^{a} \le 2\,$GeV.
To accommodate the STAR collaboration's experimental setup, the calculations are carried out in the pseudo-rapidity interval $| \eta |< 1$.
By carrying out the modified reaction-plane (MRP) method~\cite{RHIC-star-ridge-6} as employed by the STAR collaboration, particle pairs with $| \Delta\eta | <$ 0.5 are excluded from the construction of the event planes.
Correlated pairs with $| \Delta \eta | < 0.7$ between the trigger particles and the associated particles are also excluded from the analysis because the original purpose was to minimize the near-side jet contribution~\cite{RHIC-star-plane-3}.
The ZYAM method is then used to construct the correlation pattern due to the anisotropic flow.
The primary flow correlated background is from the elliptic flow, caused by the average almond shape of the superposition region of the initial energy distribution, the triangular flow, caused by the fluctuations of the initial conditions that occur on an event-by-event basis~\cite{hydro-v3-01}, as well as the quadrangular flow.
The above flow harmonics contribute to the two-particle correlations and are to be subtracted from the proper correlation pattern.
In~\cite{RHIC-star-plane-3}, the flow correlated background from the elliptic, triangular as well as quadrangular flow are expressed as follows
\begin{equation}
\begin{aligned}
&\frac{dN}{d\Delta\phi} = B \left(1+2v_{2}^{a}v_{2}^{t}\cos2\Delta\phi+2v_{3}^{a}v_{3}^{t}\cos3\Delta\phi\right. \\
&\left.+2v_{4}^{a} \{\Psi_{2}\} v_{4}^{t}\{\Psi_{2}\}\cos4\Delta\phi \right)    \label{eq-sec5-1-zya1}
\end{aligned}
\end{equation}
where $B$ is the background normalization, $v_{2}^{a}$, $v_{4}^{a}\{\Psi_{2}\}$ ($v_{2}^{t}$ and $v_{4}^{t}\{\Psi_{2}\}$) are the second and fourth harmonic coefficients of the associated (trigger) particles with respect to the event plane of the second harmonic $\Psi_{2}$, $v_{3}^{t}$ and $v_{3}^{a}$ are triangular flow of the trigger and the associated particles, calculated with respect to the event plane of the third harmonic $\Psi_{3}$.
The harmonic coefficients of the trigger particle are the average value obtained in the respective slice of the azimuthal angle $\phi_{s}$.
To evaluate the background correlation, the harmonic coefficients are obtained by the event-plane method~\cite{event-plane-method-1,event-plane-method-2,event-plane-method-3}.
Subsequently, the ZYAM method is made use of and is implemented according to Eq.~(\ref{eq-sec5-1-zya1}).
The resultant correlations are shown in FIG.~\ref{corep}.
In the plots, the solid purple curves are obtained by NeXSPheRIO, and the data are represented by the red stars, whereas the gray areas between the solid lines indicate the uncertainties.
Notice that, even though the contribution of the triangular flow is explicitly subtracted from the proper correlation as shown in (\ref{eq-sec5-1-zya1}), the resultant correlation is still featured by one peak in the away-side for the in-plane direction, with its maximum at $\Delta \phi \approx \pi$, which evolves to double peaks for the out-of-plane direction.
These results further strengthened our conclusion obtained above from the comparisons with earlier STAR data~\cite{RHIC-star-plane-1, RHIC-star-plane-2}.

In order to quantitatively understand the contribution of $v_3$ in our hydrodynamical results, we analyzed the calculated values of flow harmonics used in the ZYAM subtraction.
It turns out that the average values of $v_3$ for the associated particles are much smaller in comparison with those of $v_2$.
This is expected as the analysis is not carried out in the most central collision window, which is also shown to be the case for the trigger particles in the in-plane directions.
Also, one observes that the magnitude of $v_4\{\Psi\}$ for the associated particles follows the same trend of that of $v_2$.
However, since its size is much smaller than that of $v_2$, it does not significantly impact the resultant correlations.

In Ref.~\cite{RHIC-star-plane-3}, considering the subtraction of high order harmonics, the authors concluded that the trends of the away-side correlation might underscore the importance of path-length-dependent jet-medium interactions.
As the present hydrodynamical simulations can reproduce observed features of two-particle correlation, it strongly indicates that the observed correlations are likely to be a collective-flow effect of the system.
Moreover, it seems that the ZYAM procedure, devised to essentially subtract the contribution of collective flow from the proper two-particle correlation, has somehow failed in its purpose.
In particular, it is found that the subtraction of $v_3$ does not affect the essential feature of the resultant correlations, namely, the relative magnitudes between $V_2$ and $V_3$.
To us, this might be related to the event-by-event fluctuating initial conditions and their impact on flow harmonics.
This is because the subtracted $v_n$ in Eq.~(\ref{eq-sec5-1-zya1}) is, in fact, the event average value, $\langle v_n \rangle$.
Due to the event-by-event fluctuations, the event average value of a product of harmonic coefficients can be significantly different from the product of the corresponding average values.
In other words, not only the magnitude of the triangular flow, $v_3$, is understood to be related to the event-by-event fluctuations, its fluctuations might also play a non-trivial role in the particle correlations.
Moreover, the event planes between different harmonics might be correlated for a given event but uncorrelated among the various events, which further complicates the problem.
As a result, the average of the Fourier expansion in azimuthal angle cannot be simply approximated by a Fourier expansion in terms of the products of average harmonic coefficients, even rescaled by the ZYAM scheme.
Again, the calculations employing NeXSPheRIO give reasonable results for two different sets of data obtained by different procedures.
This indicates that the underlying physics behind the observed data is mostly hydrodynamical in nature.

It is noted that in Ref.~\cite{LHC-alice-vn-2}, two-particle correlations are analyzed by using a global fit by adjusting $v_n$ vs. $p_T$ to the data.
This result seems to imply that flow harmonics alone are sufficient to capture the observed two-particle correlations entirely.
Although they are different collision system, if one compares the results of Ref.~\cite{LHC-alice-vn-2} against those of Ref.~\cite{RHIC-star-plane-3}, one may observe that the difference comes from the definition of harmonic coefficients.
In Ref.~\cite{LHC-alice-vn-2}, the flow harmonics are obtained from the two-particle correlations, $v_n\{2\}$~\cite{hydro-corr-ph-02}.
If flow fluctuations are taken into account, it is understood that at leading order, we have $v_2\{2\}= v_2\{EP\}+\delta+\sigma_v^2$ where $\delta$ represents the non-flow~\cite{hydro-corr-ph-09}.
Indeed, the above difference probably gives rise to the non-vanishing resultant correlations found in Ref.~\cite{RHIC-star-plane-3}.
As discussed above, the latter is mostly reproduced by hydrodynamic calculations.

\subsection{Centrality dependence}

%*2. Hydrodynamic approach to the centrality dependence of di-hadron correlations, Wagner M. Castilho (Sao Paulo, IFT), Wei-Liang Qian (Sao Paulo U.), Fernando G. Gardim (Alfenas Fed. U., Pocos de Caldas), Yogiro Hama (Sao Paulo U.), Takeshi Kodama (UFRJ, Rio de Janeiro). Oct 13, 2016. 7 pp. Published in Phys.Rev. C95 (2017) no.6, 064908

\begin{figure*}[ht]
\begin{tabular}{ccc}
\begin{minipage}{160pt}
\centerline{\includegraphics[width=180pt]{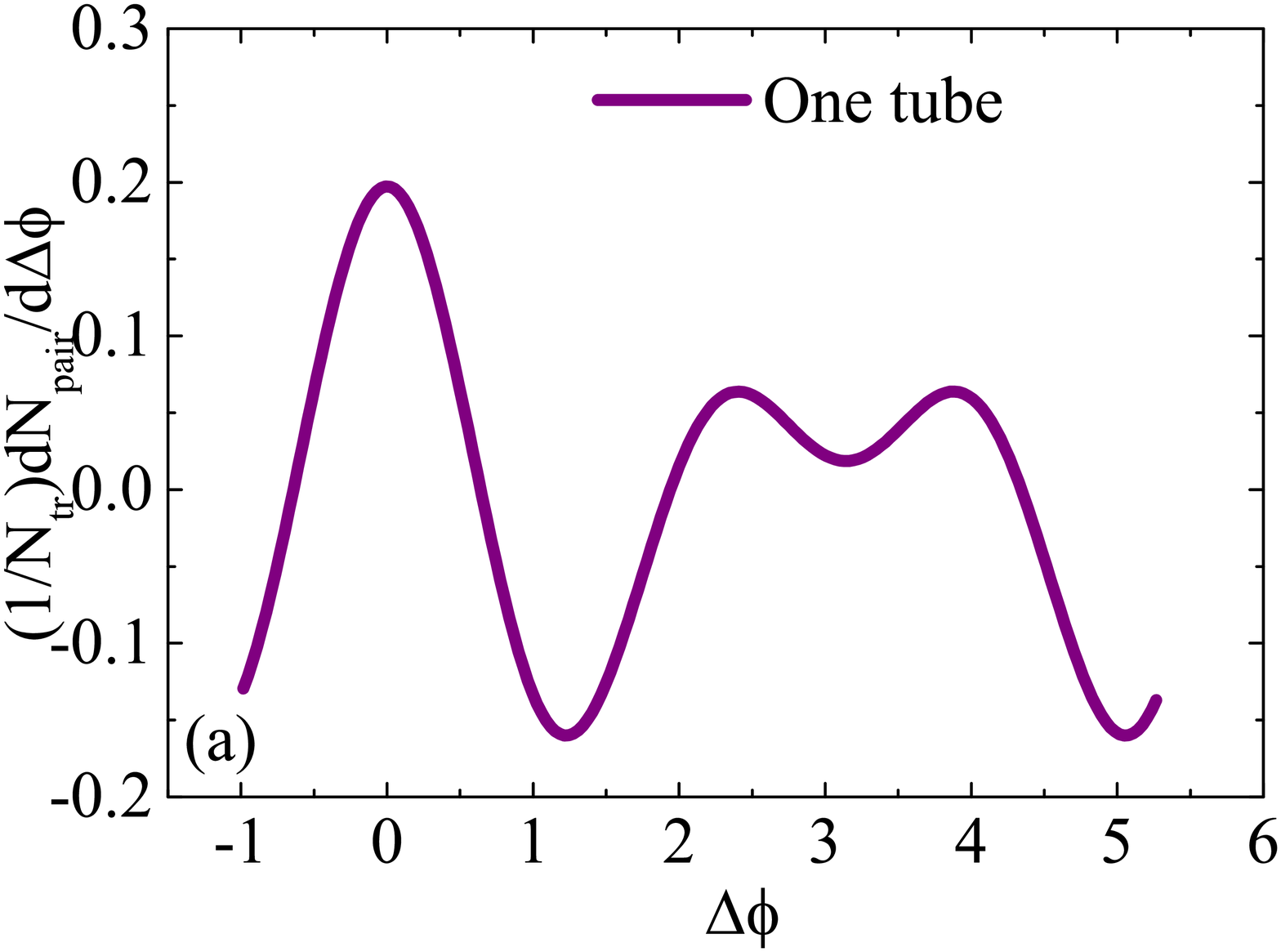}}
\end{minipage}
&
\begin{minipage}{160pt}
\centerline{\includegraphics[width=180pt]{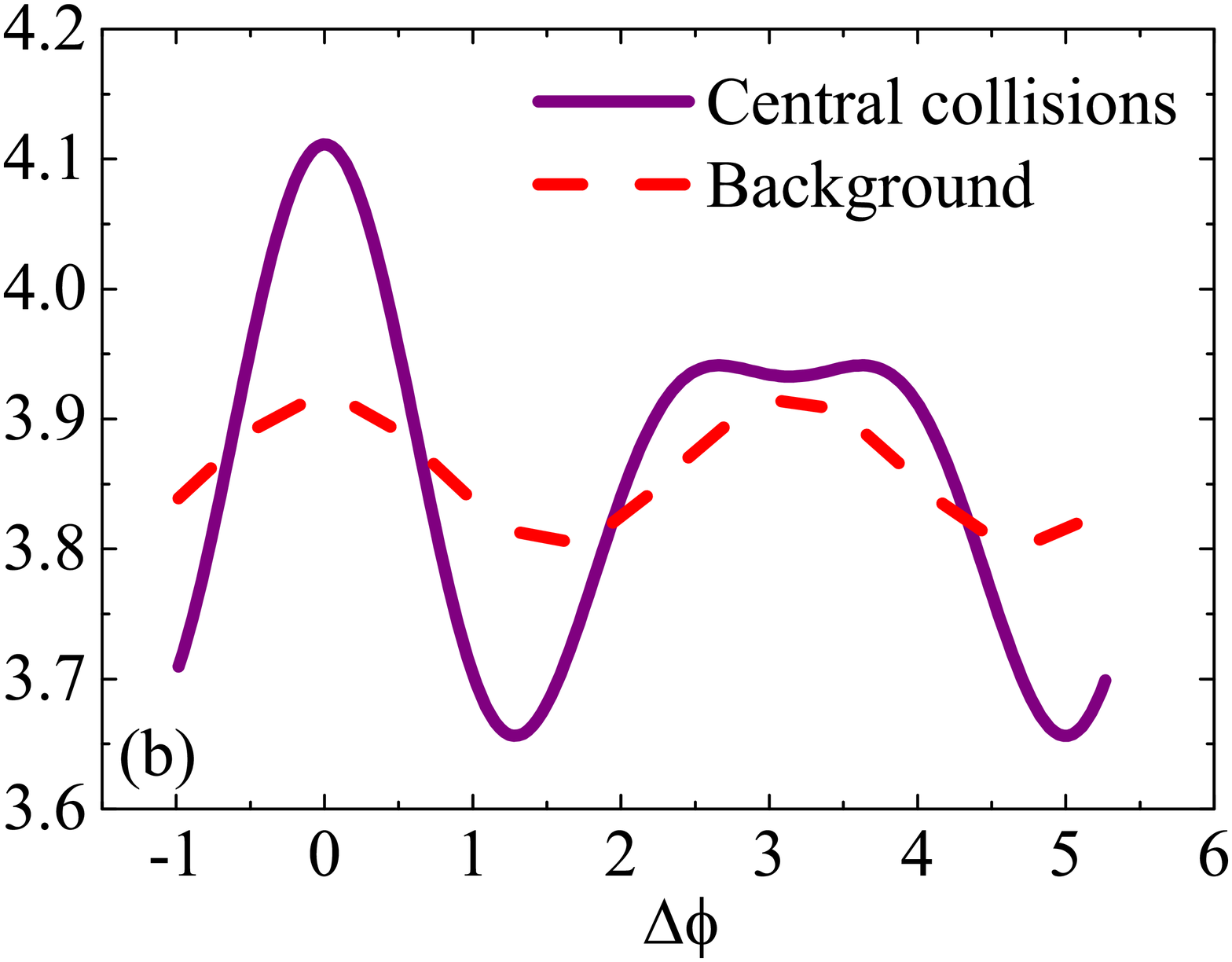}}
\end{minipage}
&
\begin{minipage}{160pt}
\centerline{\includegraphics[width=180pt]{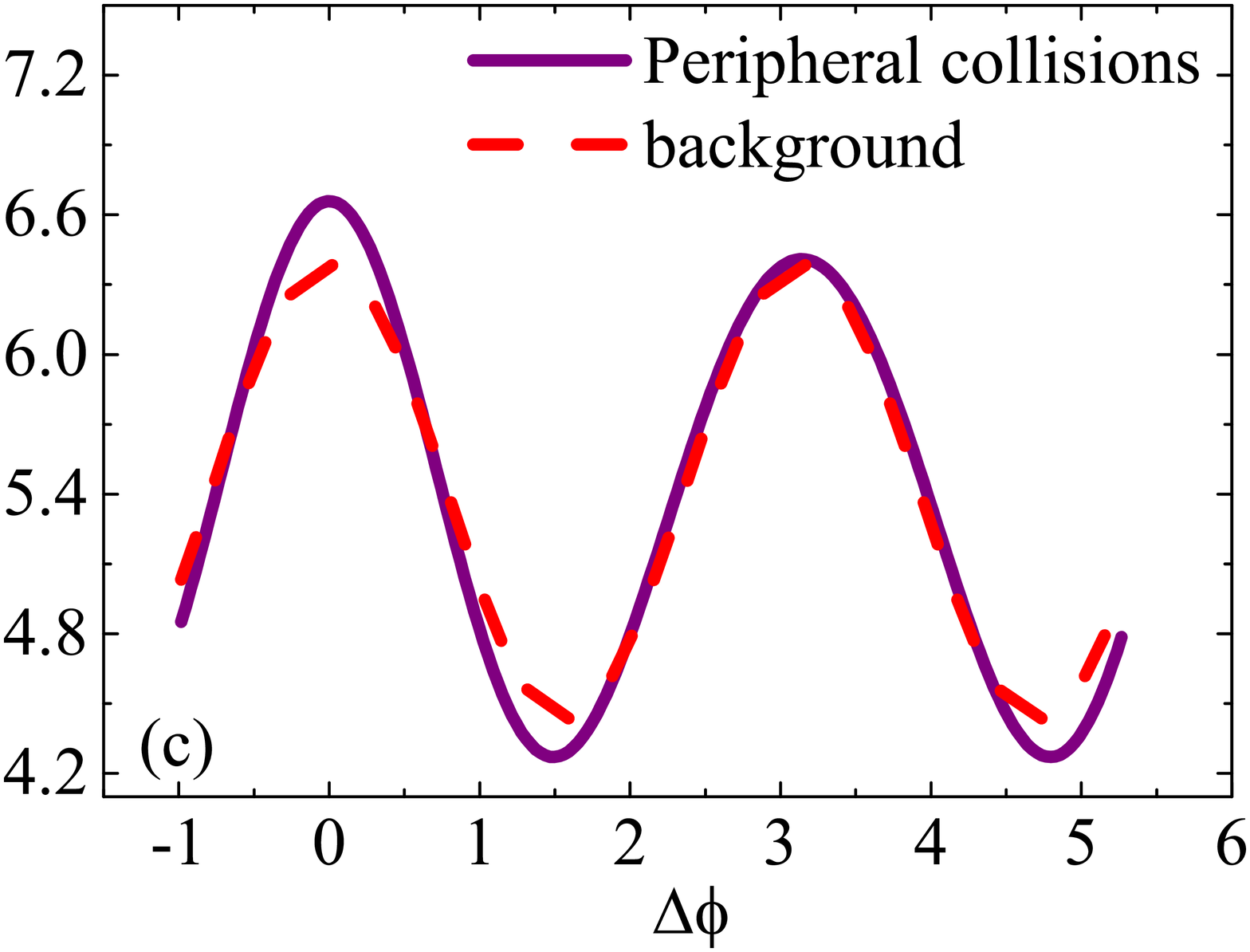}}
\end{minipage}
\end{tabular}
 \caption{ Plots of di-hadron correlations calculated by cumulant method, the correlation is normalized by the number of particles.
 (a) the tube contribution;
 (b) the one from the background (dashed line) and the resultant correlation (solid line) for central collisions, as
 given by Eq.~(\ref{ccumulant-sec5-2}); and (c) the corresponding ones for the peripheral collisions.}
 \label{fig-sec5-2-1}
\end{figure*}

\begin{figure*}[ht]
 \centerline{\includegraphics[width=400pt]{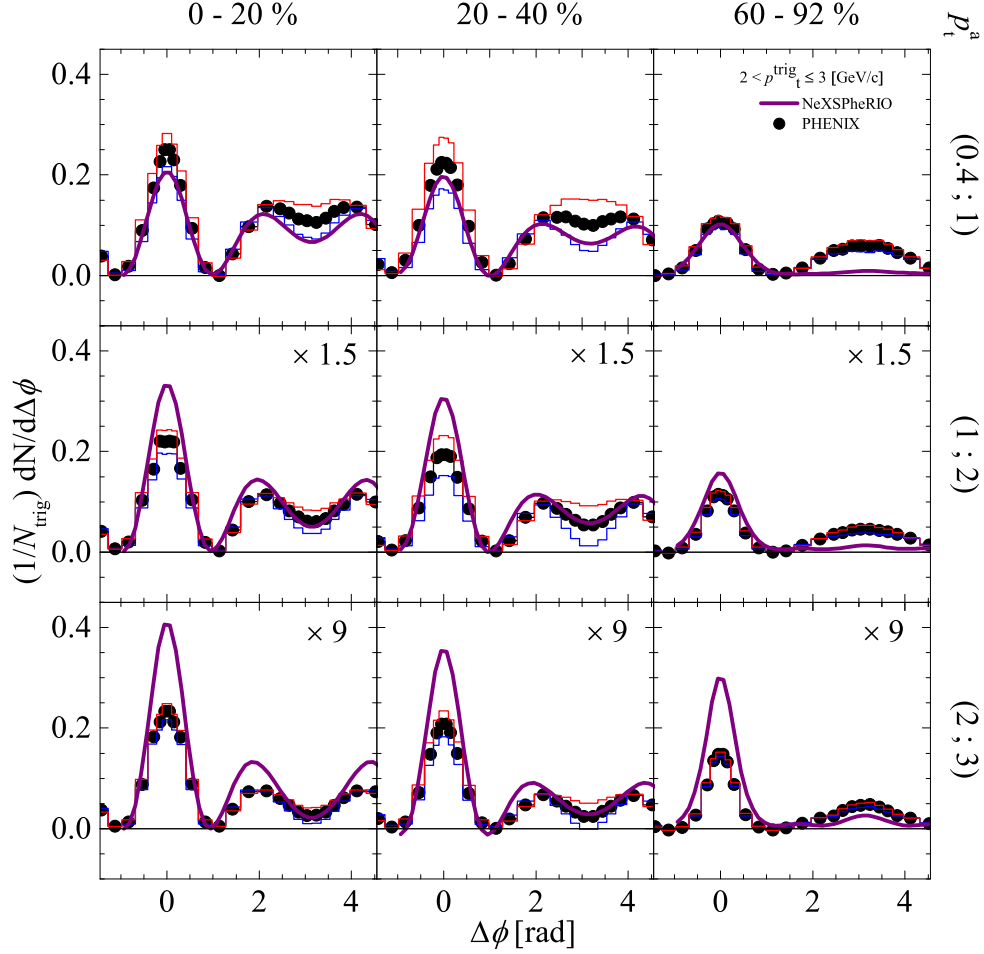}}
 \caption{Color online) The subtracted di-hadron correlations as a function of $\Delta\phi$ for different centrality windows and $p_T^{a}$ range for 200A GeV Au+Au collisions.
   NeXSPheRIO results in solid curves, obtained by using ZYAM method, are compared with PHENIX data~\cite{RHIC-phenix-ridge-5} in filled circles.}
 \label{fig-sec5-2-3}
\end{figure*}

Following the line of thought of the previous subsection, the tube model can be further utilized to study the di-hadron correlation's centrality dependence.
In order to discuss the average correlation at different centrality windows, we integrate the obtained results~\cite{sph-corr-ev-04} over the azimuthal angle of the trigger particle.
The centrality dependence comes naturally from that of the background elliptic flow, whose magnitude increases from central to peripheral collisions.
As a result, when one goes from peripheral to central collisions, the away-side correlation is expected to evolve from single- to double-peak structure, and meanwhile, the magnitude of the correlation increases.
This was precisely observed in the measurements carried out by PHENIX Collaboration~\cite{RHIC-phenix-ridge-5}, as shown in FIG.~\ref{fig-sec5-2-3}.
In what follows, we first show that the main feature of the two-particle correlations can be indeed reproduced by using the tube model.
The numerical simulations are then carried out by employing the hydrodynamical code NeXSPheRIO, and the correlations are evaluated by both cumulant and the ZYAM method~\cite{sph-corr-ev-06}.

For the present case, the purpose of the tube model is to show in a clear-cut way how several characteristics of the so-called ridge phenomena are produced. 
Being so, although extracted from the more realistic studies, for the sake of clarity, only essential ingredients are retained in the model, namely,
\begin{itemize}
\item The collective flow consists of contributions from the background and those induced by randomly distributed peripheral tubes. 
For the reasons discussed above, here we consider just one such tube. 
\item The background elliptic-flow coefficient increases from central to peripheral collisions, while the multiplicity decreases.
\item The event-by-event fluctuation is reflected in the model in two aspects: first, the tube's azimuthal location is randomized from event to event, and, second, the background multiplicity fluctuates from event to event.
\end{itemize}
Recall that experimentally the background multiplicity always fluctuates, and this is important to be considered in the correlation calculation, as will become clear later.
Again, as in Eq.~(\ref{eq-sec5-1-1}), one may write down the one-particle distribution as a sum of two contributions: the distribution of the background and that of the tube.
We note here that the overall triangular flow in the present approach is generated only by the tube and so its symmetry axis is correlated to the tube location $\phi_t$.
The azimuthal angle $\phi$ of the emitted hadron and the position of the tube $\phi_t$ are measured with respect to the event plane $\Psi_2$ of the system. Since the flow components from the background are much bigger than those generated by the tube, as discussed below, $\Psi_2$ is mainly determined by the elliptic flow of the background $v_2^b$.
For the same reason, we prefer, in this analysis, not to include the radial-flow component in the tube contributions, so $N_b$ in Eq.~(\ref{eq-sec5-1-2}) may be literally interpreted as the overall multiplicity.

Following the methods used by PHENIX Collaboration~\cite{RHIC-phenix-ridge-5}, the subtracted di-hadron correlation is given by
\begin{eqnarray}
 \left<\frac{dN_{\text{pair}}}{d\Delta\phi}\right>   =\left<\frac{dN_{\text{pair}}}{d\Delta\phi}\right>^{\text{proper}} -\left<\frac{dN_{\text{pair}}}{d\Delta\phi}\right>^{\text{mixed}}.
 \label{eq4-sec5-2}
\end{eqnarray}
In tube model, the proper correlations have the form of Eq.~(\ref{eq-sec5-1-corr-proper}).
The combinatorial background $\left<{dN_{\text{pair}}}/{d\Delta\phi}\right>^{\text{mixed}}$ can be evaluated by using either cumulant or the ZYAM method. 
In fact, both methods lend very similar conclusions.
Here, we first carry out the calculation by using the cumulant method.

By using the simplified parametrization, Eqs.~(\ref{eq-sec5-1-1}-\ref{eq-sec5-1-3}) and, by averaging over events, one obtains
\begin{eqnarray}
\left<\frac{dN_{\text{pair}}}{d\Delta\phi}\right>^{\text{(cmlt)}}
 &=&\frac{\langle N_b^2\rangle -\langle N_b\rangle ^2}{(2\pi)^2}\left(1+2(v_2^b)^2\cos(2\Delta\phi)\right) \nonumber\\
 &+&(\frac{N_t}{2\pi})^2\sum_{n=2,3}2({v_n^t})^2
 \cos(n\Delta\phi).
 \label{ccumulant-sec5-2}
\end{eqnarray}
Again, one observes that the multiplicity fluctuation gives rise to a difference between the factors multiplying the background terms of the proper- and mixed-event correlations. 
Therefore, the background elliptic flow is {\it not} canceled out but does contribute to the correlation.
From the r.h.s. of Eq.~(\ref{ccumulant-sec5-2}), one sees that the resultant correlation is a sum of two terms.
The first term is determined by the overall multiplicity fluctuations and the background elliptic flow.
Experimental measurements showed that the elliptic flow coefficient increases when one goes to more peripheral collisions.
It is noted that this fact plays a vital role in the analysis.
The second term measures the correlations from the peripheral tube, reflecting the physics of event-by-event fluctuating IC.

Despite its simplicity, the above analytic model captures the main characteristics of the centrality dependence of the di-hadron correlations. 
FIG.~\ref{fig-sec5-2-1} serves as a schematic diagram of the tube model, which reproduces the observed data's main feature. 
Here, the parameters on the r.h.s. of Eq.~(\ref{ccumulant-sec5-2}) can be estimated using the results from NeXSPheRIO calculations.
Since the first term on the r.h.s. of Eq.~(\ref{ccumulant-sec5-2}) is mostly determined by the elliptic flow of the system, its magnitude increases from central to peripheral collisions, following that of the background elliptic flow $v_2^b$.
On the away-side where $\Delta\phi = \pi$, the contribution of the first term (shown by the red dashed lines in FIG.~\ref{fig-sec5-2-1}b and FIG.~\ref{fig-sec5-2-1}c) is always positive.
Consequently, for peripheral collisions, it may be just big enough to fill up the ``valley" of the second term (shown in FIG.~\ref{fig-sec5-2-1}a), which results in a single peak on the away-side as shown by the black curve in FIG.~\ref{fig-sec5-2-1}c.
For central collisions, on the other hand, the second term (shown in FIG.~\ref{fig-sec5-2-1}a) dominates the overall shape of the di-hadron correlations, as one observes that the black curves in FIG.~\ref{fig-sec5-2-1}a and FIG.~\ref{fig-sec5-2-1}b look similar.

Now we show that very similar results will be again obtained if one evaluates the combinatorial mixed event contribution using the ZYAM method.
The spirit of the ZYAM method is first to estimate the form of correlation solely due to the average background collective flow and, then, to rescale the evaluated correlation by a factor $B$. 
This factor is determined by assuming a vanishing signal at the minimum of the subtracted correlation.
Di-hadron correlation for the background flow is given by
\begin{eqnarray}
 \left<\frac{dN_{\text{pair}}}{d\Delta\phi}\right>^{\text{mixed}\text{(ZYAM)}}
  =B\int\frac{d\phi}{2\pi}
 \frac{dN_{\mathrm{bgd}}}{d\phi}(\phi)\frac{dN_{\mathrm{bgd}}}{d\phi}(\phi+\Delta\phi),\nonumber\\
 \label{mixedz1}
\end{eqnarray}
where, according to~\cite{RHIC-phenix-ridge-5}, the elliptic flow coefficients are to be obtained by using event plane method.
A straightforward calculation gives
\begin{equation}
\begin{aligned}
 &\left<\frac{dN_{\text{pair}}}{d\Delta\phi}(\phi_s)\right>^{\text{(ZYAM)}} \\
 &=\frac{\langle N_b^2\rangle -B\langle N_b\rangle ^2}{(2\pi)^2}(1+2(v_2^b)^2\cos(2\Delta\phi)) \\
 &+(\frac{N_t}{2\pi})^2\sum_{n=2,3}2{v_n^t}^2\cos(n\Delta\phi). \label{czyam-sec5-2}
\end{aligned}
\end{equation}
As shown by numerical calculations, similar results are obtained for the case of ZYAM.

In the above, the arguments are mostly based on an analytic tube model, which merely considers a simplified IC.
However, in realistic collisions, the IC generally contain several high energy tubes whose location, size, and energy may also fluctuate.
In order to show that a hydrodynamical description indeed captures the main physical content of the observed data, we present the results of numerical simulations by using the hydrodynamic code NeXSPheRIO.
In the calculations, the di-hadron correlations are obtained by the ZYAM method and compared to the data by PHENIX Collaboration.
We calculate the correlations by using the method of PHENIX Collaboration presented in reference~\cite{RHIC-phenix-ridge-5}.
The elliptic flow coefficients for the trigger and associated particles are obtained using the event plane method~\cite{event-plane-method-2} together with corresponding acceptance cuts adopted by PHENIX Collaboration.
To evaluate the event plane, one considers hadrons within the pseudo rapidity window $|\eta| < 1$ and with transverse momentum $p_T > 0.1$ GeV.
The elliptic flow is evaluated by taking into account hadrons within the pseudo-rapidity range $|\eta| < 1 $.
The resulting di-hadron correlations are shown in FIG.~\ref{fig-sec5-2-3}.
As expected, one finds that the hydrodynamic calculations describe the data better for central collision and at small transverse momentum.
As one goes to more peripheral windows as well as a higher momentum range, deviations appear.
Also, the ratio of the amplitude at the peak to that at the valley between the two peaks is different between the data and numerical simulations."
However, the data's main features are reasonably reproduced by the calculations, which imply that the hydrodynamic model captures the main physics in the observed centrality dependence of di-hadron correlations.

Therefore, we understand that the observed centrality dependence of di-hadron correlations can be understood in terms of the tube model.
In the simple analytic model, the observed features in PHENIX data can be reproduced by a proper choice of parameters.
We note that ongoing studies on the event planes correlation and symmetric cumulant~\cite{LHC-atlas-vn-3, LHC-atlas-vn-5, LHC-alice-vn-5} might potentially provide a way to distinguish between different models.

\subsection{Extracted model parameters}

In this section, we shift the focus onto the model parameters~\cite{sph-corr-08}. 
In particular, instead of numerical calibration, we extract the model parameters according to their respective physical interpretations.
It is carried out using event-by-event simulations with various devised IC reflecting the average background and event-by-event fluctuations.
Subsequently, it is shown that those extracted values are indeed qualitatively in agreement with the observed two-particle correlation data.

As discussed previously, by further averaging out angle of the trigger particle, $\phi_s$, one obtains
\begin{equation}
\begin{aligned}
&\langle \frac{dN_\mathrm{pair}}{d\Delta\phi}\rangle ^{(\mathrm{cmlt})}  \\
&=\frac{\langle N_\mathrm{bgd}^T N_\mathrm{bgd}^A\rangle -\langle N_\mathrm{bgd}^T\rangle \langle N_\mathrm{bgd}^A\rangle }{(2\pi)^2}(1+2v_2^{\mathrm{bgd},T}v_2^{\mathrm{bgd},A}\cos(2\Delta\phi)) \\  
&+\frac{\langle N_\mathrm{tube}^T N_\mathrm{tube}^A\rangle }{(2\pi)^2}\sum_{n=2,3}2v_n^{\mathrm{tube},T}v_n^{\mathrm{tube},A} \cos(n\Delta\phi). 
 \label{ccumulant2-sec5-3}
\end{aligned}
\end{equation}
The derivation of the above equation is similar to Eq.~(\ref{ccumulant-sec5-2}) if one assumes that the trigger and associated particles satisfy distinct one particle distributions.

For the present purpose, we focus on mid-central 200 AGeV Au+Au collisions.
Now we first devise the IC according to the peripheral tube model.
This is achieved by estimating the background energy distribution by averaging over the 343 events generated by NEXUS, for the centrality window 20\%-40\%.
The obtained almond shaped energy distribution is then fitted by the following parametrization
\begin{eqnarray}
\epsilon_\mathrm{bgd} &=& (c_5+c_6 r^2+c_7 r^4)e^{-(c_8 r)^{c_9}}\ , \label{energyb-sec5-3} 
\end{eqnarray}
with
\begin{eqnarray}
r &=& \sqrt{c_{10}x^2+c_{11}y^2} \  ,   \nonumber
\end{eqnarray}
where $c_5=9.33\,$fm$^{-3}$, $c_6=7\,$fm$^{-5}$, $c_7=2\,$fm$^{-7}$, $c_8=1\,$fm$^{-1}$, $c_9=1.8$, $c_{10}=0.41$, and $c_{11}=0.186$.

The profile of the high energy tube is calibrated to that of a typical peripheral tube in NEXUS IC as follows
\begin{eqnarray}
\epsilon_\mathrm{tube}&=& c_{12} e^{-(x-x_\mathrm{tube})^2-(y-y_\mathrm{tube})^2} \label{energyt-sec5-3b}\ , 
\end{eqnarray}
where $c_{12}=12\,$fm$^{-3}$, and the tube is located at a given value of energy density close to the surface, determined by a free parameter $r_\mathrm{tube}$, so that its coordinates on the transverse plane read
\begin{eqnarray}
x_\mathrm{tube} &=& \frac{r_\mathrm{tube}\cos\theta}{\sqrt{c_{10}\cos^2\theta+c_{11}\sin^2\theta}}   \label{energyt-sec5-3c} \\
y_\mathrm{tube} &=& \frac{r_\mathrm{tube}\sin\theta}{\sqrt{c_{10}\cos^2\theta+c_{11}\sin^2\theta}} \ . \nonumber
\end{eqnarray}
Here and below, $r, x, y$ are numerical values for position parameters in the unit of fm, and $\epsilon$ is the numerical value for energy density in the unit of fm$^{-3}$.
Also, $r_\mathrm{tube}$ is used as a free parameter whose value is to be determined below, and the azimuthal angle of the tube $\theta$ is randomized among different events.

\begin{figure}
\begin{tabular}{c}
\begin{minipage}{250pt}
\centerline{\includegraphics[width=250pt]{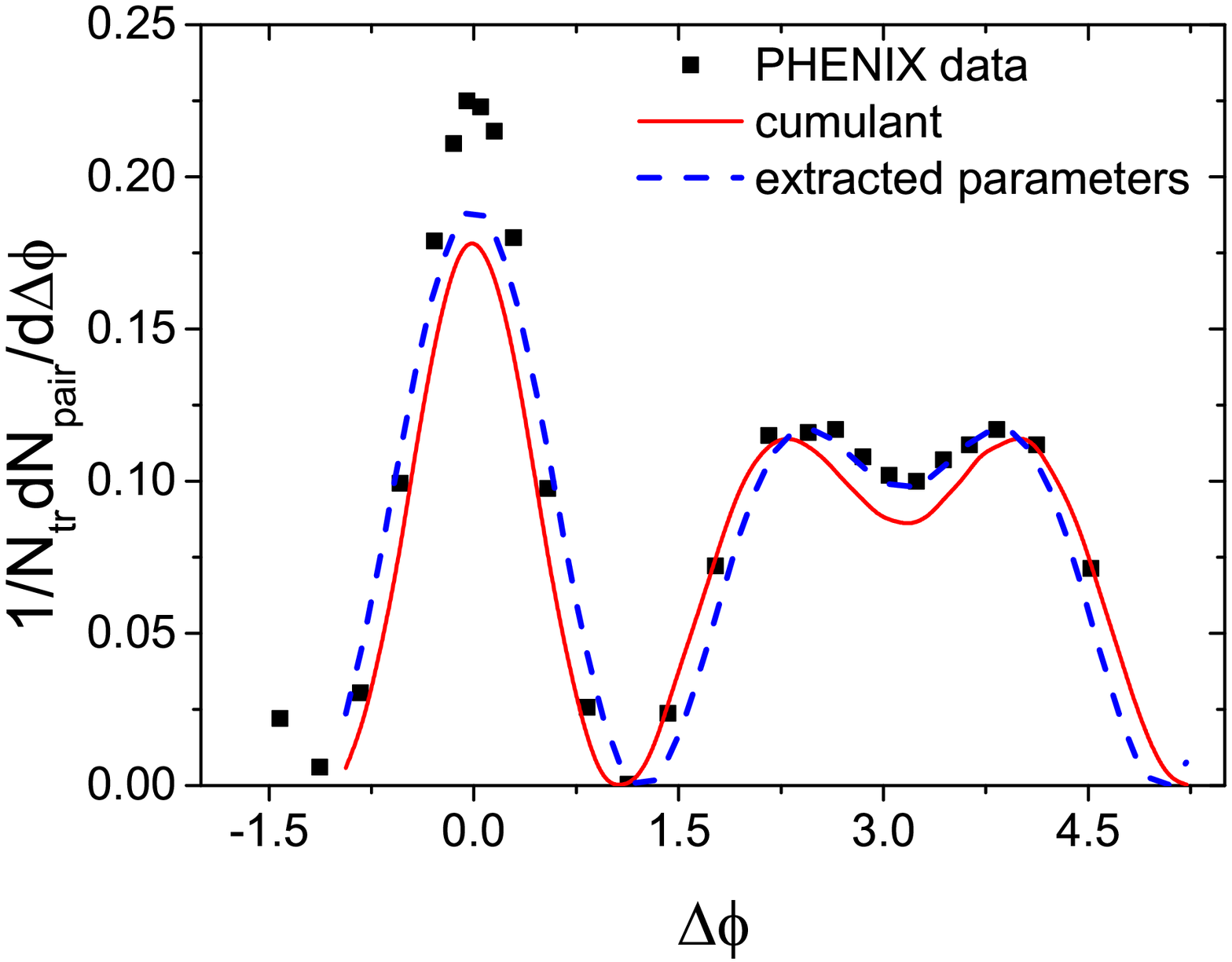}}
\end{minipage}
\\
\begin{minipage}{250pt}
\centerline{\includegraphics[width=250pt]{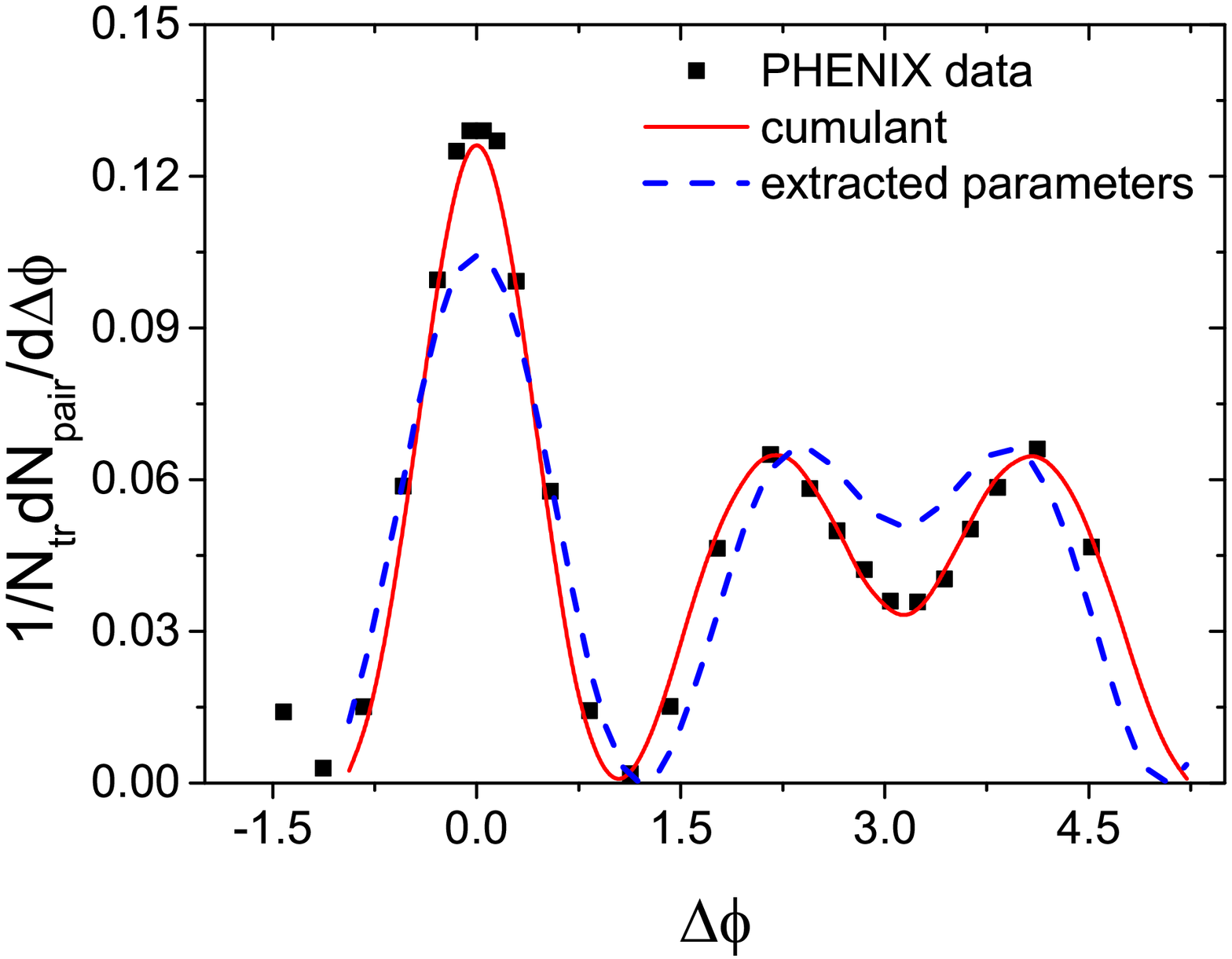}}
\end{minipage}
\end{tabular}
\caption{ The calculated two-particle correlations by using one-tube IC for $20\% - 40\%$ centrality window in comparison with the corresponding data by PHINEX Collaboration~\cite{RHIC-phenix-ridge-5}, and those obtained by using the extracted paramters~\cite{sph-corr-08}.
The SPheRIO results from using the cumulant method are shown in red solid curves, the data are shown in solid squares, and those obtained by the estimated parameters are shown by the blue hatched curves.
Top: the results for the momentum intervals $0.4<p_T^A<1$ Gev and $2<p_T^T<3$ Gev. 
Bottom: those for the momentum intervals $1<p_T^A<2$ Gev and $2<p_T^T<3$ Gev. }
 \label{fig-sec5-3-fig2p}
\end{figure}

By combining the two pieces together, the IC for the present model read
\begin{eqnarray}
\epsilon = \epsilon_\mathrm{bgd}+\epsilon_\mathrm{tube} .
\end{eqnarray}

Subsequently, we carry out hydrodynamical simulations using the SPheRIO code, with a total of 2000 events generated according to the IC profile discussed above.
The resultant two-particle correlations, evaluated by cumulant method, are shown in FIG.~\ref{fig-sec5-3-fig2p}, compared with the PHENIX data~\cite{RHIC-phenix-ridge-5}.
The paramter $r_\mathrm{tube}$ is chosen to be $2.3$ fm in the calculations.
As the hydrodynamic simulations are of two-dimension, the obtained correlations are multiplied by a factor related to the system's longitudinal scaling.

Now, the goal is to verify that the parameters given in Eq.~(\ref{ccumulant2-sec5-3}) are indeed quantitatively meaningful.
To achieve this, we extract the model parameters in Eq.~(\ref{ccumulant2-sec5-3}) using mostly the same arguments leading to very expression.
We first estimate those for the background distribution $\frac{dN_{\mathrm{bgd}}}{d\phi}$.
The background flow coefficients $v_2^\mathrm{bgd}$ can be obtained directly by investigating the hydrodynamic evolution of IC solely determined by $\epsilon_\mathrm{bgd}$.
A total of 2000 events with 200 Monte Carlo each are considered in the evaluation.
To estimate the multiplicity fluctuations of the background, $\langle N_\mathrm{bgd}^TN_\mathrm{bgd}^A\rangle-\langle N_\mathrm{bgd}^T\rangle\langle N_\mathrm{bgd}^A\rangle$, we count, on an event-by-event basis, the number of particles of corresponding momentum intervals. 
The events in question are those generated by NEXUS of 20\% - 40\% centrality window, and we made use of a total of 1000 events.
By using Fourier expansion of the two particle correlation, and extracting the second and third order coefficients, one obtains the parameters related to $v_2^{\mathrm{tube},T}$, $v_2^{\mathrm{tube},A}$, $v_3^{\mathrm{tube},T}$, and $v_3^{\mathrm{tube},A}$.

As a matter of fact, some of the parameters can also be inferred straightforwardly from the experimental data.
The overall elliptic flow, $v_2^\mathrm{all}$, obtained by simulations of event-by-event fluctuating IC consisting of the background and tube, should be consistent with the collisions of the same centrality window.
This is confirmed by comparing the value of $v_{2}^\mathrm{all}$ against the data of 20\%-60\% Au+Au collisions obtained by PHENIX~\cite{RHIC-phenix-v2-4}.

Finally, by substituting the obtained model parameters back into Eq.~(\ref{ccumulant2-sec5-3}), one obtains the two-particle correlation as also shown in FIG.~\ref{fig-sec5-3-fig2p}.
It is found that the two approaches are in good agreement with each other.

\subsection{Further speculations}

\begin{figure}
\begin{tabular}{c}
\vspace{0pt}
\begin{minipage}{150pt}
\centerline{\includegraphics[width=180pt]{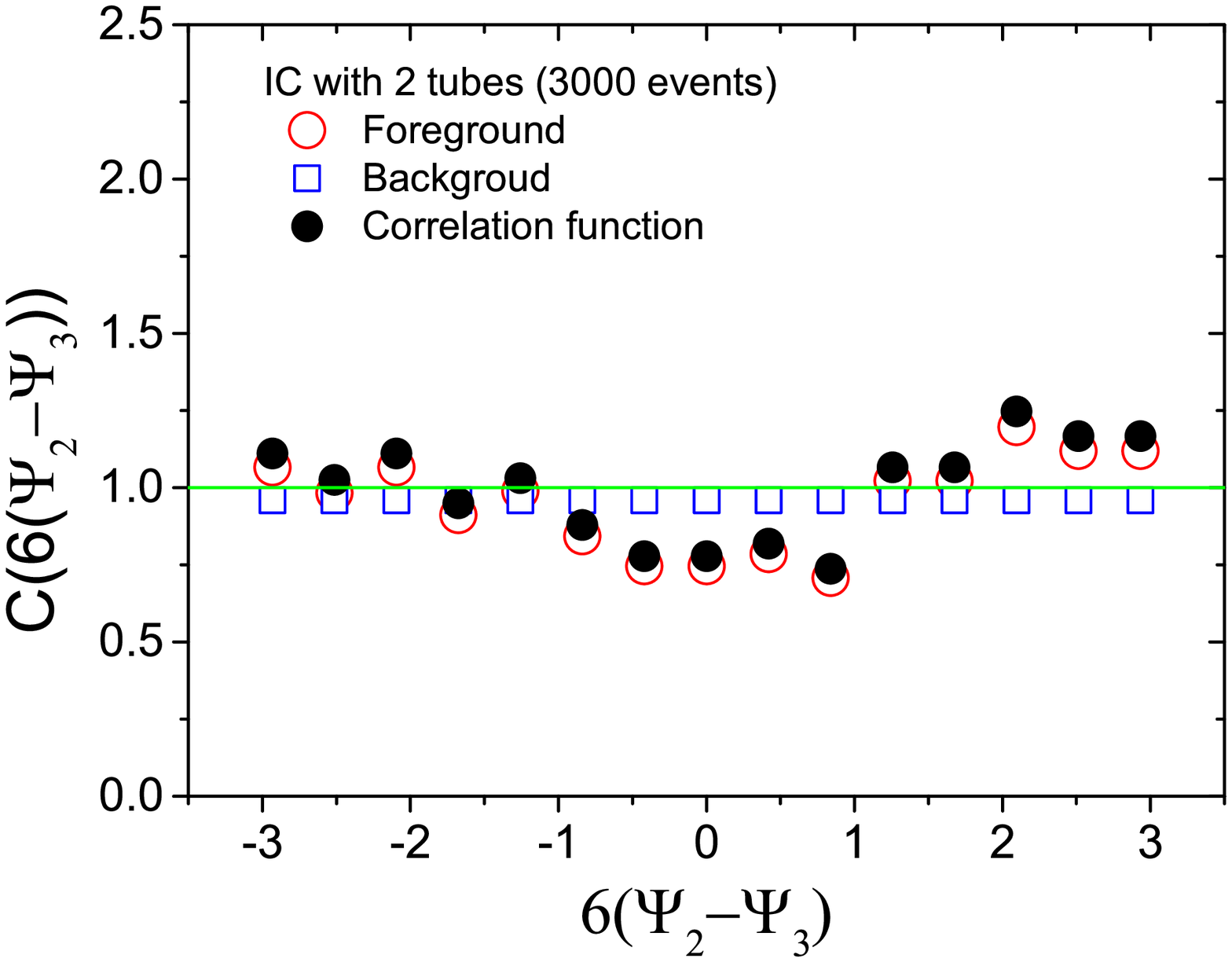}}
\end{minipage}
\\
\begin{minipage}{150pt}
\centerline{\includegraphics[width=180pt]{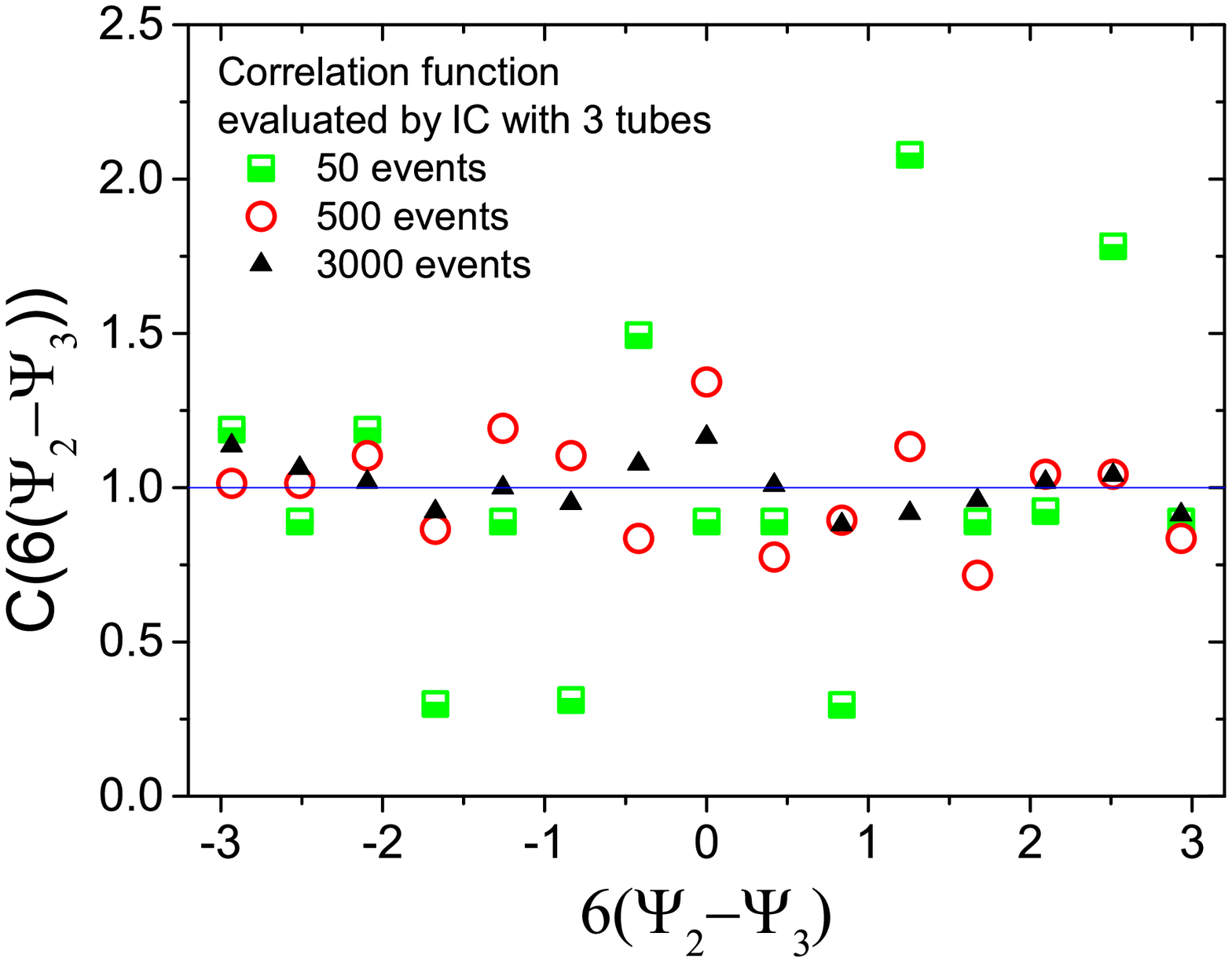}}
\end{minipage}
\\
\begin{minipage}{150pt}
\centerline{\includegraphics[width=180pt]{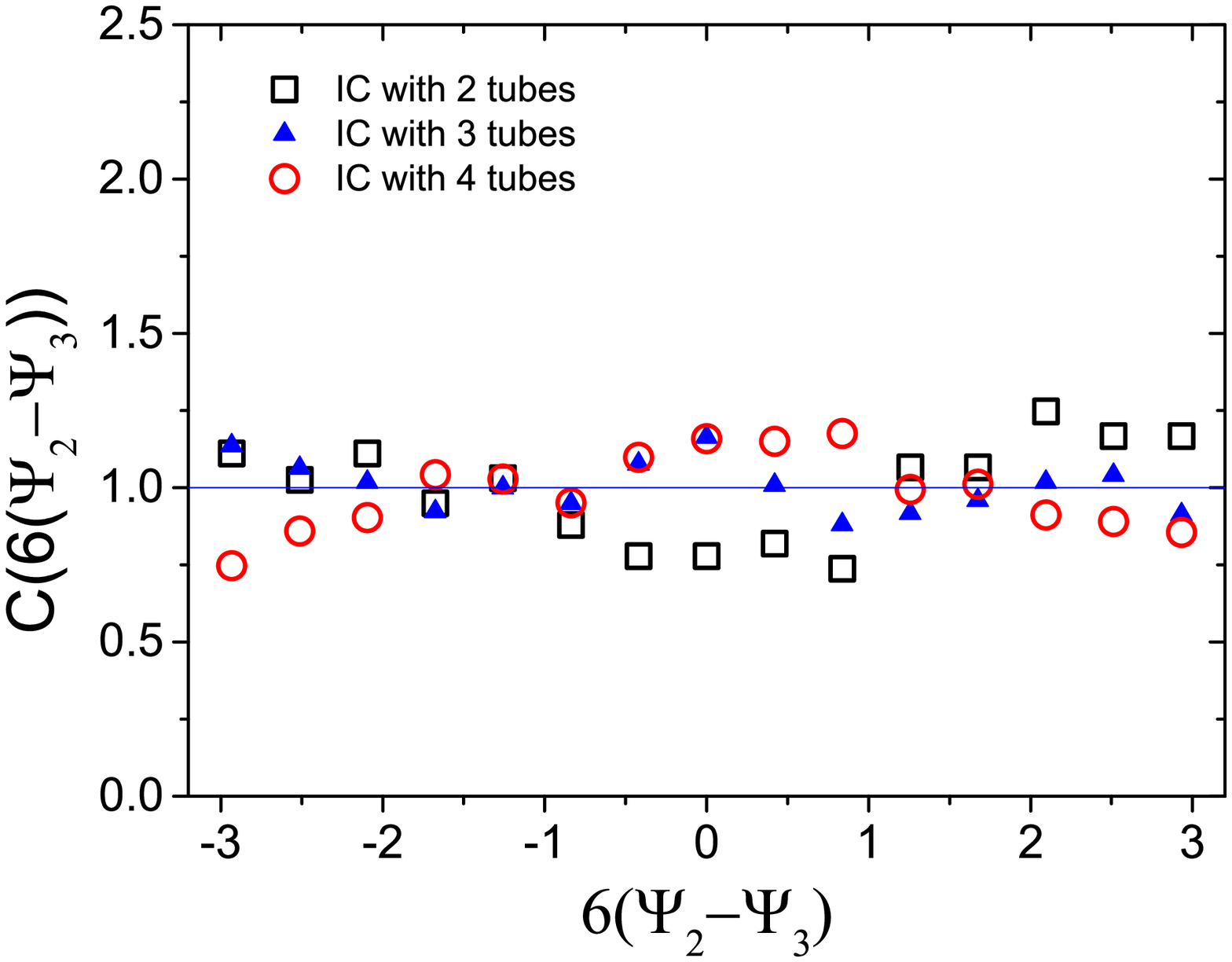}}
\end{minipage}
\\
\vspace{0pt}
\end{tabular}
\caption{(Color online) The relative angle distributions between $\Psi_2$ and $\Psi_3$ in tube model.
Top: The relative angle distribution (filled black circles) evaluated by the ratio of foreground (empty red circles) to the background (empty blue squares) signals. 
The results are for IC with two high energy tubes.
Middle: The results evaluated by using different numbers of IC with three high energy tubes.
Bottom: The relative angle distributions calculated using IC with different number of tubes, where each curve is obtained by using 3000 events.}
 \label{fig-sec5-4-figureCPsi23}
\end{figure}

In previous sections, we show that when fluctuations are represented by a peripheral tube, the event planes of the elliptic and triangular flows from the tube are correlated to the former's azimuthal angle.
However, this feature is quite different from the analysis one carries out in terms of flow harmonics.
To be specific, the origin of $v_2$ is the almond shape of the averaged IC, while that of $v_3$ is mostly the event-by-event fluctuations.
Therefore, there is no particular reason why their event planes should be correlated.
Moreover, the correlations between $\Psi_2$ and $\Psi_3$ have been studied experimentally, for instance, by ATLAS~\cite{LHC-atlas-vn-3}.
It was shown that event planes are mostly uncorrelated.
In terms of the peripheral tube model, it is therefore interesting to verify whether the overall event planes $\Psi_2$ and $\Psi_3$ are indeed correlated.
We subsequently carry out simulations for IC with a different number of tubes in accordance with Ref.~\cite{LHC-atlas-vn-3}.
The results are presented in FIG.~\ref{fig-sec5-4-figureCPsi23}.

From FIG.~\ref{fig-sec5-4-figureCPsi23}, one observes that the correlation between event planes is not manifested, especially as the number of events increases.
This is probably because the observable in question, the calculated two-particle correlation, is an event-average quantity.
According to the peripheral tube model, let us consider the case where two high-energy tubes are added onto an isotropic background.
For simplicity, one assumes that the resulting distribution is the superposition of the contributions of individual tubes.
The resultant event planes can be shown to depend nonlinearly on the relative angle of the tubes. 
To be specific, one denotes the two tubes' azimuthal angles by $0$ and $\phi_t$, the harmonic coefficients by $v_n^{(i)}$ with $i=1, 2$.
It is straightforward to find that
\begin{eqnarray} 
\Psi_2 = \frac12\arctan\frac{v_2^{(2)}\sin 2\phi_t}{v_2^{(1)}+v_2^{(2)}\cos 2\phi_t} \ ,\\
\Psi_3 = \frac13\arctan\frac{v_3^{(2)}\sin 3\phi_t}{v_3^{(1)}+v_3^{(2)}\cos 3\phi_t} \ .
\end{eqnarray}
Indeed, it is confirmed numerically by considering the scenario with different numbers of tubes and an appropriate number of events.

Moreover, it is still interesting to investigate whether the correlations between different harmonics embedded in the model might lead to any observable effect in the multi-particle correlations.
One possible way to study the $k-$particle correlation is to employ the mixed harmonics introduced in Ref.~\cite{hydro-corr-ph-03,hydro-corr-ph-06}, 
\begin{equation} 
\begin{aligned}  
&v(n_1,n_2,\cdots,n_k) \equiv <k>_{n_1,n_2,\cdots,n_k} \\
&= <\cos(n_1\varphi_1+n_2\varphi_2+\cdots+n_k\varphi_k)> , \label{vnk}
\end{aligned}  
\end{equation} 
where $n_1,\cdots,n_k$ are integers, $\varphi_1,\cdots,\varphi_k$ are azimuthal angles of particles belonging to the same event, and angular brackets indicate average over different particle combinations as well as different events. 
The condition $n_1+n_2+\cdots+n_k=0$ must be satisfied so the defined quantity is not vanishing.
In fact, the above definition was introduced in the first place as a generalization to study the linearity between collective flow and eccentricity in terms of mixed harmonics.
However, if one assumes pure flow dynamics, particle emissions are entirely independent and therefore multi-particle correlation only comes from one-particle correlation, one may obtain further results.
This is because one particle correlation follows the probability distribution
\begin{equation} 
\begin{aligned}  
P(\phi)=\frac{1}{2\pi}\sum_n V_n e^{-in\phi}=\frac{1}{2\pi}\sum_n v_n e^{-in(\phi-\Psi_n)} , \label{oneP}
\end{aligned}  
\end{equation}
where one adopts the complex notation for $V_n$~\cite{hydro-corr-03}.
Then it can be used to evaluate the average in Eq.~(\ref{vnk}) and one finds
\begin{equation} 
\begin{aligned} 
&v(n_1,n_2,\cdots,n_k)=\left<V_{n_1}^1 V_{n_2}^2\cdots V_{n_k}^k\right> \\
&=\left<v_{|n_1|}^1 v_{|n_2|}^2\cdots v_{|n_k|}^k\cos(n_1\Psi_{n_1}^1+n_2\Psi_{n_1}^2+\cdots+n_1\Psi_{n_k}^k)\right> .
\end{aligned}  
\end{equation}
Here in the last step the $\sin$ term vanishes because the distribution of the difference of any two event plane is an even function due to symmetry, and therefore it justifies the complex notation is meaningful and compact.
And for instance, some specific cases one has~\cite{hydro-corr-ph-02}
\begin{equation} 
\begin{aligned} 
&-v_n(4)^4=v(n,n,-n,-n)-2v(n,-n)^2 \\
&=<v_n^2>-2<v_n^2>^2 ,
\end{aligned}  
\end{equation}
which is completely determined by flow coefficients. 
Following the same arguments as one only assumes flow dynamics, many coefficients of mixed harmonics vanish because the event-by-event average $\left<\cos(n_1\Psi_{n_1}^1+n_2\Psi_{n_1}^2+\cdots+n_1\Psi_{n_k}^k)\right>$ evaluates to zero due to symmetry.
Nevertheless, since all of these derivations start from the assumption Eq.~(\ref{oneP}), it is not difficult to see the peripheral tube model may have several terms survive the event average, which should be zero otherwise.
Following this line of thought, several other quantities that might be interested are the correlations between the event planes~\cite{glauber-en-1,glauber-en-2} and higher harmonics~\cite{hydro-vn-10,hydro-corr-04}.
Recently, these quantities have been measured by ATLAS~\cite{LHC-atlas-vn-3} and CMS~\cite{LHC-cms-vn-3} Collaborations.

%\input peripheraltube_Weiliang_o.tex
%\section{Some observed properties and relation with the peripheral tubes}
%\subsection{Trigger-angle dependence}
%\subsection{Centrality dependence}
%\subsection{Ridge in pp}

\newpage

\newpage

\newpage

\section{Conclusions and outlooks}

Hydrodynamics provides a successful description of many experimental data in relativistic heavy-ion collisions, mainly in terms of the emergence of collective flow dynamics~\cite{hydro-review-01, hydro-review-02, hydro-review-03, hydro-review-04, hydro-review-05, hydro-review-06, hydro-review-07, hydro-review-08, hydro-review-09, hydro-review-10}. 
The relevant dynamics is expressed by a set of flow harmonics, including correlations among them~\cite{hydro-corr-ph-02,hydro-corr-ph-04,hydro-corr-ph-06,hydro-corr-ph-07,hydro-corr-ph-08} on an event-by-event basis.  
This paper has carried out a survey of the peripheral-tube model, which has been developed to give a dynamical and intuitive view of the long-range two-particle correlations in high-energy nuclear collisions. 
Let us expose below some of the main conclusions.

Our primary tool was the simplest version of ideal hydrodynamics with fluctuating IC, with tube-like structures. 
This approach has shown to be quite appropriate for studying several aspects of high-energy nuclear collisions, including the goal, the so-called ridge effect, which appears in long-range two-particle correlations at low and intermediate $p_T$.
It successfully reproduces the experimental data and plays a vital role in understanding the physical mechanism of the phenomena.

As well known, the elliptical flow $v_2$ is causally connected with the almond shape of the IC, namely, $\epsilon_2$ of the energy-density distribution, produced by the collision geometry, with some small fluctuations.
On the other hand, a similar relation between the triangular flow $v_3$ and some eccentricity component does not seem straightforward.
Indeed, $v_3$ is intimately connected to the long-range two-particle correlations, and it may appear as a consequence of IC fluctuation. 
However, such a fluctuation is random and, on the other hand, to correctly reproduce the observed two-particle correlation, one needs some $<v_2>$, well correlated with $<v_3>$, even in the most central collisions. 
How do such correlations appear, starting from both random and independent $\epsilon_2$ and $\epsilon_3$?
Thus, establishing a relation $<v_3>\sim<\epsilon_3>$ does not seem to solve the main problem, namely to physically understand how ridge+shoulders structure appears.
As pointed out, differently from $\epsilon_2$, $\epsilon_3$ comes from the fluctuations in the initial distribution. 
Moreover, it does not have information on what kind of fluctuation it originates. 
Therefore, without specifying the physical source of $\epsilon_3$, the determination of EoS and transport coefficients is not feasible.
This is because such a simple event averaged relation does not carry the information on the real event-by-event dynamics faithfully.
Subsequently, it does not suffice to identify the properties of strong granularity, which present itself crucially in terms of the event-by-event IC.

We claim that we found an easily identifiable causal element in the time evolution of the hot matter, namely the evolution of the flow stem from the substantial space irregularity in the hot matter.
The peripheral-tube model makes use of such an element. 
At the beginning of a high-energy nuclear collision, many high-density tubes are randomly produced. 
However, the evolution of the flow starting from such a tube is causal.
Subsequently, if some of them are peripheral, that is, formed in those sufficiently rarefied regions, they evolve in a similar fashion by following the same law and producing the same structure. 
Therefore, when averaged over many events, these structures are overlapped and enhanced.
Fortunately, such a structure appears strong enough, as we can observe it experimentally as the ridge in correlation.

As described in Section IV, especially in Subsection IV-B, the peripheral-tube model is a simplified model of IC, based on more realistic IC in high-energy nuclear collisions.
It is explicitly devised to clarify the ridge formation mechanism. 
As so, we simplified as much as possible the seemingly unnecessary complexity of the more realistic model, namely the NEXUS generator. 
The essential elements in the single-tube version of this model are: first, a typical high-energy-density tube which often appears on the surface of the high-density matter in the realistic model; and second, the averaged background matter. 
The latter produces the averaged radial flow, which is deviated by the violent explosion of the former.
Subsequently, the resultant single-particle distribution is characterized by an angular separation of $\Delta\phi\sim2\,rad$, which, in turn, produces the characteristic three-ridge structure of two-particle correlation. 
Here, we emphasize that this mechanism naturally produces, in a unified way, the observed three-ridge structure as experimentally measured.
The observed structure consists of a high central ridge and two symmetrically placed lower ridges with about half the central one's height. 
Remark that, in terms of anisotropic flow parameters, the flow described above naturally contains well correlated $v_2$ and $v_3$.

In Subsection IV-C, we studied the parameter dependence of the correlation predicted by the single-tube model. 
We saw that some parameters affect much and others less the correlation, but a remarkable characteristic is that the shape of the correlation, with one high and two low symmetrical ridges, is very robust. 
As naturally expected, higher tubes give stronger correlations, and so do fatter ones. 
For these two parameters (height $\epsilon_t$ and radius $\Delta r$), what imports is approximately the combination $\epsilon_t\Delta r^2$, which gives the energy content in the tube. 
As for the position of the tube, the peripherality is more or less a well-defined property, namely, if the tube is inside the cylindrical surface of $r\simeq4.0\,$fm, in Au+Au collisions, it does not produce any effect we are discussing (see FIG.~\ref{rlimit} in comparison with FIG.~\ref{r_dependence}). 
When the tube moves further from the center, the correlation increases more or less linearly with the distance, until $r\simeq5.5\,$fm, where the selected typical tube was located.
The strength of the correlation maintains this value as the tube dislocates further towards the edge.

This is why we refer to these tubes as {\it peripheral tubes}. 
The two-particle correlation is only weakly dependent on the initial transverse velocity, as well as on the background height. 
However, it is very sensitive to the shape of the background energy distribution. 
As shown in FIG.~\ref{bg-shape}, if the background is long-tailed, typically Gaussian, instead of the more edgy one we are using, as given by the first term of Eq.~(\ref{par}), the correlation decreases considerably.

In the realistic NEXUS IC employed in this work, one usually encounters more than one peripheral tube. 
So, in Subsection IV-D, we extended the study to a multi-peripheral tube model, which includes, instead of just one tube, 2, 3, and 4 randomly distributed ones. 
Although limited to only one kind of tubes, as described in Eq.~(\ref{parmulti}), and statistically limited (50 events for each number of tubes), the results are quite impressive.
That is, although the azimuthal flow parameters $v_n$ and their relative angles $\Psi_n-\Psi_m$ are random and not connected as expected, the final two-particle correlations are more or less constant, similar to the one produced by just one peripheral tube (see FIG.~\ref{correlations}). 
Observe that the similarity goes further to the correlation produced by the original complete NEXUS. 
These results clearly mean that the three-ridge structure of the two-particle correlation is produced independently by each peripheral tube and not by the global distribution of the matter, which is random. 

As mentioned above, this hydrodynamic process, starting from a peripheral tube, is causal, and the superposition of the flows arising from different tubes may cause interference. 
However, the results show that 50 events are already enough to wash out these interference terms.
The causal connection of the ridge, with respect to the longitudinal distribution, has been discussed elsewhere~\cite{dumitru}. 
Now, we are sure that the ridge structures, including near-side and away-side ones, are entirely connected by causality.

In Section V, ridge phenomena involving ellipticity, such as centrality dependence and trigger-angle dependence, are discussed. 
There, we understand that the resultant particle correlations are led by two independent causal phenomena, namely, the elliptic expansion due to the averaged almond matter distribution and the shadowed expansion originated from the high-energy tubes.
By devising an analytic model, various features found in the experimental data can be successfully explained.
In particular, in the framework of the peripheral tube model, the extracted model parameters regarding collective flow and multiplicity fluctuations from the measurements are shown to provide a consistent estimation of the measured two-particle correlation.

Before closing this paper, we give a few final comments from a more general perspective.
When we proceed to explore the properties of the QCD matter from a hydrodynamic viewpoint, one has to address the physical concepts such as the degree of local thermal equilibrium, the equation of state, and transport coefficients.
In practice, the chosen tool must reflect one particular characteristic of a long-wavelength limit of a strongly interacting system, which is, by and large, genuinely nonlinear.  
In this context, quantities accessible both from theoretical and experimental sides, such as higher-order correlators~\cite{hydro-corr-ph-07,hydro-corr-ph-08,hydro-vn-10}, are meaningful observables up to the task.
Such quantities may demonstrate the intrinsical nonlinearity of the system in an untrivial way, and meanwhile, an intuitive physical interpretation of the results might not be straightforward. 
That is the reason that we do not favor the more popular viewpoint regarding {\it small deviations from the linear response}. 
Moreover, some observed universal properties are seemingly beyond the conventional picture of hydrodynamics, termed ``hydrodynamization" by some authors~\cite{hydro-review-10}.  
Besides inspiring results from the AdS/CFT approaches, we note that several traditional transport approaches, such as AMPT or PHSD, have shown to possess similar properties as viscous hydrodynamic calculations~\cite{ampt-4,ampt-5,hydro-review-06}.
In particular, it is worth mentioning that in the framework of the event-by-event fluctuations, a state close to the local thermal equilibrium only corresponds to a tiny space-time domain during the entire dynamical evolution (Ref.~\cite{hydro-review-06,phsd-01}). 
To clarify up to what degree the genuine event-by-event hydrodynamics is still feasible, one needs a new set of observables up to the task.
Instead of focusing on the deviations from the linearity, these observables must be intrinsically sensitive to the nonlinear nature of the system.

Regarding the peripheral tube model discussed in this paper, it generates the correlations among different higher-order eccentricity and flow components. 
As a result, the analysis of fluctuations in terms of higher order correlations may reveal specific signals of the genuine nonlinear hydrodynamics. 
It is, therefore, interesting to propose such observables that carry such signals and investigate them. 
Relevant quantities associated with the event plane correlations~\cite{LHC-atlas-vn-3,LHC-atlas-vn-5,LHC-alice-vn-5} and the 2+1 correlations~\cite{RHIC-star-corr-2plus1-1} might be meaningful candidates. 
We believe that the present approach, to be proven right or not, will offer a reference for the study of the nonlinear regime of the flow dynamics.

\section*{Appendix}

In this Appendix, we numerically study the temporal evolution and the resultant two-particle correlations, of the peripheral tube model by using MUSIC~\cite{hydro-music-01,hydro-music-02,hydro-music-03}.
As a state-of-the-art 3+1D relativistic second-order viscous hydrodynamical model, MUSIC simulates the evolution of the plasma formed in heavy ion collisions.
The algorithm of the code is based on the fixed grid finite difference method.
The spatial derivatives of the stress-energy tensor are evaluated by using the Kurganov-Tadmor method~\cite{finite-difference-algorithm-KT-01} regarding cell average, while the temporal evolution employs the Runge-Kutta method.
The freeze-out surface is handled by the Cornelius method~\cite{hydro-fz-02}.
Both the thermal spectra of hadrons and the decay of unstable resonance are implemented by phase space integration.

\begin{figure}[ht]
\begin{tabular}{c}
  \includegraphics[width=9.5cm]{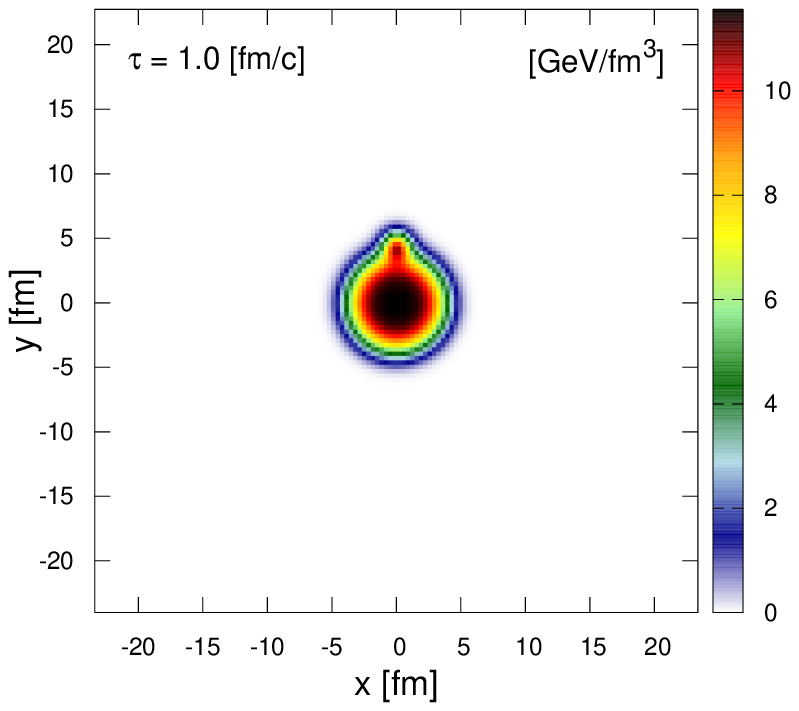}
    \vspace*{-0.7cm}
  \\
  \includegraphics[width=9.5cm]{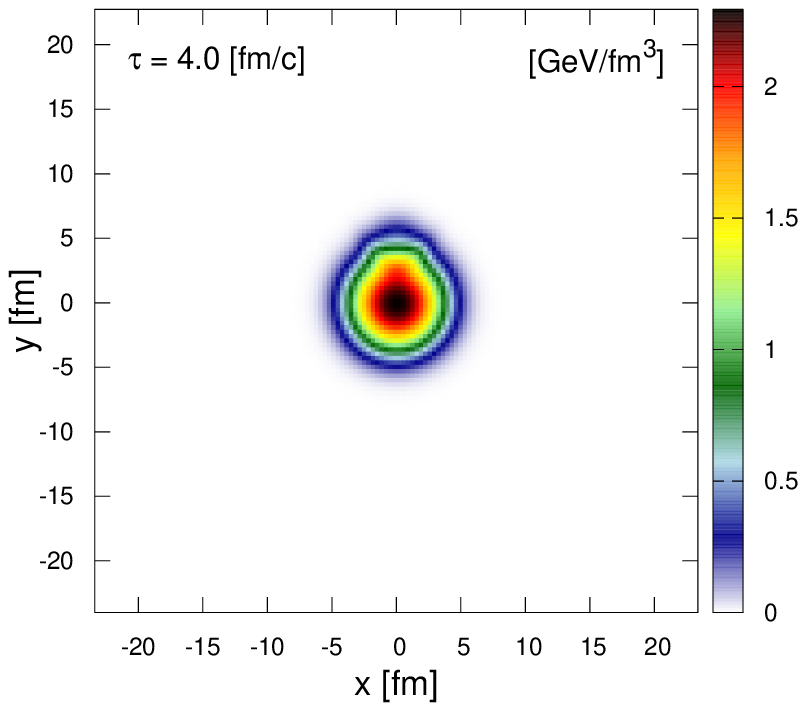}
    \vspace*{-0.7cm}
  \\
  \includegraphics[width=9.5cm]{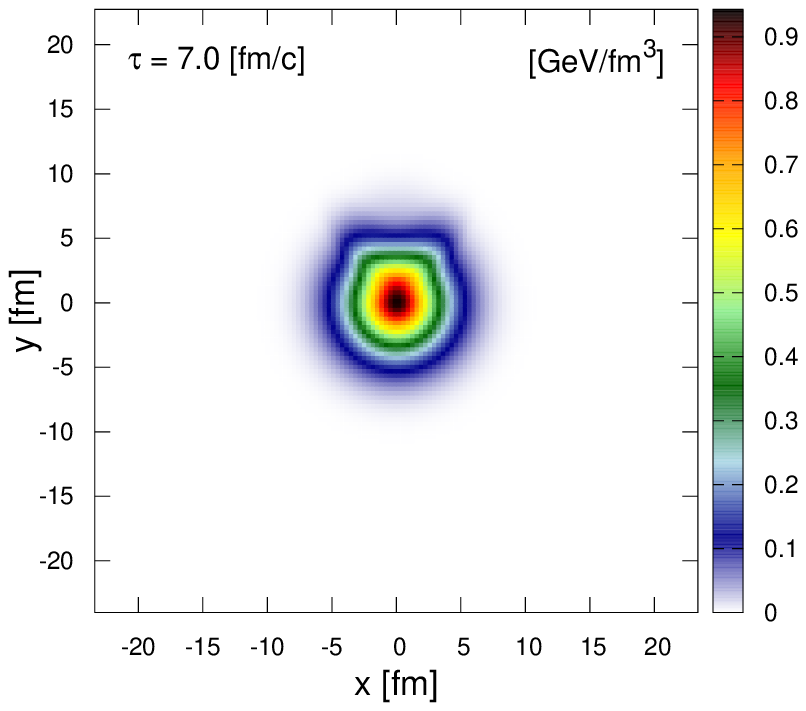}
    \vspace*{-0.7cm}
\end{tabular}
   \vspace*{-0.0 cm}

  \caption{\label{music-evo-vis} Temporal evolution obtained by viscous hydrodynamics with $\eta/s=0.12$ using MUSIC for the IC given in Eq.~\eqref{par}. }
\end{figure}

\begin{figure}[ht]
 \begin{center}
  \includegraphics[width=8.cm]{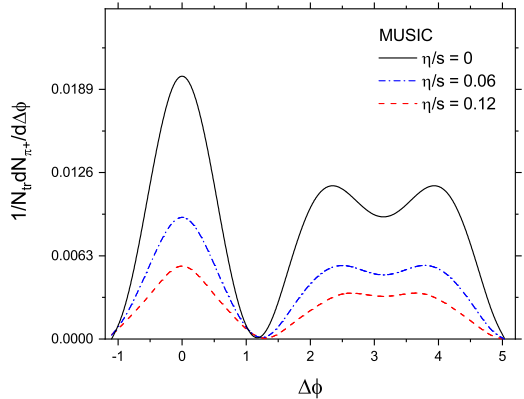}
  \caption{\label{music-2p-corr} The resultant two particle correlations for the IC given in Eq.~\eqref{par} evaluated by using MUSIC.
  The calculations are carried out for $\eta/s=0, 0.06$, and $0.12$ and for the momentum intervals $1.32 < p^A_T < 2.25$ GeV and $2.44 < p^T_T < 3.0$ Gev.
  The resultant correlations have been shifted to attain zero at their minima.}
 \end{center}
\end{figure}

In this Appendix, the calculations are carried out by using MUSIC 3.0, released recently\footnote{http://www.physics.mcgill.ca/music/}.
We use the IC given in Eq.~\eqref{par}, a lattice QCD-based EoS with a phenomenological critical point~\cite{eos-latt-18}, and Cooper-Frye freeze-out.
The code is run in 2D mode with longitudinal invariance, and the size of the arena is chosen to be 40 fm $\times$ 40 fm with 201 grid points in each direction.
The results are obtained by using different strength of shear viscosity $\eta/s$.
The temporal evolution of the IC given by Eq.~\eqref{par} is obtained and shown in FIG.~\ref{music-evo-vis}. 
It is observed that the background is deflected into two well-defined directions.
Subsequently, as shown in FIG.~\ref{music-2p-corr}, the three-ridge structure is formed in the resultant two-particle correlations.
It is noted that the overall strength of the correlation is suppressed, as $\eta/s$ increases, while the ridge structure remains visible.

\begin{acknowledgments}
We are thankful for fruitful discussions with Philipe Mota, Chong Ye, and Dan Wen.
We are indebted to Klaus Werner and his Nantes group for the collaboration and the offer to use NEXUS and EPOS.
We also wish to express our gratitude to Chun Shen, Matthew Luzum, and Gabriel Denicol for the aids in running MUSIC. 
Some figures in the paper are reproduced under the Creative Commons License CC BY-NC-ND\footnote{https://creativecommons.org/licenses/by-nc-nd/4.0/}, proper credit and citation to the original work are given in the text.
We gratefully acknowledge the financial support from
Funda\c{c}\~ao de Amparo \`a Pesquisa do Estado de S\~ao Paulo (FAPESP),
Funda\c{c}\~ao de Amparo \`a Pesquisa do Estado do Rio de Janeiro (FAPERJ),
Conselho Nacional de Desenvolvimento Cient\'{\i}fico e Tecnol\'ogico (CNPq),
Coordena\c{c}\~ao de Aperfei\c{c}oamento de Pessoal de N\'ivel Superior (CAPES),
and National Natural Science Foundation of China (NNSFC).
\end{acknowledgments}

\bibliographystyle{h-physrev}
%\bibliography{references_hama}{}
%\bibliography{references_hama,references_qian}
\bibliography{references_hama,references_qian}

\end{document}